%% file: main.tex
\documentclass[aps,prc,reprint,superscriptaddress,twocolumn]{revtex4-2}
\usepackage{amsfonts}
\usepackage{textcomp}
\usepackage{amsmath}
\usepackage{amssymb}
\usepackage{bm}
\usepackage{xcolor}
\usepackage[hyperindex,breaklinks]{hyperref}
\hypersetup{
    colorlinks,
    linkcolor={red!50!black},
    citecolor={blue!50!black},
    urlcolor={blue!80!black}
}
\usepackage[export]{adjustbox}
\usepackage[section]{placeins}
\usepackage{csquotes}
\usepackage{graphicx}
\usepackage{url}
\usepackage[utf8]{inputenc}
\usepackage{comment}
\usepackage{lineno}
\usepackage{orcidlink}
\usepackage{lipsum}
\usepackage{gensymb}
\usepackage[normalem]{ulem}
\usepackage{rotating}
\usepackage{multirow}
\usepackage{booktabs}
\usepackage{placeins}
\usepackage{float}
\usepackage{tikz}

\newcommand*\fullcirc{\tikz\fill (0,0) circle (0.6ex);} 
\newcommand*\mycbox{\tikz\fill (0,0) rectangle (1.2ex,1.2ex);}

\setcitestyle{square,comma, numbers, sort&compress}

\begin{document}


\title{Evolution of high-spin states in neutron-rich $^{195-202}$Au isotopes \protect\\approaching the $N=126$ shell closure}

\input{AuthorList}

\date{\today}

\begin{abstract}

    \textbf{Background: } The nuclear structure in the region southwest of the doubly magic $^{208}$Pb is important to benchmark theoretical models relevant for neutron-rich heavy element formation and for the origin of the $A\approx195$ peak in the mass abundance distribution. 
    Although neutron-rich Pb, Tl, and Hg ($Z=80-82$) isotopes are relatively well studied, experimental data for neutron-rich Au isotopes ($Z=79$) remain largely limited due to experimental challenges.
    
    \textbf{Purpose: } The purpose of this work is to investigate the evolution of the high-spin structure in neutron-rich Au isotopes near the $N=126$ shell closure and to characterize the underlying shell-model configurations, with a particular focus on the roles of the high-$j$ unique-parity $\pi h_{11/2}$ and $\nu i_{13/2}$ orbitals.
    
    \textbf{Methods: } Neutron-rich $^{195-202}$Au isotopes were produced using multi-nucleon transfer reactions of $^{198}$Pt($^{136}$Xe, $^{x}$I)$^{y}$Au at a beam energy of 7 MeV/u.  
    Prompt and delayed $\gamma$-ray spectroscopy of isotopically identified reaction products was performed with a unique experimental setup combining the VAMOS++ large acceptance magnetic spectrometer, the AGATA HPGe $\gamma$-ray tracking array, and the CATLIFE detection system.
    
    \textbf{Results: } In even-$A$ $^{196,\,198,\,200,\,202}$Au isotopes, prompt $\gamma$-ray transitions built on the $12^-$ isomer have been observed for the first time. The odd-$J$ band built on the $11^-$ state vanishes for $A\ge196$. 
    In odd-$A$ $^{195,\,197,\,199,\,201}$Au isotopes, new prompt and delayed $\gamma$-ray transitions above the $11/2^-$ isomer were identified, confirming and extending the known level schemes. New isomeric states assigned to ($25/2^+$) have been established in $^{199}$Au and $^{201}$Au.  
    The minima in the excitation energies of the bands as a function of $N$ built on the $11/2^-$ and $12^-$ isomers occur at $^{198}$Au ($N=119$), showing a dip against an otherwise smooth evolution. 
    
    \textbf{Conclusions: } Large-Scale Shell-Model calculations based on the Kuo-Herling hole effective interaction show an overall good agreement with experimental level schemes. However, the present calculations do not reproduce the minima in excitation energies at $N=119$ observed in both Au and Hg isotopes. A systematic analysis of $\pi h_{11/2}$ and $\nu i_{13/2}$ hole occupancies, together with spin contributions of protons and neutrons, sheds light on the role of these high-$j$ unique-parity orbitals in the generation of high-spin states. Part of the results of this work have been reported in a letter publication~\cite{Cho2026_PLB}.

\end{abstract}

\maketitle



\section{Introduction}
\label{introduction}

The southwest region of the doubly-magic nucleus $^{208}$Pb ($Z=82,\,N=126$) serves as a key testing ground for nuclear structure models for neutron-rich heavy nuclei. Spectroscopy of excited states in these rare isotopes refines our understanding of the shell structure and effective interactions used for Large-Scale Shell-Model (LSSM) calculations \cite{herling1972, warburton1991, yuan2022, Steer2011, PODOLYAK2009}. In particular, high-spin excitations, arising from a few proton holes below $Z=82$ and neutron holes below $N=126$, are mainly characterized by high-$j$ unique-parity orbitals $\pi h_{11/2}$ and $\nu i_{13/2}$. These orbitals drive the systematic occurrence of long-lived isomers in odd-$A$ (odd-even) and even-$A$ (odd-odd) Au isotopes and shape the structure built on these isomeric states \cite{Dracoulis2013_isomer}. Therefore, a detailed knowledge of the level schemes above these isomers is important to elucidate the role of the high-$j$ unique-parity orbitals in generating the high-spin structure. 
Furthermore, these spectroscopic data can provide valuable inputs for constraining the astrophysical r-process \cite{Burbidge1957}, such as theoretical predictions of (first-forbidden) $\beta$-decay rates \cite{Mumpower2016, Mumpower2024, kumar2024, Suzuki2012}. 

The odd-$A$ Au ($Z=79$) isotopes, with three proton holes relative to $Z=82$, present various structures associated with different proton single-particle configurations (See Fig.~1 of Ref.~\cite{Venhart2017}). The $3/2^+$ ground state of the isotopes for $A\ge191$ arises from the $\pi d_{3/2}$ orbital. A low-lying $11/2^-$ state involving a hole in the $\pi h_{11/2}$ orbital, often isomeric, has been observed in many of these isotopes. In even-$A$ isotopes, the coupling of this $11/2^-$ state to a $i_{13/2}$ neutron hole produces the isomer with a spin-parity of either $11^-$ (for $A\le194$) or $12^-$ (for $A\ge196$). The collective features of these structures have been interpreted using particle-plus-rotor models~\cite{gono1977, gono1979, TJOM1974, kolscbach1985, neskakis1979, neskakis1982, Bucurescu2017, Chen2006}, and more recently, through cranking shell-model calculations that highlight the role of $\pi h_{11/2}$ and $\nu i_{13/2}$ orbitals in driving deformation as well as rotation alignment of neutron pairs~\cite{Gueorguieva2003, Oktem2007, WANG2012, Gueorguieva2004, oktem2012, gao2012}. In $^{205}$Au, several levels above the $11/2^-$ isomer were well reproduced by the spherical LSSM calculations based on a three-proton hole configuration \cite{Steer2011, PODOLYAK2009}.  

Heavy-ion fusion-evaporation reactions have mainly been used to study high-spin states in neutron-deficient Au isotopes up to $A=195$. In odd-$A$ $^{189-195}$Au isotopes, band structures built on the $11/2^-$ isomer have been consistently identified ~\cite{gono1977, gono1979, TJOM1974, kolscbach1985, Gueorguieva2003, Oktem2007, WANG2012}. Similarly, studies of even-$A$ $^{190-194}$Au isotopes have revealed band structures above the $11^-$ isomer~\cite{neskakis1979, neskakis1982, Gueorguieva2004, oktem2012, gao2012}. 
Neutron-rich Au nuclei beyond $A=195$ have been investigated using different methods. The $11/2^-$ isomers of $^{197}$Au and $^{199}$Au were studied via neutron capture reactions \cite{Fotiades2005, miyahara2004} or few-nucleon transfer reactions \cite{MUNGER1978}. Similarly, the $11/2^-$ isomers of $^{201}$Au and $^{203}$Au \cite{flynn1981}, along with the $12^-$ isomers of $^{196}$Au \cite{WAPSTRA1967}, $^{198}$Au \cite{Cunnane1972_198Au}, and $^{200}$Au \cite{Cunnane1972_200Au}, have also been identified using few-nucleon transfer reactions. 
The $\beta$-decay study of the ground states of  $^{201}$Au and $^{202}$Au has been performed using spallation reactions on a UCx target \cite{Bowry2011, BARZAKH2022}. 
These approaches have been currently limited to probing only the low-spin states below the isomeric states.
Delayed $\gamma$-ray spectroscopy following fragmentation of a $^{208}$Pb beam was used to explore high-spin states in $^{201}$Au and $^{205}$Au. For $^{201}$Au, two $\gamma$ rays cascading from the 5.6 $\mu$s isomeric state to the $11/2^-$ isomer were observed \cite{Steer2011}. In $^{205}$Au, several levels above the $11/2^-$ isomer were experimentally established \cite{Steer2011, PODOLYAK2009}. However, systematic studies of high-spin structures have been limited, and consequently spectroscopic data on high-spin states in neutron-rich Au isotopes near $N=126$ remain very sparse~\cite{Podolyak2023}.

Multi-Nucleon Transfer (MNT) reactions have revived interest as a powerful tool for producing neutron-rich nuclei near the $N=126$ shell \cite{Dasso1994, zagrebaev2008}. Earlier studies have shown that the $^{136}$Xe $+$ $^{198}$Pt MNT reaction is particularly effective for the population of neutron-rich nuclei in this region \cite{watanabe2015, barrett2015, desai2019}. In addition, heavy-ion-induced MNT reactions are known to favor the population of high-spin yrast states~\cite{Cocks1997, Kraus1988, GUID85}. Various nuclei in the Re-Pt region were recently studied by $\gamma$-ray spectroscopy \cite{WATANABE2021, HWatanabe2018, John2017, John2014} and laser spectroscopy \cite{WATANABE2025, Hirayama2024, Hirayama2022, Choi2020} following MNT reactions. A few in-beam $\gamma$-ray spectroscopy investigations of $^{197}$Au \cite{Wheldon2006} and $^{195}$Au \cite{DRACOULIS2013} were carried out with high-fold $\gamma$-$\gamma$ coincidence techniques.
However, the difficulty in isotopic identification of MNT products has left a majority of Au isotopes with $A\ge196$ largely unexplored. The scarce experimental data, together with improved spectroscopic capability, motivated a systematic exploration of high-spin states in neutron-rich Au isotopes via MNT reactions. 

\begin{figure*}[t]
    \centering
    \includegraphics[width=0.95\textwidth]{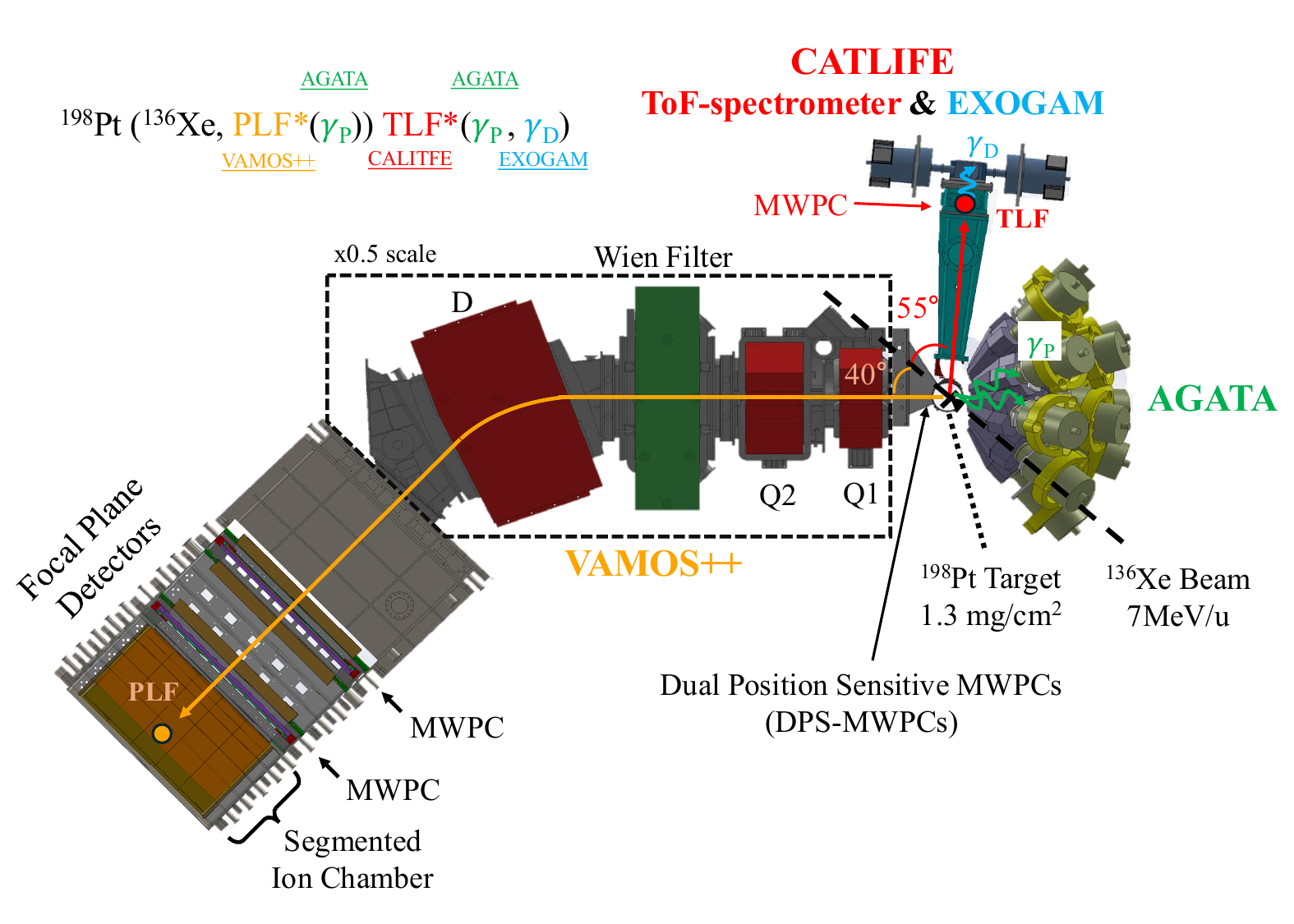}
    \caption{
    Schematic top view of the experimental setup showing the various components of VAMOS++ along with the Complementary Arm for the Target-Like FragmEnts (CATLIFE) and the AGATA $\gamma$-ray detector array.
    }
    \label{fig:Expt_Setup}
\end{figure*}

The present work reports on high-spin spectroscopy of neutron-rich Au isotopes ($^{195-202}$Au) using MNT reactions. Combining prompt-delayed $\gamma$-ray spectroscopy with isotopic identification of the reaction partners, we significantly expand our knowledge of their high-spin excited states. Several new $\gamma$-ray transitions above $11/2^-$ isomers are identified in odd-$A$ $^{195,\,199,\,201}$Au, including the half-life of new isomers in $^{199,\,201}$Au. In even-$A$ $^{196,\,198,\,200,\,202}$Au isotopes, new prompt $\gamma$-ray transitions built upon the $12^-$ isomers are established for the first time. The experimental level schemes are interpreted using LSSM calculations. The evolution of excitation energies for selected states in Au isotopes is discussed in the context of the corresponding states in Hg isotopes and the shell-model calculations. The hole occupancies of high-$j$ orbitals and contributions of proton and neutron to high-spin states are systematically investigated. Part of these results have been reported in a recent related Letter~\cite{Cho2026_PLB}. The present manuscript includes new results including a comprehensive description of the experimental and theoretical analysis for the complete data for $^{195-202}$Au isotopes.


\section{Experimental setup and data analysis}
\label{experiment}

\begin{figure}[t]
    \centering
    \includegraphics[width=0.48\textwidth]{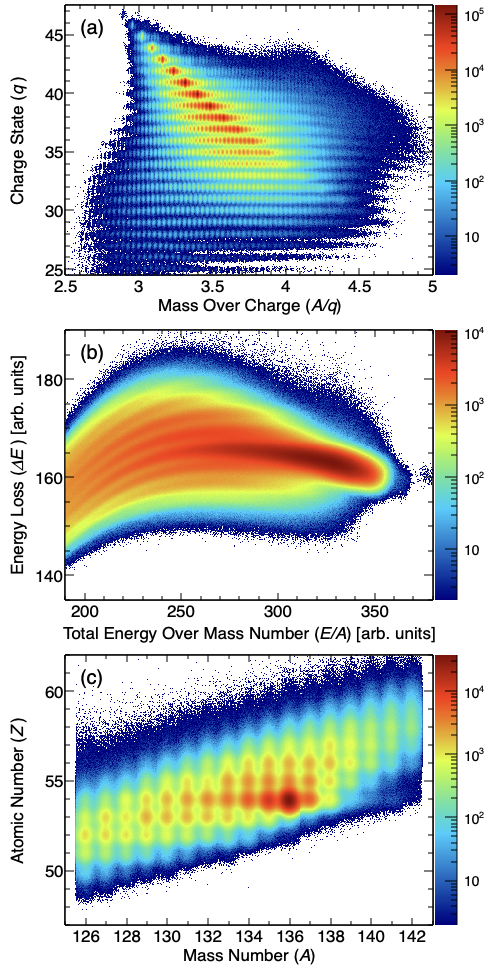}
    \caption{
     Particle identification of the Projectile-Like Fragments (see text). 
     (a) The distribution of charge state ($q$) as a function of mass number over charge ($A/q$).
     (b) The distribution of energy loss ($\Delta E$) as a function of total energy over mass number ($E/A$) for $126\le A\le 142$.
     (c) The distribution of atomic number ($Z$) as a function of mass number ($A$) of the ions detected by VAMOS++.
     }
    \label{fig:PID}
\end{figure}

Multi-Nucleon Transfer (MNT) reactions at energies above the Coulomb barrier were used to produce neutron-rich nuclei ranging from W to Pb in the southwest region of $^{208}$Pb. A 7 MeV/u $^{136}$Xe beam at the GANIL facility, with a typical beam intensity of $\approx$1.5 pnA, impinged on a 1.3 mg/cm$^2$-thick $^{198}$Pt target~\cite{DataE806}. Fig.~\ref{fig:Expt_Setup} shows a schematic view of the experimental setup. The VAMOS++ large acceptance magnetic spectrometer \cite{REJMUND_NIMA2011}, placed $40\degree$ near the grazing angle with respect to the beam axis, was used to detect the Projectile-Like Fragments (PLFs) produced in MNT reactions. The assembly of two Dual Position-Sensitive Multi-Wire Proportional Counters (DPS-MWPCs)~\cite{VANDEBROUCK_NIMA2016} at the entrance of VAMOS++ provided the start signal for the time-of-flight (ToF) and the angle of the PLFs detected in VAMOS++. The detection system at the VAMOS++ focal plane included two Multi-Wire Proportional Counters (MWPCs) for ion tracking and provided the stop signal for ToF measurements. The magnetic rigidity and flight path of the ions were reconstructed from their measured positions and angles at the focal plane \cite{LEMASSON_NIMA2023}. The suitable combinations of energy loss in an ionization chamber with seven segments were used to measure the kinetic energy and $\Delta E-E$ of the fragments. Event-by-event identification of the charge state ($q$), mass number ($A$), and atomic number ($Z$) of the ions was achieved by multi-parameter analysis using the measured variables \cite{REJMUND_NIMA2011, LEMASSON_NIMA2023}. 
The distribution of $q$ as a function of $A/q$ is shown in Fig.~\ref{fig:PID}(a). A machine learning-based calibration of the kinetic energy measurement was implemented, yielding an improvement in $q$ resolution of about 8\% \cite{CHO2023}. Fig.~\ref{fig:PID}(b) presents the $\Delta E-E$ plot for the selected mass range of 126$\leq A \leq$142. To further improve the $Z$ identification of the PLFs in these reactions where (quasi) elastic scattering dominates, a narrow contour around the calculated centroid for each identified $Z$ in the $\Delta E-E$ was taken for a large number of small $E$ bins and for each of the measured mass numbers. Fig.~\ref{fig:PID}(c) illustrates the particle identification plot for the ions detected in VAMOS++. The distribution of $Z$ is shown as a function of $A$ for the mass range of 126$\leq A \leq$142. 
The achieved resolutions are $\Delta Z/Z \simeq 1/60$, $\Delta A/A \simeq 1/210$, and $\Delta q/q \simeq 1/86$.

The Complementary Arm for Target-LIke FragmEnts (CATLIFE) \cite{NIMB2023_SON}, a newly developed detection system, was employed to aid the identification of Target-Like Fragments (TLFs). It consisted of a large MWPC and an aluminum stopper, followed by four EXOGAM HPGe Clover detectors \cite{SIMPSON_Acta2000} covering the area of the MWPC. The MWPC, positioned 1230 mm from the target, provided the horizontal and vertical positions of the TLFs (see Fig.~\ref{fig:Expt_Setup}). The time-of-flight of TLFs was determined from the time difference between the DPS-MWPC of VAMOS++ and the MWPC in CATLIFE. The latter covered a geometrical acceptance of 46.5 msr, spanning horizontal angles between 51$\degree$ and 59$\degree$ around the grazing of the TLFs. This coverage corresponds to about $58\%$ of the acceptance of VAMOS++ \cite{REJMUND_NIMA2011}. After traversing the CATLIFE’s MWPC, ions were implanted into a 2 mm-thick aluminum stopper. 

\begin{figure}[t]
    \centering
    \includegraphics[width=0.5\textwidth]{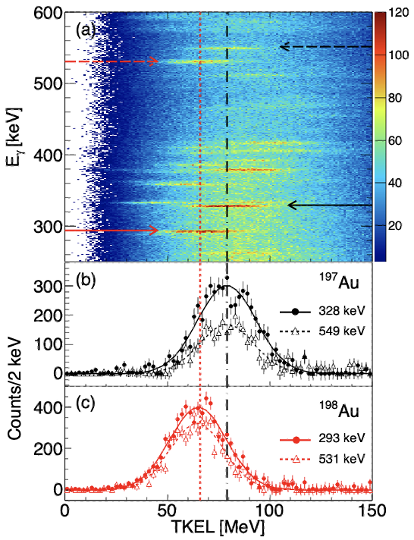}
    \caption{
    (a) The Doppler-corrected prompt $\gamma$-ray spectra of the Au isotopes in coincidence with the $^{131}$I partner as a function of the Total Kinetic Energy Loss (TKEL). The solid (dashed) arrows on the left and right side indicate the newly identified 293 (531) keV transition of $^{198}$Au (red) and the known 328 (549) keV transitions of $^{197}$Au (black), respectively. 
    (b) The TKEL distribution for the 328 keV (filled circle) and 549 keV (open triangle) transitions of $^{197}$Au. The Gaussian functions fitted for the corresponding TKEL distributions are shown as solid and dotted curves, respectively.
    (c) Same as (b) for 293 keV (filled circle) and 531 keV (open triangle) transitions of $^{198}$Au. The Gaussian functions fitted for the corresponding TKEL distributions are shown as solid and dotted curves, respectively.
    The vertical red dotted and black dashed lines represent the optimal TKEL values used for the spectroscopy of the $^{198}$Au and $^{197}$Au, respectively (see text).
    }
    \label{fig:PGxTKEL_Gate131I}
\end{figure}

Direct identification of TLFs is very challenging due to their high ($A_{\text{TLF}},\,Z_{\text{TLF}}$) and low energy ($<$ 1.5 MeV/u).
Instead, the atomic number of TLFs is deduced from the identified PLF partner, assuming a binary reaction with no proton evaporation. Their mass number cannot be determined directly, since neutron evaporation can occur whenever the fragment’s excitation energy exceeds the neutron separation energy. The number of neutrons emitted from a fragment has a statistical nature that depends on the excitation energy (neglecting the relatively small angular momentum dependence). Consequently, TLFs associated with a given PLF species ($A_{\text{PLF}},\,Z_{\text{PLF}}$) generally comprise a mixture of several mass numbers smaller than the binary partner. In the following analysis, the excitation energy of the reaction products is treated in terms of Total Kinetic Energy Loss (TKEL) \cite{Brown1951, Szilner2007}, derived from two-body kinematics. The TKEL represents the difference in total kinetic energy between the initial system and the final fragments in the center-of-mass frame.

The four EXOGAM detectors, positioned 6 mm behind the Al stopper in CATLIFE, were used to detect $\gamma$ rays emitted from the isomeric states of the TLFs. These delayed $\gamma$ rays were recorded within a time window ($\Delta t$) ranging from 0.1 $\mu$s to 100 $\mu$s with respect to the DPS-MWPC of VAMOS++, which allowed half-life measurements of the isomers.

The AGATA $\gamma$-ray tracking array \cite{AKKOYUN_NIMA2012} was used to measure the prompt $\gamma$ rays in coincidence with a PLF detected by the DPS-MWPC at the target position and the MWPC at the VAMOS++ focal plane. The array, comprised of 36 segmented HPGe crystals covering a solid angle of about $1\pi$, was positioned in the backward direction with respect to VAMOS++ at a distance of 235 mm (nominal configuration) \cite{NIMA2017_CLEMENT}. 
A 140 ns time window was applied between the VAMOS++ trigger from DPS-MWPC and AGATA to select prompt $\gamma$-ray events.

The $\gamma$ rays emitted from the isotopically selected TLFs were identified based on their correlation with the TKEL for a given PLF. The Cubix software \cite{Cubix}, a ROOT-based program \cite{ROOT} developed specifically for $\gamma$-ray spectroscopy, was suitably extended and used for $\gamma$-TKEL analysis. The optimal TKEL window for a certain TLF species, associated with a specific PLF partner, was determined from the yield of its characteristic $\gamma$-ray transitions as a function of TKEL.  

The Doppler correction for the prompt $\gamma$ rays of the TLFs was performed using the measured velocity of their partner and the first interaction point of the $\gamma$ rays in AGATA \cite{AKKOYUN_NIMA2012}. The velocity of TLFs is determined on an event-by-event basis under the assumption of a binary reaction. 
Uncertainties in $\gamma$-ray energies are 1 keV unless explicitly mentioned.
Fig.~\ref{fig:PGxTKEL_Gate131I}(a) displays the Doppler-corrected $\gamma$-ray spectrum of Au isotopes as a function of TKEL for the case of the $^{131}$I isotope as the PLF partner. The nascent binary partner of $^{131}$I is $^{203}$Au, assuming no neutron evaporation. The projected TKEL distributions of two known $\gamma$-ray transitions from $^{197}$Au \cite{Wheldon2006}, the $6n$ evaporation channel, are shown in Fig.~\ref{fig:PGxTKEL_Gate131I}(b). The filled circle and open triangle histograms represent the projected TKEL distributions for the 328 and 549 keV transitions, respectively. The TKEL contributions arising from the $\gamma$-ray background were subtracted. Gaussian functions fitted to these distributions are shown as the solid line for the 328 keV transition and the dotted line for the 549 keV transition. Both transitions exhibit a common maximum near 80 MeV (marked by the vertical dashed line), which corresponds to the optimal TKEL value ($<$TKEL$>$) of $^{197}$Au for a PLF partner of $^{131}$I.

\begin{figure}[t]
    \centering
    \includegraphics[width=0.5\textwidth]{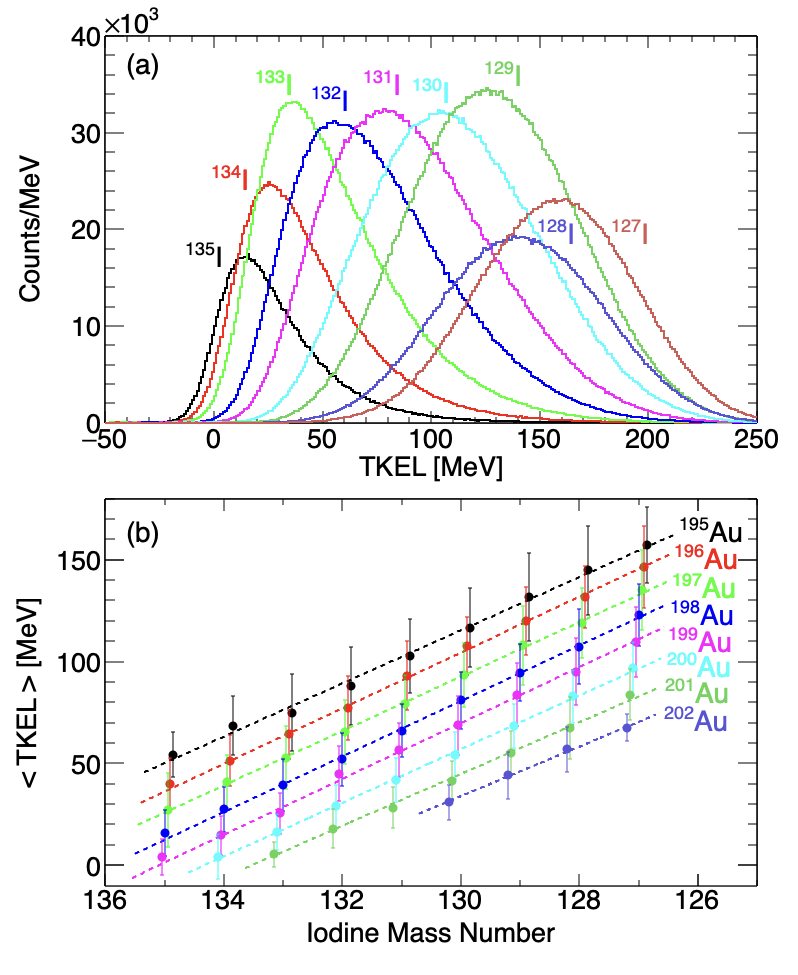}
    \vspace{-8 mm}
    \caption{
    (a) Total kinetic Energy Loss (TKEL) distribution for the $^{127-135}$I ($Z_{\text{PLF}}=53$) isotopes identified in VAMOS++. 
    (b) The evolution of optimal TKEL value ($<$TKEL$>$) for each Au isotope ($195\leq A \leq202$) as a function of iodine mass number. The vertical bars represent the width of the TKEL distribution for each Au isotope (one standard deviation). The dashed lines are drawn to guide the eye.
    }
    \label{fig:TKEL_Syst}
\end{figure}

Fig.~\ref{fig:PGxTKEL_Gate131I}(c) presents the TKEL distributions of two newly observed $\gamma$-ray transitions at 293 keV and 531 keV, both showing maximum yields at a common TKEL value of approximately 67 MeV (marked by the vertical dotted line). Since the mean TKEL value of 67 MeV is lower than that of 80 MeV of $^{197}$Au, these $\gamma$ rays are emitted from a reaction channel with a lower number of neutron evaporation. No other $\gamma$-ray transitions were observed with peak yields between these two TKEL values. Thus, the 293 keV and 531 keV transitions can be assigned to $^{198}$Au. 

Similar analyses can be performed for the delayed $\gamma$ rays using the additional correlation with the half-life of the isomeric states. The prompt and delayed $\gamma$-TKEL matrices for given iodine isotopes were systematically analyzed to identify $\gamma$-ray transitions and $<$TKEL$>$ for each Au isotope. The background-subtracted TKEL distribution of each iodine isotope ($127\le A \le 135$) identified in VAMOS++ is shown in Fig.~\ref{fig:TKEL_Syst}(a). This shows a wide TKEL range for each iodine isotope, implying that several Au isotopes are generally associated with a single iodine isotope due to neutron evaporation. 
Fig.~\ref{fig:TKEL_Syst}(b) illustrates the evolution of $<$TKEL$>$ for the $^{195-202}$Au isotopes as a function of the iodine mass number. The error bars indicate one standard deviation of the TKEL distribution. In general, the $<$TKEL$>$ of a given Au isotope increases linearly as more neutrons are transferred during the reactions (lower iodine mass number). 

The $A_{\text{PLF}},\,Z_{\text{PLF}}$ gated \enquote{singles} $\gamma$-ray spectra of individual Au isotopes for a given iodine isotope were obtained by projection of the $\gamma$-TKEL matrix for the optimal region around $<$TKEL$>$, typically $\pm$ 25 MeV for determining the relative transition intensities (I$_\gamma$) ($\approx\pm$ 15 MeV was used for generating $\gamma$-$\gamma$ matrices). This approach maximizes the yield for a given Au isotope, ensuring that its characteristic $\gamma$ rays dominate the spectrum. To suppress contamination from adjacent Au isotopes, $\gamma$-ray spectra from neighboring TKEL regions were appropriately subtracted.
The total \enquote{singles} $\gamma$-ray spectrum focused on a single Au isotope was constructed by summing the individual \enquote{singles} spectra obtained from each iodine isotope. Relative $\gamma$-ray intensities were determined from the total \enquote{singles} $\gamma$-ray spectra. The $\gamma$-$\gamma$ matrices were generated for both the prompt and delayed transitions. 

\section{Results}
\label{results}

\begin{sidewaysfigure*}[]
	\centering
    \vspace*{9.0 cm}
	\includegraphics[angle=270, width=0.95\textheight]{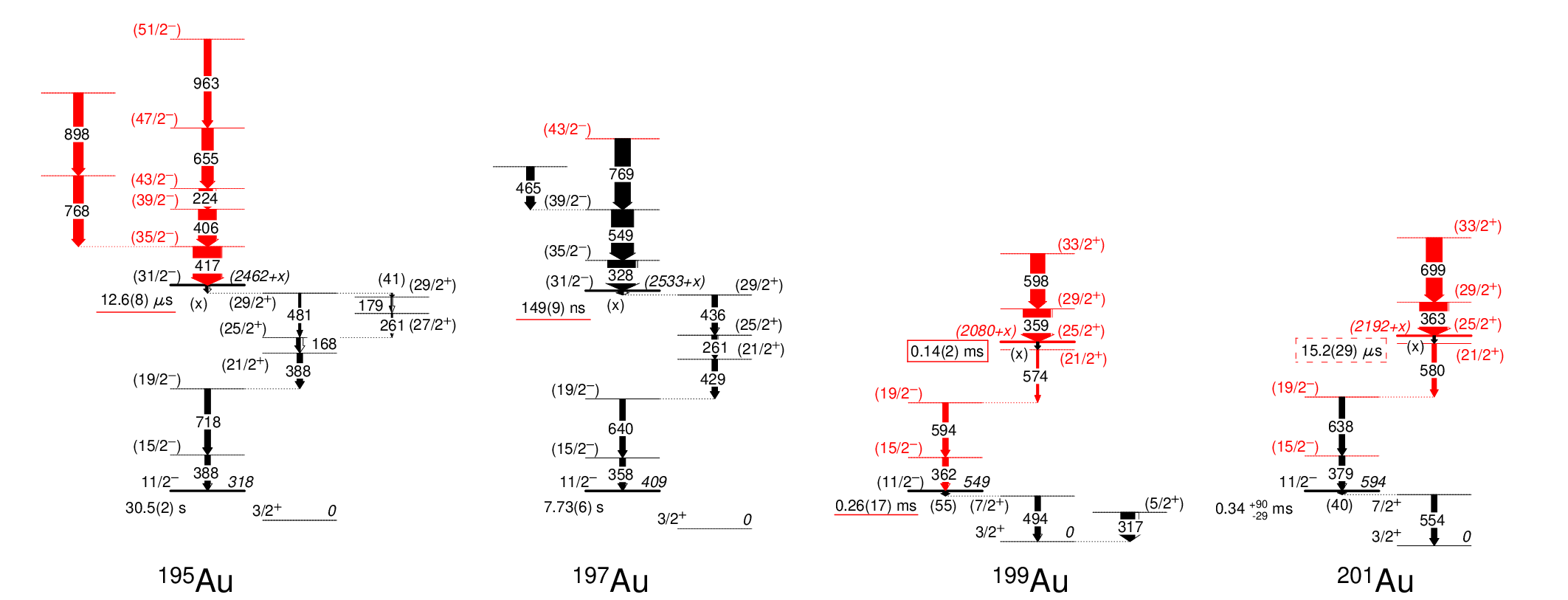}
    \includegraphics[angle=270, width=0.95\textheight]{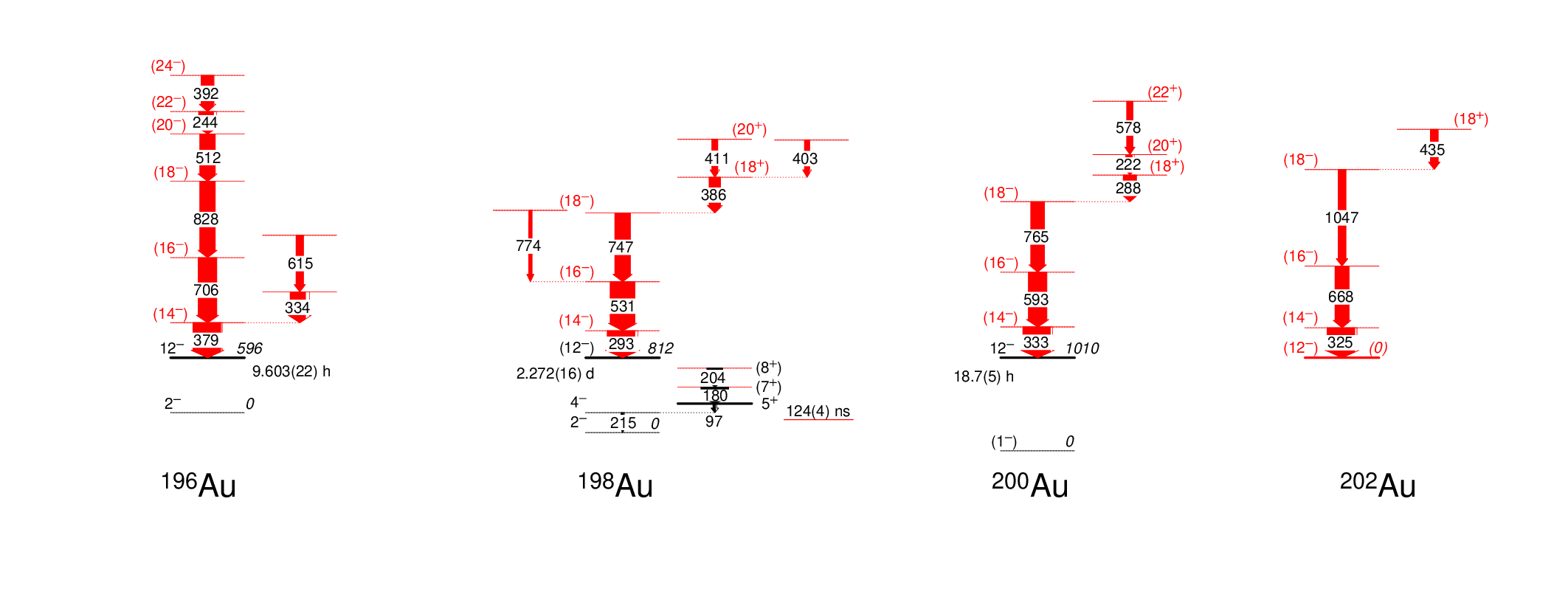}
    \vspace*{-1.0 cm}
	\caption{The level schemes of $^{195-202}$Au isotopes obtained from the present work. The newly observed $\gamma$-ray transitions and placement of the levels are indicated in red. The thickness of the arrows indicates the intensity relative to the lowest transition of the band. 
    Isomeric levels are highlighted with thick horizontal lines. Half-lives previously reported but remeasured here are underlined in red. Newly determined half-lives are enclosed in solid red boxes, while values in dashed red boxes indicate disagreement with earlier measurements. Half-lives not remeasured in this study are taken from Ref.~\cite{ENSDF} and are also shown.
    Energies of unobserved transitions are indicated in parentheses. 
    For the even-$N$(odd-$N$) isotopes, the excitation energies are aligned with respect to the $11/2^-$($12^-$) level.} 
	\label{fig:Au_LS_All}%
\end{sidewaysfigure*}

A summary of the experimental level schemes of the $^{195-202}$Au isotopes obtained in the current work is presented in Fig.~\ref{fig:Au_LS_All}. The level schemes were built by selecting excitation energy gates using TKEL, analyzing $\gamma$-$\gamma$ coincidence relationships, and $\gamma$-ray intensities. The intensities are corrected for internal conversion using BrIcc calculation~\cite{NIMA2008_KIBEDI}. Spin-parity assignments were made based on the available systematics and the Large-Scale Shell-Model (LSSM) calculations discussed in Sec.~\ref{discussion}. A one-component exponential fit ($A e^{-t/T}+C$, where $A, T, C$ are the fitting parameters) was used to obtain the half-life of the isomeric states.

\subsection{Odd-$A$ Au Isotopes}

In the odd-$A$ isotopes $^{195,\,197,\,199,\,201}$Au studied here, we observe a small number of transitions below the 11/2$^{-}$ isomer and high-spin bands built upon this state. Above the 11/2$^{-}$ isomer, additional isomeric states are identified, with half-lives ranging from a few hundred nanoseconds to several hundred microseconds. 

\subsubsection{$^{195}$Au}

\begin{figure}[t]
  \centering
  \includegraphics[width=0.5\textwidth]{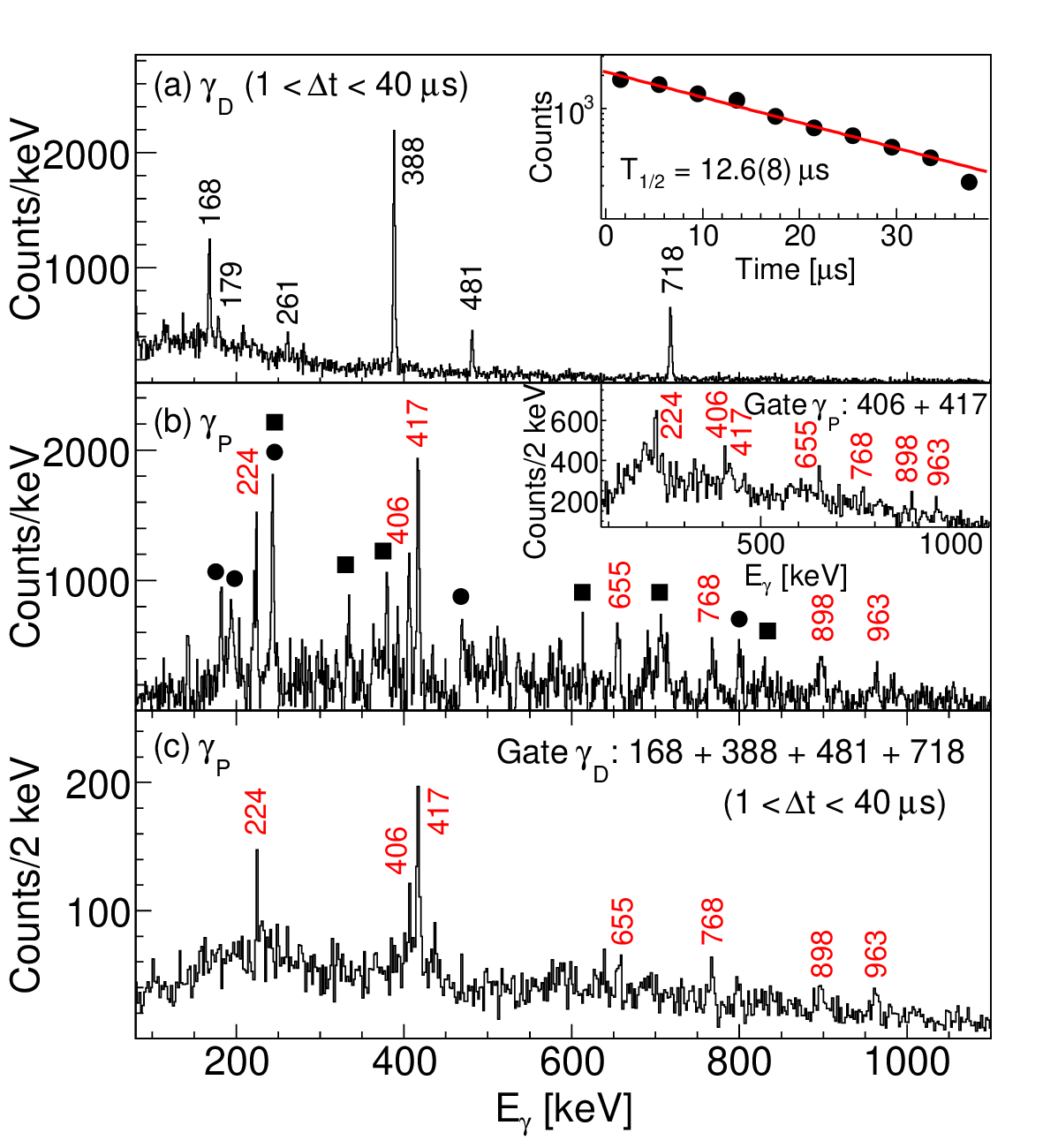}
  \caption{
  $\gamma$-ray spectra for $^{195}$Au obtained with gates on $^{127\text{-}135}$I within the respective optimal TKEL regions for each isotope. 
  The new transitions are labeled in red. 
  (a) The “singles” delayed $\gamma$-ray spectrum ($\gamma_{D}$) for 1 $< \Delta t <$ 40 $\mu$s. The inset shows the decay curve with the exponential fit (single component) with a sum gate on the three isomeric transitions (388, 481, and 718 keV). 
  (b) The “singles” Doppler-corrected prompt $\gamma$-ray spectrum ($\gamma_{P}$). Transitions from $^{194}$Au are marked with \protect \fullcirc $\space$ and from $^{196}$Au with \protect \mycbox. The inset shows the $\gamma_{P}$-$\gamma_{P}$ coincidence spectrum with a sum gate on the 406 and 417 keV transitions.
  (c) The $\gamma_{P}$ spectrum in coincidence with a sum gate on the $\gamma_{D}=168,\,388,\,481,\, \text{and}\,\,718$ keV transitions for $1< \Delta t <40$ $\mu$s.}
  \label{fig:GammaSpec_195Au}
\end{figure}

Previous $\gamma$-ray spectroscopy studies for the high-spin states of $^{195}$Au have been reported in Refs.~\cite{TJOM1974, WANG2012, DRACOULIS2013}. A ($31/2^-$) isomer with $T_{1/2}=12.89(21)\,\mu$s was identified \cite{DRACOULIS2013}. 
In the present work, we confirm the known isomeric transitions at 168, 179, 261, 388, 481, and 718 keV, as shown in the delayed “singles” $\gamma$-ray spectrum (1 $< \Delta t <$ 40 $\mu$s; Fig.~\ref{fig:GammaSpec_195Au}(a)). 
The inset in Fig.~\ref{fig:GammaSpec_195Au}(a) shows that a fit to the decay curve gated on the 388, 481, and 718 keV transitions results in $T_{1/2}=12.6(8)\,\mu$s, matching the literature value. 
The proposed low-energy transition directly depopulating the ($31/2^-$) isomer \cite{DRACOULIS2013} (marked “x” in Fig.~\ref{fig:Au_LS_All}) was not observed in the present experiment. 

Fig.~\ref{fig:GammaSpec_195Au}(b) presents the “singles” Doppler-corrected prompt $\gamma$-ray spectrum ($\gamma_{P}$) of $^{195}$Au. Seven new transitions are observed at 224, 406, 417, 655, 768, 898, and 963 keV and are shown in red. The inset in Fig.~\ref{fig:GammaSpec_195Au}(b) shows a $\gamma_{P}$-$\gamma_{P}$ coincidence spectrum with a sum gate on the 406 keV and 417 keV transitions, confirming their mutual coincidence. Further, as illustrated in Fig. \ref{fig:GammaSpec_195Au}(c), each of these prompt $\gamma$ rays is in coincidence with the delayed $\gamma$ rays that de-excite the ($31/2^-$) isomer. On this basis, the cascade decay chain of the $\gamma_{P}$ transitions is placed above the (31/2$^{-}$) isomer. Since the 768 and 898 keV transitions have nearly equal intensities, it is possible to interchange their ordering.

Table~\ref{tab:GammaSummary_195Au} in the Appendix gives the details of all the transitions of $^{195}$Au observed in this work. 

\subsubsection{$^{197}$Au}

\begin{figure}[t]
  \centering
  \includegraphics[width=0.5\textwidth]{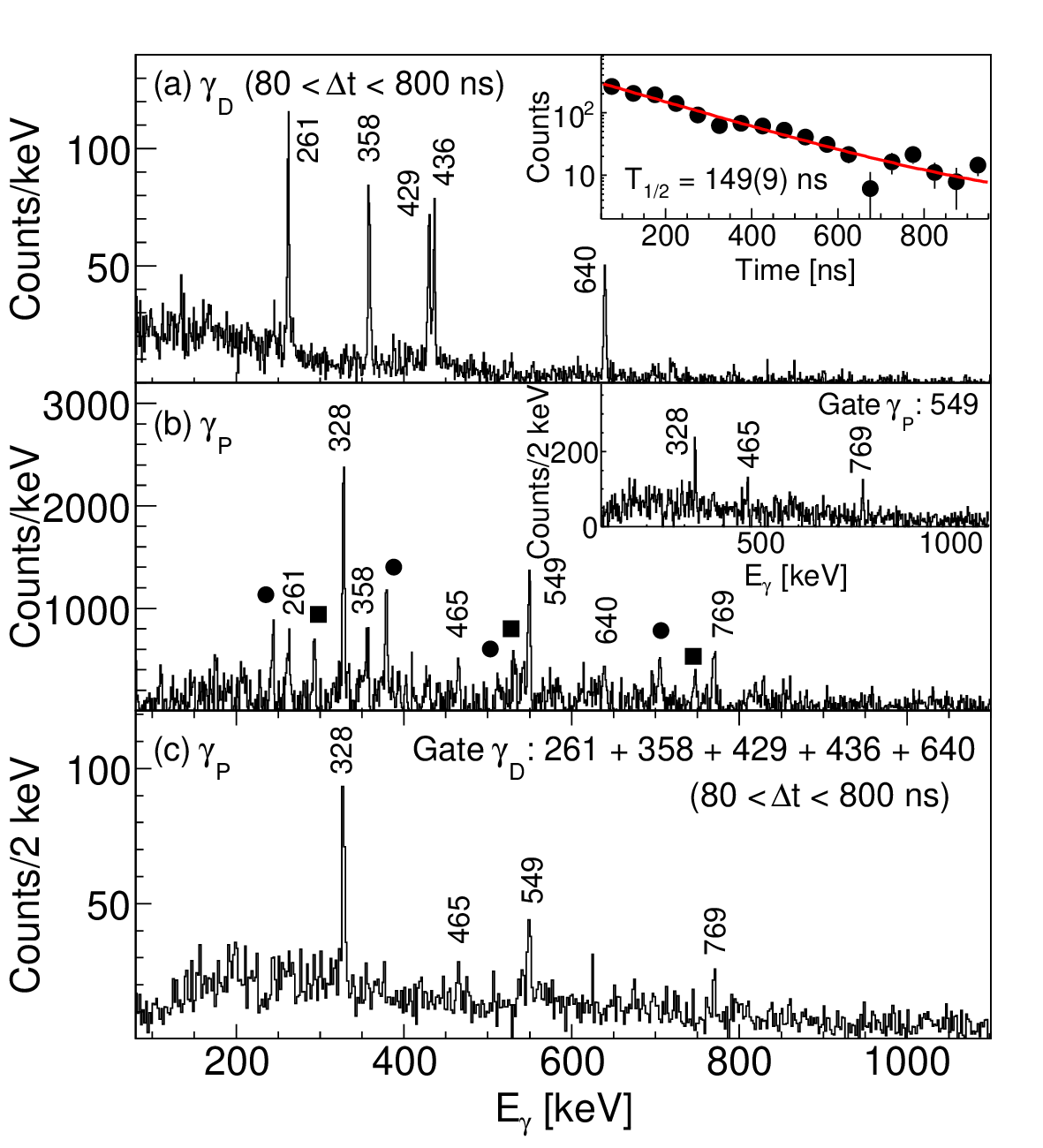}
  \caption{
  $\gamma$-ray spectra for $^{197}$Au obtained with gates on $^{127\text{-}135}$I within the respective optimal TKEL regions for each isotope
  (a) The “singles” delayed $\gamma$-ray spectrum ($\gamma_{D}$) for 80 $< \Delta t <$ 800 ns. The inset shows the decay curve with the exponential fit (single component) for the 436 keV isomeric transition. 
  (b) The “singles” Doppler-corrected prompt $\gamma$-ray spectrum ($\gamma_{P}$). Transitions from $^{196}$Au are marked with \protect \fullcirc $\space$ and from $^{198}$Au with \protect \mycbox. The inset shows the $\gamma_{P}$-$\gamma_{P}$ coincidence spectrum with a gate on the 549 keV transition.   
  (c) The $\gamma_{P}$ spectrum in coincidence with a sum gate on the $\gamma_{D}=261,\,358,\,429,\,436,\,\text{and}\,\,640$ keV transitions for 80 $< \Delta t <$ 800 ns.
  }
  \label{fig:GammaSpec_197Au}
\end{figure}

The ($31/2^-$) isomer in $^{197}$Au, with a half-life of 150(5) ns, was first reported in Ref.~\cite{Wheldon2006}. Fig.~\ref{fig:GammaSpec_197Au}(a) shows the delayed “singles” $\gamma$-ray spectrum for 80 $< \Delta t <$ 800 ns, in which the known isomeric transitions at 261, 358, 429, 436, and 640 keV are clearly observed. The inset in Fig.~\ref{fig:GammaSpec_197Au}(a) presents the decay spectrum of the 436 keV transition, from which a half-life of 149(9) ns is extracted, in good agreement with earlier results~\cite{Wheldon2006}. In analogy to $^{195}$Au \cite{DRACOULIS2013}, the unobserved low-energy E1 transition, labeled “x” in Fig.~\ref{fig:Au_LS_All}, is attributed to direct decay of the ($31/2^-$) isomer to the ($29/2^+$) state.

In Ref.~\cite{Wheldon2006}, several transitions above the ($31/2^-$) isomer of $^{197}$Au were also reported. Among these transitions, the 328, 465, 549, and 769 keV transitions are observed in the “singles” Doppler-corrected prompt $\gamma$-ray spectrum ($\gamma_{P}$) of $^{197}$Au, as shown in Fig.~\ref{fig:GammaSpec_197Au}(b). 
These prompt transitions are in coincidence with the delayed $\gamma$ rays from the decay of the ($31/2^-$) isomer (See Fig.~\ref{fig:GammaSpec_197Au}(c)). 

The inset in Fig.~\ref{fig:GammaSpec_197Au}(b) presents the $\gamma_{P}$-$\gamma_{P}$ coincidence spectrum gated on the 549 keV transition, confirming its coincidence with the 769 keV transition. This places the 769 keV transition above the 549 keV transition, in contrast to the earlier level scheme of Ref.~\cite{Wheldon2006}, which assigns the 769 keV $\gamma$-ray as a direct feeding transition of the ($35/2^-$) state (See Fig.~\ref{fig:Au_LS_All}).
ß
Table~\ref{tab:GammaSummary_197Au} in the Appendix gives the details of all the transitions of $^{197}$Au observed in this work. 

\subsubsection{$^{199}$Au}

\begin{figure}[t]
  \centering
  \includegraphics[width=0.5\textwidth]{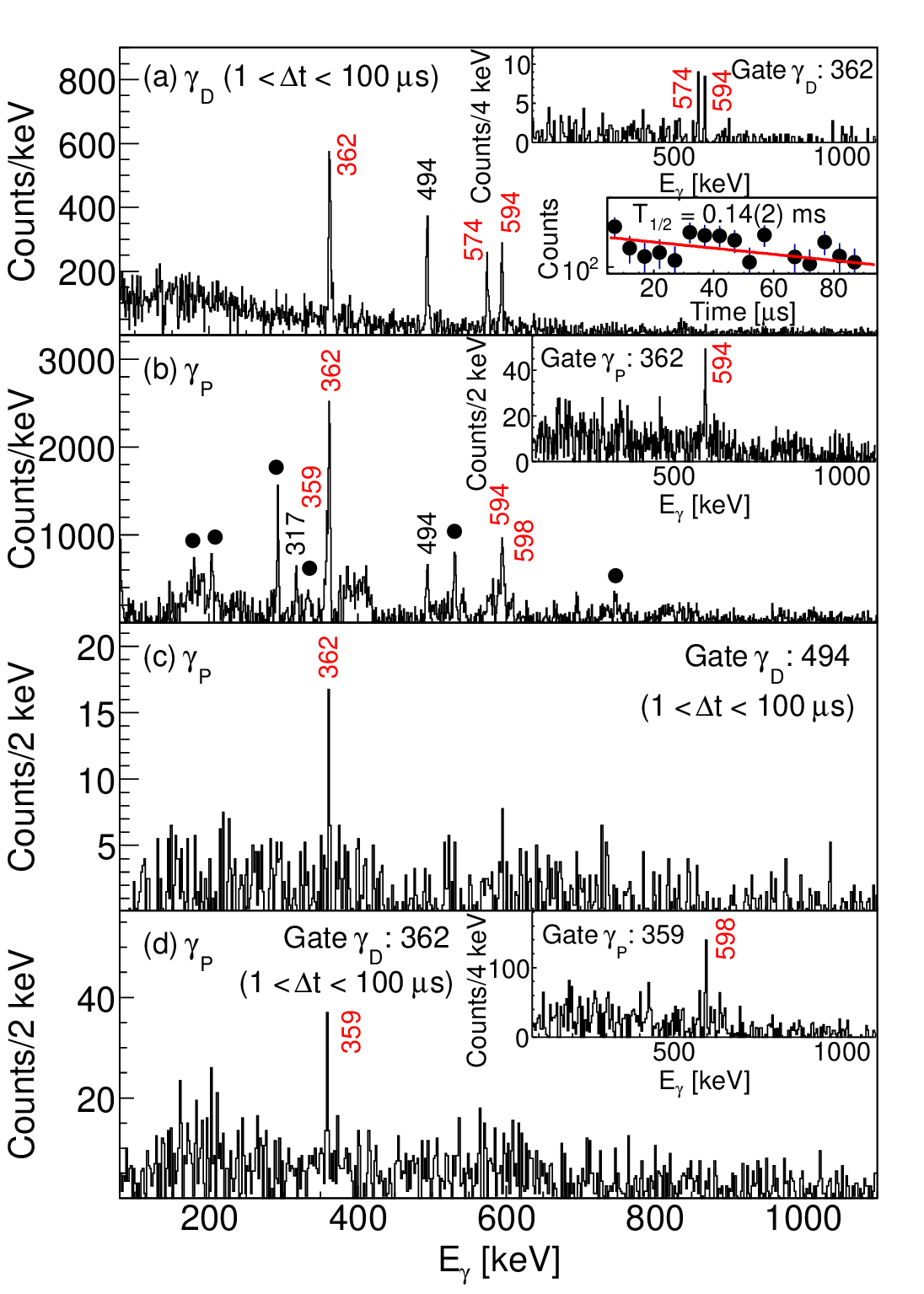}
  \caption{
  $\gamma$-ray spectra for $^{199}$Au obtained with gates on $^{127\text{-}135}$I within the respective optimal TKEL regions for each isotope. 
  The new transitions are labeled in red. 
  (a) The “singles” delayed $\gamma$-ray spectrum ($\gamma_{D}$) for 1 $< \Delta t <$ 100 $\mu$s. 
  The upper inset shows the delayed $\gamma$-$\gamma$ coincidence spectrum ($\gamma_D$-$\gamma_D$) with a gate on the 362 keV transition. 
  The lower inset presents the decay curve along with the fit with a sum gate on the 362 and 574 keV transitions.  
  (b) The “singles” Doppler-corrected prompt $\gamma$-ray spectrum ($\gamma_{P}$). Transitions from $^{198}$Au are marked with \protect \fullcirc. The inset shows the $\gamma_{P}$-$\gamma_{P}$ coincidence spectrum with a gate on the 362 keV transition. 
  (c) The $\gamma_{P}$ spectrum in coincidence with the $\gamma_{D}=494$ keV transition for 1 $< \Delta t <$ 100 $\mu$s. 
  (d) The $\gamma_{P}$ spectrum in coincidence with the $\gamma_{D}=362$ keV transition for 1 $< \Delta t <$ 100 $\mu$s. The inset shows the Doppler-corrected $\gamma_P$-$\gamma_P$ coincidence spectrum with a gate on the 359 keV transition.
  }
  \label{fig:GammaSpec_199Au}
\end{figure}

The low-spin states and their decay transitions in $^{199}$Au isotope were previously reported~\cite{ENSDF}. 
The 55 keV $M2$ transition that directly de-excites the $(11/2^-)$ isomer \cite{BORGGREEN1968} was not detected in this experiment due to its high conversion coefficient ($\alpha_{\text{tot}}=283$ calculated using  BrIcc~\cite{NIMA2008_KIBEDI}) and low detection efficiency. However, the known 494 keV isomeric transition is observed in our “singles” delayed $\gamma$-ray spectrum ($\gamma_{D}$) in the time range of 1 $< \Delta t <$ 100 $\mu$s (See Fig.~\ref{fig:GammaSpec_199Au}(a)). The extracted half-life of 0.26(17) ms (not shown in the figure) has considerable uncertainty due to low statistics and relatively long half-life, but is in agreement within the error of the literature value ($T_{1/2}=0.44(3)$ ms) \cite{BORGGREEN1968}. 
The 494 keV transition and another known transition at 317 keV \cite{OKANO1969} are also clearly observed in the “singles” Doppler-corrected prompt $\gamma$-ray spectrum ($\gamma_{P}$), as shown in Fig.~\ref{fig:GammaSpec_199Au}(b).

The newly observed isomeric transitions at 362, 574, and 594 keV in the $\gamma_{D}$ spectrum are shown in red in Fig.~\ref{fig:GammaSpec_199Au}(a). As shown in the upper inset in Fig.~\ref{fig:GammaSpec_199Au}(a), these transitions are in mutual coincidence. The lower inset of Fig.~\ref{fig:GammaSpec_199Au}(a) shows the decay curve with a sum gate on the 362 and 574 keV transitions. The half-life of the new isomer is determined as $T_{1/2}$ = 0.14(2) ms. The 362 and 594 keV transitions are also observed in the $\gamma_{P}$ spectrum (Fig.~\ref{fig:GammaSpec_199Au}(b)), and their coincidence is shown in the inset. In addition, the prompt 362 keV transition is in coincidence with the delayed 494 keV transition, as shown in Fig.~\ref{fig:GammaSpec_199Au}(c). Therefore, a decay chain of the 362, 594, and 574 keV transitions is placed above the ($11/2^-$) isomeric state.

In analogy to the neighboring $^{197}$Au isotope, which shows a similar structure, the 574 keV transition is assumed to be an E1 decay from the $(21/2^+$) state. The long-lived ($25/2^+$) state is then assigned as an isomer that feeds the $(21/2^+)$ state via an unobserved low-energy E2 transition (marked “x” in Fig.~\ref{fig:Au_LS_All}) based on the shell-model calculations and the measured half-life. If the 574 keV transition directly depopulates the isomer, its relative intensity compared to the 362 keV and 594 keV transitions suggests a M4 character through internal-conversion analysis. However, such a high multipolarity contradicts its 0.14(2) ms lifetime and has no precedent in this mass region.

Two new prompt $\gamma$ rays, 359 and 598 keV transitions, are observed in the $\gamma_{P}$ spectrum (Fig.~\ref{fig:GammaSpec_199Au}(b)). The 359 keV transition is in coincidence with the delayed 362 keV $\gamma$-ray, as shown in Fig.~\ref{fig:GammaSpec_199Au}(d). The inset in Fig.~\ref{fig:GammaSpec_199Au}(d) also shows the coincidence between the two prompt transitions at 359 and 598 keV. Based on these correlations, the cascade decay of the 359 and 598 keV transitions placed above the ($25/2^+$) isomeric state. 

Table~\ref{tab:GammaSummary_199Au} in the Appendix gives the details of all the transitions of $^{199}$Au observed in this work.

\subsubsection{$^{201}$Au}

\begin{figure}[t]
  \centering
  \includegraphics[width=0.5\textwidth]{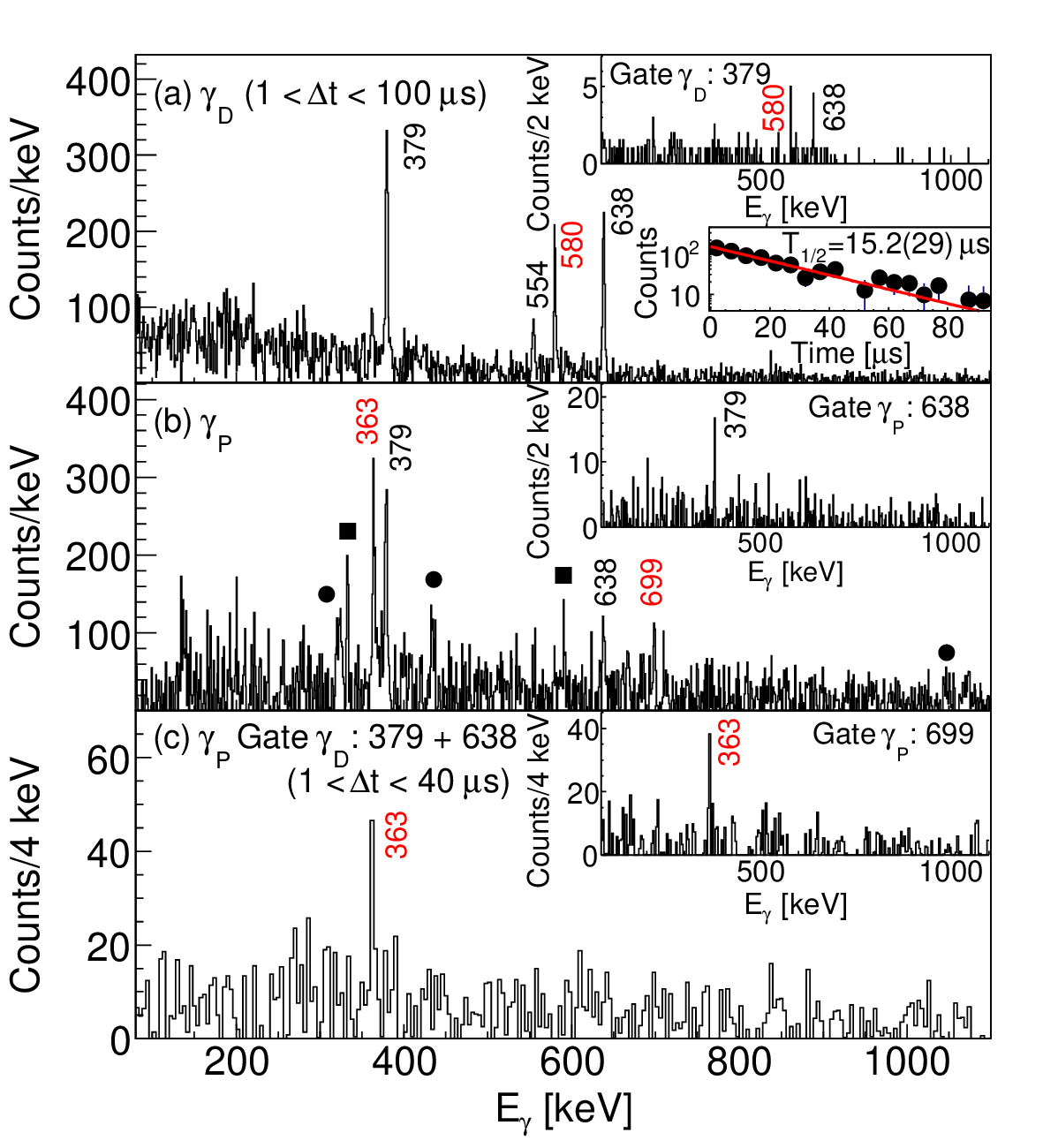}
  \caption{
  $\gamma$-ray spectra for $^{201}$Au obtained with gates on $^{127\text{-}135}$I within the respective optimal TKEL regions for each isotope. 
  The new transitions are labeled in red.   
  (a) The “singles” delayed $\gamma$-ray spectrum ($\gamma_{D}$) for 1 $< \Delta t <$ 100 $\mu$s. 
  The upper inset shows the delayed $\gamma$-$\gamma$ coincidence spectrum ($\gamma_D$-$\gamma_D$) with a gate on the 379 keV transition. 
  The lower inset presents the decay curve along with the fit with a sum gate on delayed 379, 580, and 638 keV transitions.  
  (b) The “singles” Doppler-corrected prompt $\gamma$-ray spectrum ($\gamma_{P}$). Transitions from $^{200}$Au are marked with \protect \fullcirc $\space$ and from $^{202}$Au with \protect \mycbox. The inset shows the $\gamma_{P}$-$\gamma_{P}$ coincidence spectrum with a gate on the 638 keV transition. 
  (c) The $\gamma_{P}$ spectrum in coincidence with a sum gate on the $\gamma_{D}=379,\,638$ keV transitions for 1 $< \Delta t <$ 100 $\mu$s. The inset shows the $\gamma_P$-$\gamma_P$ coincidence spectrum with a gate on the 699 keV transition. 
  }
  \label{fig:GammaSpec_201Au}
\end{figure}

For $^{201}$Au, the earlier work reported the 554 keV transition from the decay of the $11/2^-$ isomer, with $T_{1/2}=0.34^{+90}_{-29}$ ms, and the transitions of 379 and 638 keV from a higher-lying isomer with $T_{1/2}=5.6(24)\,\mu$s \cite{Steer2011}. In the “singles” delayed $\gamma$-ray spectrum ($\gamma_{D}$) for the 1 $< \Delta t <$ 100 $\mu$s time window, all three transitions are clearly observed (See Fig.~\ref{fig:GammaSpec_201Au}(a)). The 40 keV $M2$ transition that directly de-excites the $(11/2^-)$ isomer was not detected in this experiment due to its high conversion coefficient ($\alpha_{\text{tot}}=1191$ calculated using BrIcc~\cite{NIMA2008_KIBEDI}) and low detection efficiency. A new delayed $\gamma$-ray at 580 keV is observed, which is in coincidence with the 379 keV transition (upper inset in Fig.~\ref{fig:GammaSpec_201Au}(a)). The half-life obtained from the decay curve with a sum gate on the 379, 580, and 638 keV transitions is 15.2(29) $\mu$s (lower inset in Fig.~\ref{fig:GammaSpec_201Au}(a)). The half-lives obtained using each transition were consistent. This value is considerably longer than the reported value of 5.6(24) $\mu$s~\cite{Steer2011}.

The 379 and 638 keV transitions are also observed in the “singles” Doppler-corrected prompt $\gamma$-ray spectrum ($\gamma_{P}$) of $^{201}$Au, as shown in Fig.~\ref{fig:GammaSpec_201Au}(b). The $\gamma_{P}$-$\gamma_{P}$ coincidence confirms their cascade decay (inset of Fig.~\ref{fig:GammaSpec_201Au}(b)). The intensity of the prompt 379 keV transition is larger than that of the 638 keV $\gamma$-ray (See Table \ref{tab:GammaSummary_201Au} in the Appendix). Therefore, the order of the 379 keV and 638 keV transitions has been swapped, compared to the suggested decay scheme in Ref.~\cite{Steer2011}. 

The 580 keV transition, which is not observed in the $\gamma_{P}$ spectrum, is placed at the top of the isomeric decay sequence. Based on analogy with the level structure of $^{199}$Au, the transition is assigned as an E1 transition depopulating the $(21/2^+$) state. Then, the ($25/2^+$) state is proposed to be isomeric and to feed into the (21/2$^+$) state via an unobserved low-energy E2 transition (denoted “x” in Fig.~\ref{fig:Au_LS_All}). 
If the 580 keV transition directly depopulates the isomer, its relative intensity compared to the 379 keV and 638 keV transitions would imply E2 or M2 character from internal conversion analysis. However, neither assignment is found to be compatible with the measured half-life of 15.2(29) $\mu$s.

Two new prompt $\gamma$ rays at 363 and 699 keV are identified in the $\gamma_{P}$ spectrum, as shown in Fig.~\ref{fig:GammaSpec_201Au}(b). The coincidence $\gamma_P$ spectrum with a sum gate on the 379 keV and 638 keV delayed transitions (Fig.~\ref{fig:GammaSpec_201Au}(c)) confirms that the 363 keV transition feeds the ($25/2^+$) isomer. The inset of Fig.~\ref{fig:GammaSpec_201Au}(c) demonstrates the coincidence between the 363 keV and 699 keV transitions, placing both of them in a cascade above the $(25/2^+)$ isomer.

Table \ref{tab:GammaSummary_201Au} in the Appendix gives the details of all the transitions of $^{201}$Au observed in this work.

\subsection{Even-$A$ Au Isotopes}\label{sec:GammaSpec_Result_Au_EvenA}

High-spin states built on top of the 12$^{-}$ isomer are commonly observed in the even-$A$ $^{196,\,198,\,200,\,202}$Au isotopes. Some previously known transitions below the isomer are also identified, while most of the observed transitions are newly established from prompt $\gamma$-ray spectra. 
    
\subsubsection{$^{196}$Au}

The $12^-$ isomer in $^{196}$Au ($T_{1/2}=9.603(22)$ h) was reported in Ref.~\cite{Moody_2020}. Because of its long half-life, isomeric transitions were not observed in the current experiment. Later studies, including Ref.~\cite{Metz1999}, primarily focused on low-spin excitations in this nucleus. However, most of the transitions identified in Ref.~\cite{Metz1999} were not observed in the present work, as MNT reactions predominantly populate yrast states with higher spin. 

\begin{figure}[t]
  \centering
  \includegraphics[width=0.5\textwidth]{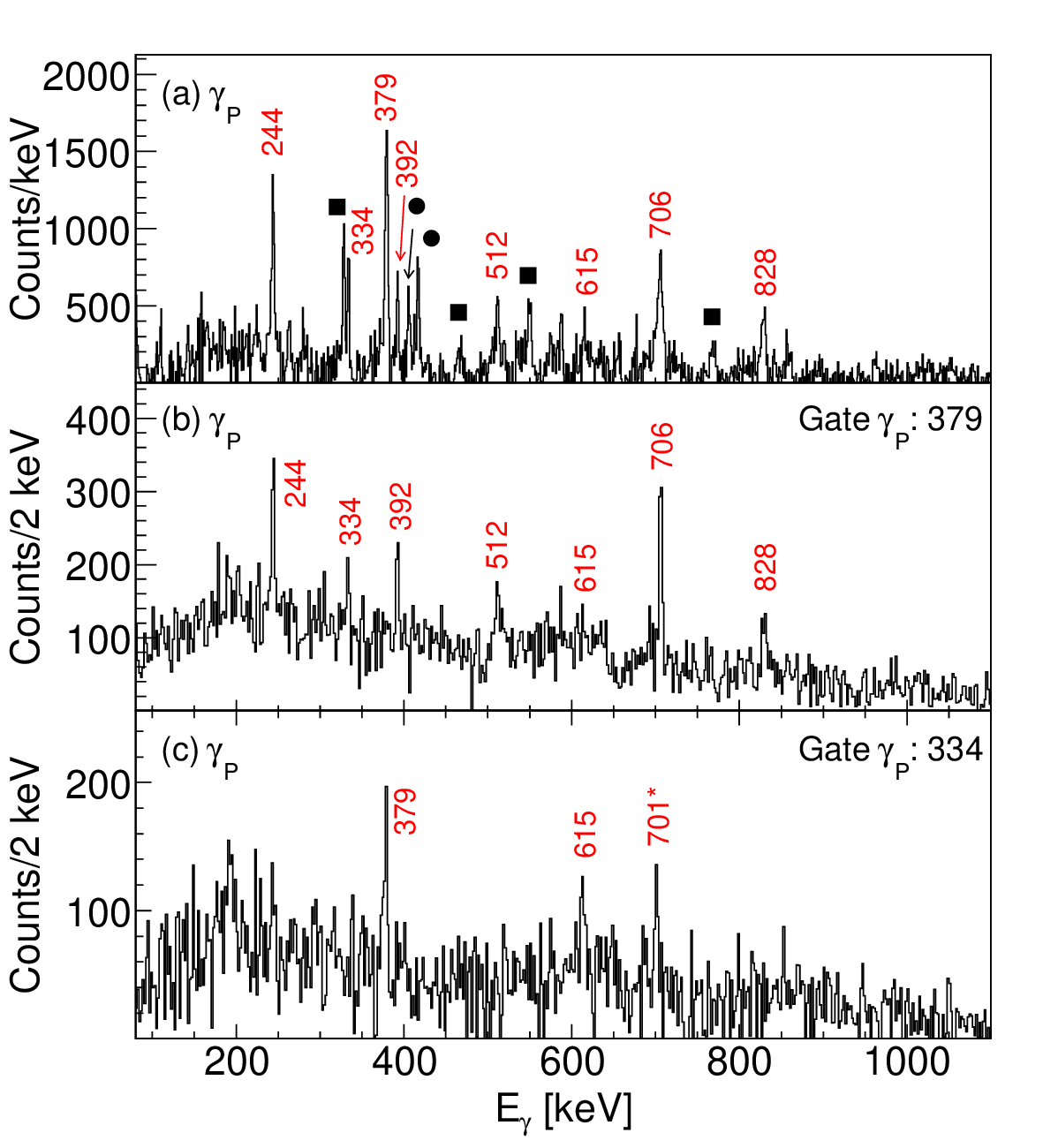}
  \caption{
  $\gamma$-ray spectra for $^{196}$Au obtained with gates on $^{127\text{-}135}$I within the respective optimal TKEL regions for each isotope. 
  The new transitions are labeled in red. 
  (a) The “singles” Doppler-corrected prompt $\gamma$-ray spectrum ($\gamma_{P}$). 
  Transitions from $^{195}$Au are marked with \protect \fullcirc $\space$ and from $^{197}$Au with \protect \mycbox.
  (b) The $\gamma_{P}$-$\gamma_{P}$ coincidence spectrum with a gate on the 379 keV transition. 
  (c) The $\gamma_{P}$-$\gamma_{P}$ coincidence spectrum with a gate on the 334 keV transition.
  }
  \label{fig:GammaSpec_196Au}
\end{figure}

The “singles” Doppler-corrected prompt $\gamma$-ray spectrum ($\gamma_{P}$) for $^{196}$Au is presented in Fig.~\ref{fig:GammaSpec_196Au}(a). Several new transitions, 244, 334, 379, 392, 512, 615, 706, and 828 keV, have been identified. The $\gamma_{P}$-$\gamma_{P}$ coincidence spectrum gated on the 379 keV transition, shown in Fig.~\ref{fig:GammaSpec_196Au}(b), demonstrates coincidence among the transitions. These transitions are placed above the known $12^-$ isomeric state at 596 keV \cite{WAPSTRA1967}. 

The coincidence $\gamma_{P}$ spectrum gated on the 334 keV transition is shown in Fig.~\ref{fig:GammaSpec_196Au}(c), where the coincidence with the 379 and 615 keV transitions are seen. In contrast, no coincidence is observed with the 706 keV transition, indicating that the 334 keV transition directly feeds the ($14^-$) state. A transition at 701 keV is visible, but it lacks a clear signature in the prompt “singles” spectrum and is therefore not included in the proposed level scheme.

Table~\ref{tab:GammaSummary_196Au} in the Appendix gives the details of all the transitions of $^{196}$Au observed in this work. 

\subsubsection{$^{198}$Au}

\begin{figure}[t]
  \centering
  \includegraphics[width=0.5\textwidth]{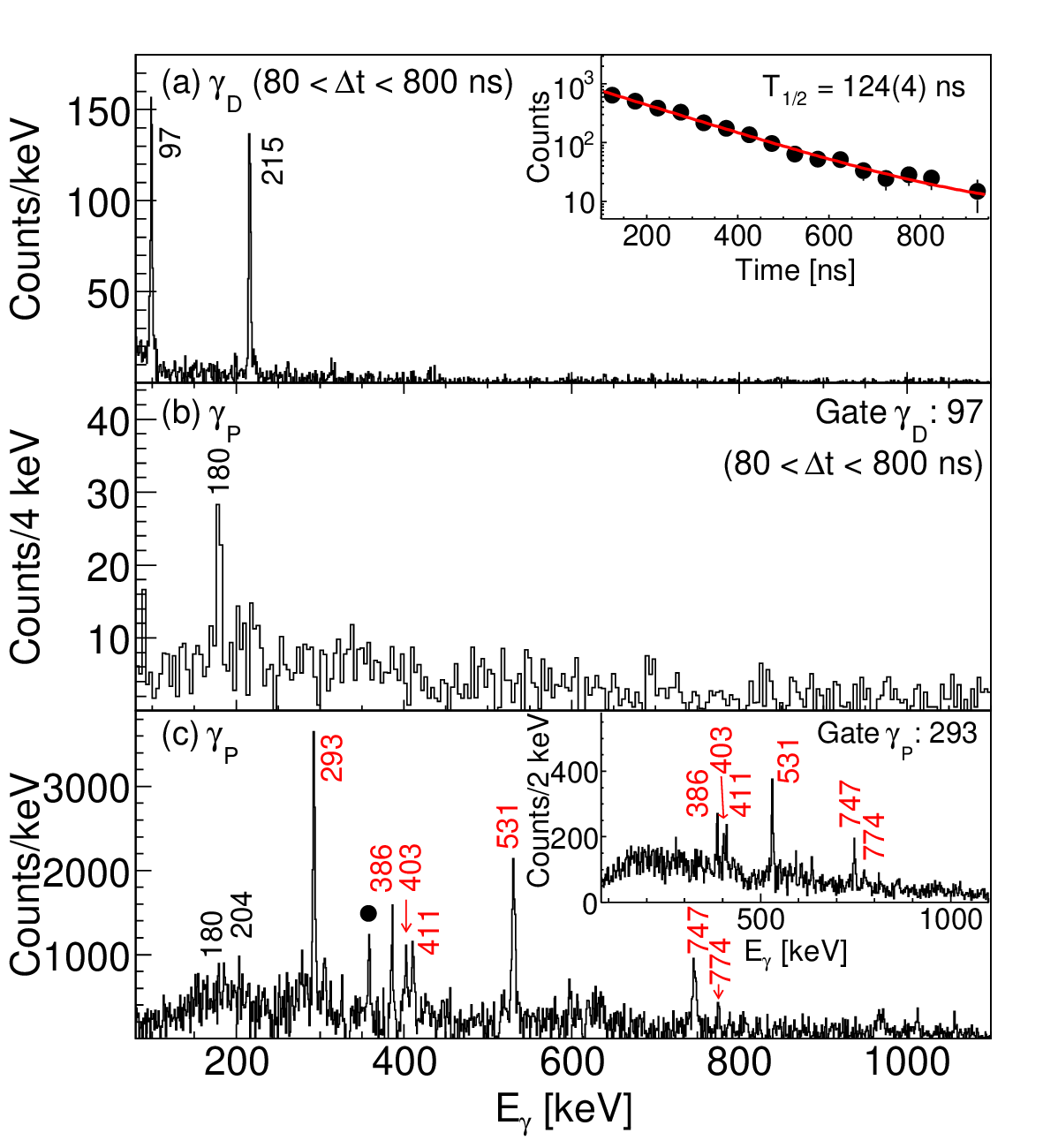}
  \caption{
  $\gamma$-ray spectra for $^{198}$Au obtained with gates on $^{127\text{-}135}$I within the respective optimal TKEL regions for each isotope. 
  The new transitions are labeled in red. 
  (a) The “singles” delayed $\gamma$-ray spectrum ($\gamma_{D}$) for 80 $< \Delta t <$ 800 ns. The inset shows the decay curve of a sum gate on the 97 and 215 keV transitions with the exponential fit. 
  (b) The “singles” Doppler-corrected prompt $\gamma$-ray spectrum ($\gamma_{P}$). Transitions from $^{197}$Au are marked with \protect \fullcirc. The inset shows the $\gamma_{P}$-$\gamma_{P}$ coincidence spectrum with a gate on the 293 keV transition. 
  (c) The $\gamma_{P}$ spectrum in coincidence with the $\gamma_{D}=97$ keV transition for 80 $< \Delta t <$ 800 ns.
  }
  \label{fig:GammaSpec_198Au}
\end{figure}

\begin{figure}[t]
  \centering
  \includegraphics[width=0.5\textwidth]{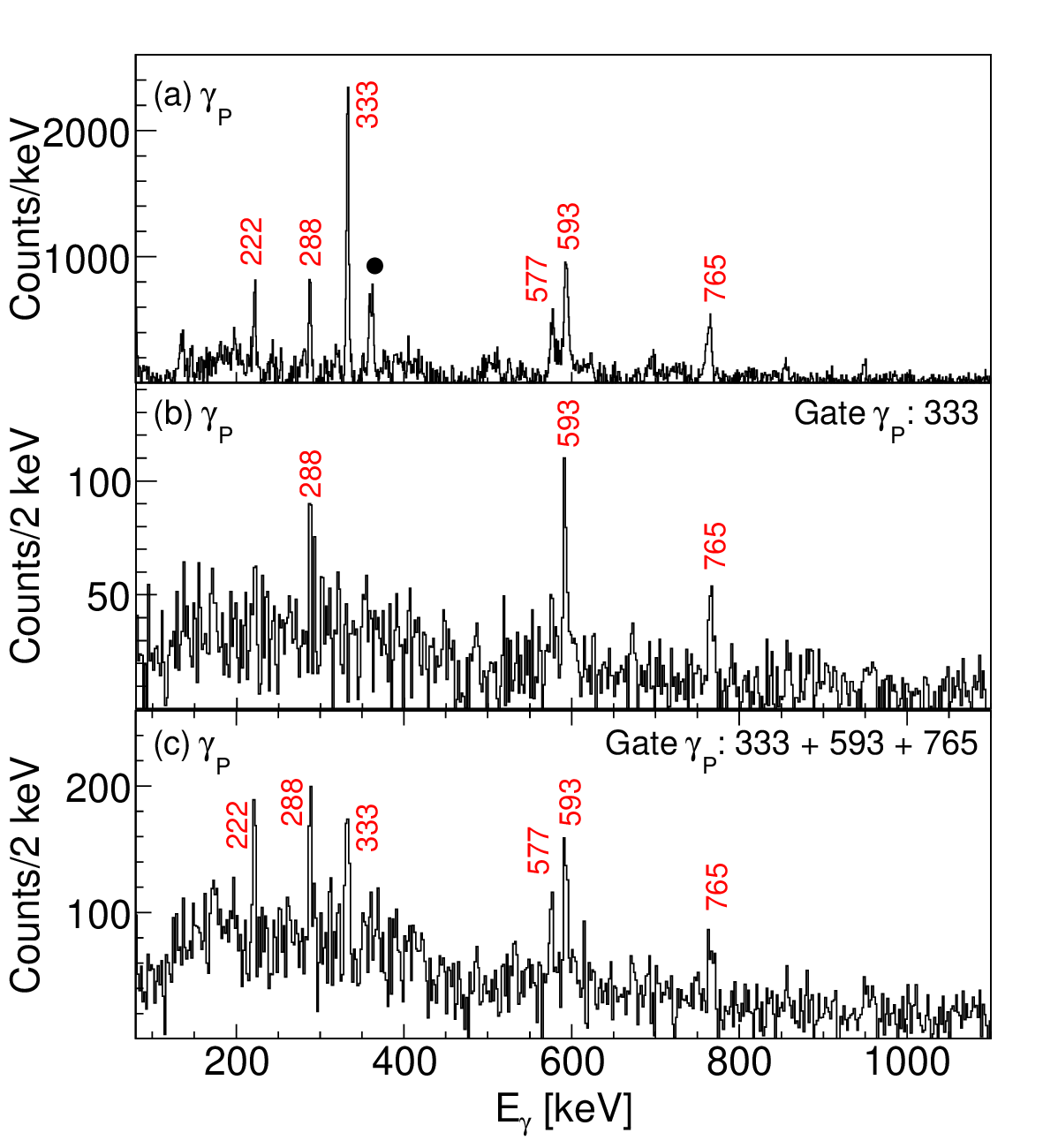}
  \caption{
  $\gamma$-ray spectra for $^{200}$Au obtained with gates on $^{127\text{-}135}$I within the respective optimal TKEL regions for each isotope. 
  The new transitions are labeled in red. 
  (a) The “singles” Doppler-corrected prompt $\gamma$-ray spectrum ($\gamma_{P}$). 
  Transitions from $^{199}$Au are marked with \protect \fullcirc.
  (b) The $\gamma_{P}$-$\gamma_{P}$ coincidence spectrum with a gate on the 333 keV transition. 
  (c) The $\gamma_{P}$-$\gamma_{P}$ coincidence spectrum with a sum gate on the 333, 593, and 765 keV transitions.
  }
  \label{fig:GammaSpec_200Au}
\end{figure}

\begin{figure}[t]
  \centering
  \includegraphics[width=0.5\textwidth]{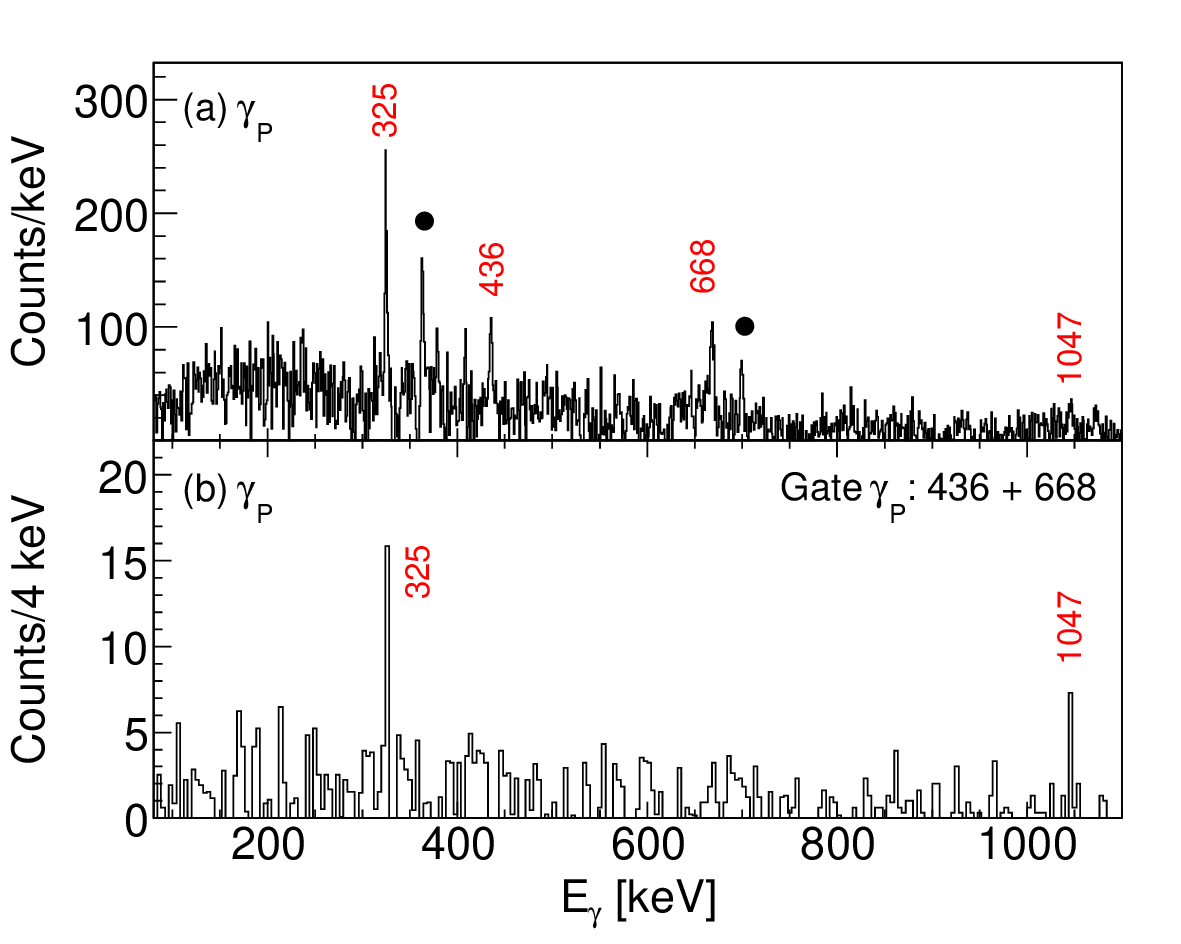}
  \caption{
  $\gamma$-ray spectra for $^{202}$Au obtained with gates on $^{127\text{-}135}$I within the respective optimal TKEL regions for each isotope. 
  The new transitions are labeled in red. 
  (a) The “singles” Doppler-corrected prompt $\gamma$-ray spectrum ($\gamma_{P}$). 
  Transitions from $^{201}$Au are marked with \protect \fullcirc.
  (b) The $\gamma_{P}$-$\gamma_{P}$ coincidence spectrum with a gate on the 325 keV transition. 
  (c) The $\gamma_{P}$-$\gamma_{P}$ coincidence spectrum with a sum gate on the 436 and 668 keV transitions.
  }
  \label{fig:GammaSpec_202Au}
\end{figure}

The $5^+$ isomer ($T_{1/2}=124(4)$ ns)~\cite{Mirza1975} and the ($12^-$) isomer ($T_{1/2}=2.28(2)$ d)~\cite{ENSDF} of $^{198}$Au were previously reported. The half-life of ($12^-$) isomeric state  lies beyond the delayed $\gamma$-ray time window in the present work, but the 97 and 215 keV transitions de-exciting the $5^+$ isomer are observed in the delayed “singles” delayed $\gamma$-ray spectrum ($\gamma_{D}$) for the 80 $< \Delta t <$ 800 ns time window (Fig.~\ref{fig:GammaSpec_198Au}(a)). The obtained half-life of the $5^+$ isomer is 124(4) ns (inset of Fig.~\ref{fig:GammaSpec_198Au}(a)), in agreement with the earlier measurement \cite{Mirza1975}.

Two known 180 keV and 204 keV transitions~\cite{Cunnane1972_198Au} are observed in the “singles” Doppler-corrected prompt $\gamma$-ray spectrum ($\gamma_{P}$), as shown in Fig.~\ref{fig:GammaSpec_198Au}(b). The prompt 180 keV $\gamma$-ray is also in coincidence with the delayed 97 keV transition (Fig.\ref{fig:GammaSpec_198Au}(c)), confirming that it feeds the $5^+$ isomer. 
The ordering of the 180 and 204 keV transitions, previously reported as tentative in Ref.~\cite{Cunnane1972_198Au}, is clarified in the present work. The prompt $\gamma$-ray intensities indicate that the 180 keV transition lies below the 204 keV transition (See Table~\ref{tab:GammaSummary_198Au}).  

Several new prompt $\gamma$-ray transitions are identified; namely, 293, 386, 403, 411, 531, 747, and 774 keV (Fig.~\ref{fig:GammaSpec_198Au}(b)). The inset in Fig.~\ref{fig:GammaSpec_198Au}(c) shows the $\gamma$$_{P}$-$\gamma$$_{P}$ coincidence spectrum with a gate on the 293 keV transition, which shows the mutual coincidence between the new transitions. A cascade decay of the transitions is placed above the known ($12^-$) state \cite{Cunnane1972_198Au}.

The spin-parity of the ($18^+$) state in $^{198}$Au, which de-excites through the 386 keV transition, is tentatively assigned through a comparison with its isotone $^{199}$Hg. The level scheme of $^{198}$Au observed in the present work closely resembles the structure built on the $13/2^+$ state in $^{199}$Hg \cite{negi2019}. Notably, the 386 keV transition in $^{198}$Au corresponds to the 381 keV transition in $^{199}$Hg, which decays from the $25/2^-$ state to the $25/2^+$ state. Given that Au isotopes can be interpreted as a proton hole in the $\pi h_{11/2}$ orbital coupled to the Hg core (See Sec.~\ref{sec:config}), the 386 keV transition in $^{198}$Au is attributed to the de-excitation of the ($18^+;\, h_{11/2}^{-1}\otimes 25/2^-$) state. This systematics serves as the basis for assigning spin-parity to the corresponding states in the heavier even-$A$ $^{200,\,202}$Au isotopes.

Table~\ref{tab:GammaSummary_198Au} in the Appendix gives the details of all the transitions of $^{198}$Au observed in this work.

\subsubsection{$^{200}$Au}

Previous $\gamma$-ray spectroscopy studies of the $^{200}$Au mainly investigated low-spin excited states via $\beta$-decay spectroscopy~\cite{sakai_1968, JCHill1976}. These transitions are not observed in the present measurement, which predominantly populates yrast states due to the nature of multi-nucleon transfer reactions~\cite{Kraus1988,GUID85}. 

The “singles” Doppler-corrected prompt $\gamma$-ray spectrum ($\gamma_{P}$) for $^{200}$Au is shown in Fig.~\ref{fig:GammaSpec_200Au}(a), where newly observed transitions 222, 288, 333, 578, 593, and 765 keV are identified. The prompt $\gamma_{P}$-$\gamma_{P}$ coincidence spectrum gated on the 333 keV transition confirms that the 593 and 765 keV transitions are in coincidence with the 333 keV transition (Fig.~\ref{fig:GammaSpec_200Au}(b)). Therefore, the 333, 593, and 765 keV transitions are placed in a cascade above the known $12^-$ isomer previously reported in Ref.~\cite{sakai_1968}. Additional prompt coincidence relationships among the 222, 288, and 578 keV transitions with this cascade are established from the spectrum shown in Fig.~\ref{fig:GammaSpec_200Au}(c). In analogy with $^{198}$Au, the state emitting the 288 keV transition is tentatively assigned a spin-parity of ($18^+$).

Table~\ref{tab:GammaSummary_200Au} in the Appendix gives the details of all the transitions of $^{200}$Au observed in this work.

\subsubsection{$^{202}$Au}

The previous $\gamma$-ray spectroscopy study of the $^{202}$Au was reported in Ref.~\cite{Steer2011}. In the present work, previously reported isomeric transitions of 138 and 414 keV~\cite{Steer2011} are not observed. These transitions were attributed to the decay of a low-lying isomer with $T_{1/2}$ = 13.1(5) ns~\cite{Steer2011}, whose half-life lies outside the sensitivity of the current delayed $\gamma$-ray measurement. 

The “singles” Doppler-corrected prompt $\gamma$-ray spectrum ($\gamma_{P}$) for $^{202}$Au is presented in Fig.~\ref{fig:GammaSpec_202Au}(a). Several new transitions at 325, 436, 668, and 1047 keV have been identified. The prompt $\gamma$-$\gamma$ coincidence spectrum ($\gamma_{P}$-$\gamma_{P}$), gated on a sum of the 436 and 668 keV transitions, is shown in Fig.~\ref{fig:GammaSpec_202Au}(b). It indicates that the 436 keV transition is in coincidence with the 325 keV transition, while the 668 keV transition is in coincidence with the 1047 keV transition. Although the coincidence between the 325 and 668 keV transitions is not clearly observed, the 668 keV $\gamma$-ray is placed above the 325 keV transition. This placement is supported by their relative intensities, the systematic trends in neighboring even-$A$ Au isotopes, and theoretical calculations (see Fig.~\ref{fig:Levels_Au_OddN}). The absence of a clear coincidence between the 325 and 668 keV transitions is likely due to the limited statistics of $^{202}$Au.

Table~\ref{tab:GammaSummary_202Au} in the Appendix gives the details of all the transitions observed in this work.

\begin{figure}[t]
  \centering
  \includegraphics[width=0.5\textwidth]{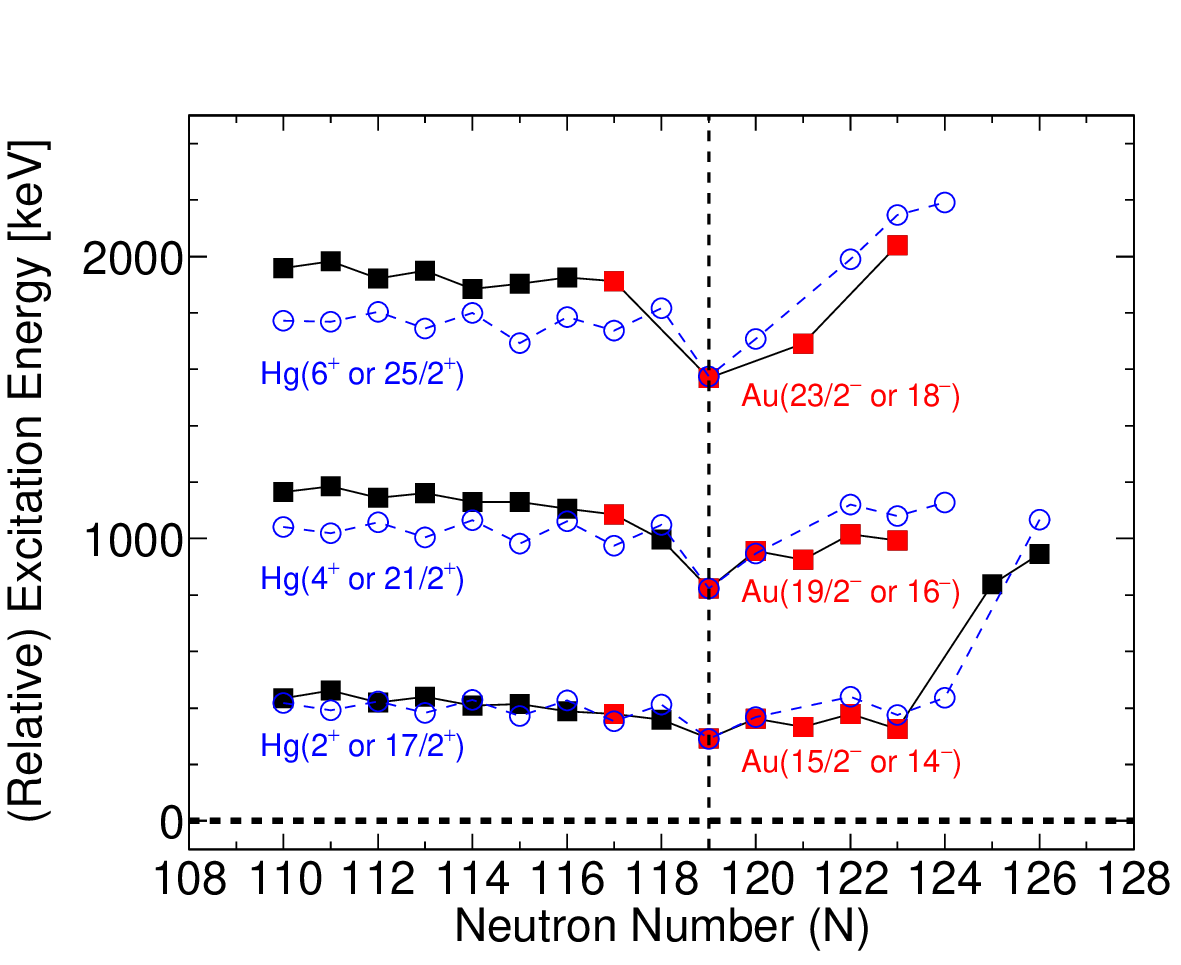}
  \caption{Experimental excitation energies of selected $\Delta J=2$ sequences in Au (solid black and red squares) and Hg (open blue circles) isotopes. Newly measured levels in Au ($N\geq117$) are shown in red, while all other values are taken from Ref.~\cite{ENSDF}. Excitation energies of Au(Hg) are plotted relative to the $11/2^-$($0^+$) state for even-$N$ and the $12^-$($13/2^+$) state for odd-$N$, respectively.}
  \label{fig:Syst_Energy_AuHg_N119}
\end{figure}

\begin{figure*}[t]
  \centering
  \includegraphics[width=0.95\textwidth]{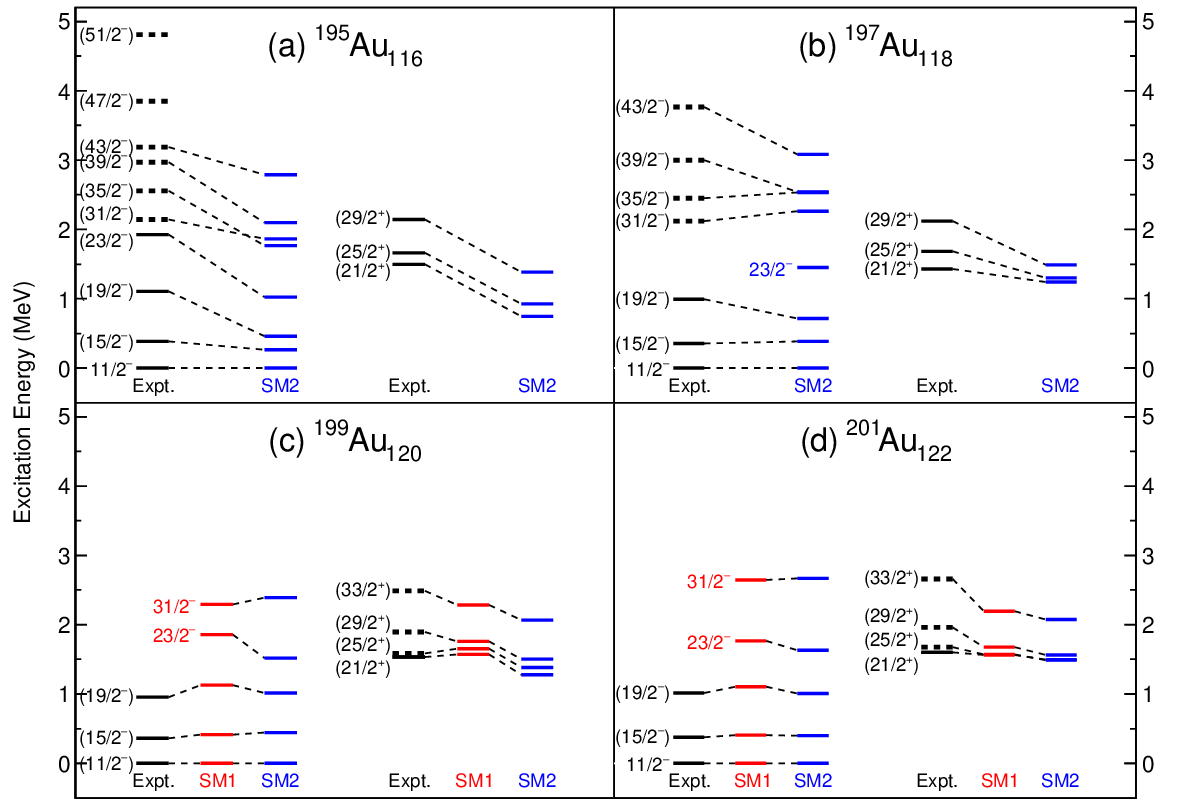}
  \caption{Comparison of experimental (Expt., black) and shell model calculations using KHHE interaction with full model space (SM1, red) or with monopole-based truncation (SM2, blue) for level schemes of odd-$A$ $^{195-201}$Au isotopes. The excitation energies are plotted relative to the $11/2^-$ states. Dashed levels denote that a small energy correction has been applied to account for unobserved transitions. (see Fig.~\ref{fig:Au_LS_All}). Experimental and shell-model states sharing the same spin–parity are connected by the dashed lines to guide the eye.}
  \label{fig:Levels_Au_EvenN}
\end{figure*}

\subsection{Features of Experimental Results}
\label{sec:Syst_Expt}

The main points of the new experimental results can be summarized as follows.

\begin{enumerate}
    \item  \textbf{The appearance of a dip like structure at $N=119$ in the energies of low-lying states above the $11/2^-$ and $12^-$ isomer in the Au isotopes.}\\ 
    In even-$N$ $^{195,\,197,\,199,\,201}$Au isotopes, negative-parity bands built on the $11/2^-$ isomer are consistently observed, while in odd-$N$ $^{196,\,198,\,200,\,202}$Au nuclei, analogous bands are built on the $12^-$ isomer. Fig.~\ref{fig:Syst_Energy_AuHg_N119} shows the systematics of excitation energies for high-spin states built on the $11/2^-$ and $12^-$ isomer in Au isotopes. In even-$N$ Au nuclei, members of the $\pi h_{11/2}^{-1}$ band ($15/2^-$, $19/2^-$, and $23/2^-$; black and red squares) are plotted with respect to the $11/2^-$ level. In odd-$N$ isotopes, the $\pi h_{11/2}^{-1}\otimes\nu i_{13/2}^{-1}$ band ($14^-$, $16^-$, and $18^-$; black and red squares) is shown with respect to the $12^-$ state. The levels newly identified in the present work are shown in red. Up to $N=118$, the energy spacings between successive states remain nearly constant. 
    At $N=119$, a drop in the level energies is observed, which becomes more pronounced at higher spin. Beyond this point, the trend recovers and then increases as the $N=126$ shell closure is approached.

    The level energy differences for Au isotopes are compared with those of the corresponding Hg nuclei in Fig.~\ref{fig:Syst_Energy_AuHg_N119}. Blue circles denote the energies of the even-$N$ Hg ground-state band ($2^+$, $4^+$, $6^+$) and the odd-$N$ $\nu i_{13/2}^{-1}$ band ($17/2^+$, $21/2^+$, $25/2^+$) with respect to the $13/2^+$ state. The excitation-energy pattern of these states in Au isotopes closely follows those of the corresponding positive-parity levels in Hg isotopes. The regularity of Au level energies and their inheritance from the corresponding Hg core are discussed in detail in Ref.~\cite{Cho2026_PLB}.
    
    \item  \textbf{New $(25/2^+)$ isomers in the $N>119$ Au isotopes}. \\
    Newly identified long-lived isomers  in $^{199}$Au ($0.14 (2)$ ms) and $^{201}$Au ($15.2 (29)$ $\mu$s) are tentatively attributed to the $(25/2^+)$ states, based on their delayed $\gamma$-ray spectra and prompt–delayed coincidence relationships (Figs. \ref{fig:GammaSpec_199Au} and \ref{fig:GammaSpec_201Au}). The spectra show that only three $\gamma$-ray transitions feed the $11/2^-$ isomer, indicating a decay path $21/2^+ \to 19/2^- \to 15/2^- \to 11/2^-$. To explain the half-lives of the isomers, we invoke an unobserved low-energy E2 feeding transition into the $21/2^+$ state. For the measured half-lives, this implies a $B(E2)$ of the order of 10 W.u., comparable to the reduced transition strengths seen for the corresponding $7^-\to5^-$ decays in neighboring Hg isotones \cite{suman2021}. In contrast, the positive-parity states of $^{195}$Au and $^{197}$Au built on the $(21/2^+)$ level end at the $(29/2^+)$ states, and the observed high-spin isomers have been tentatively assigned as $(31/2^-)$ \cite{DRACOULIS2013}. These $(31/2^-)$ isomers likely decay to the $(29/2^+)$ state via an unobserved low-energy E1 transition \cite{DRACOULIS2013}. 
    
    \item \textbf{Non observation of the $11^-$ band in $N>117$ Au isotopes.} \\
    Previous studies of even-$A$ $^{190,\,192,\,194}$Au isotopes using fusion-evaporation reactions have revealed two band structures built on the $11^-$ isomer and the $12^-$ state \cite{neskakis1979, Gueorguieva2004, oktem2012, gao2012}. In each case, the $11^-$ isomer lies at a lower excitation energy than the $12^-$ state, and the two bands display similar level spacings with pronounced inter-band transitions linking these bands. In the present work, no clear signatures of these odd-$J$ bands built on the $11^-$ state are observed for $A \geq 196$. 
\end{enumerate}.

\begin{figure*}[t]
  \centering
  \includegraphics[width=0.95\textwidth]{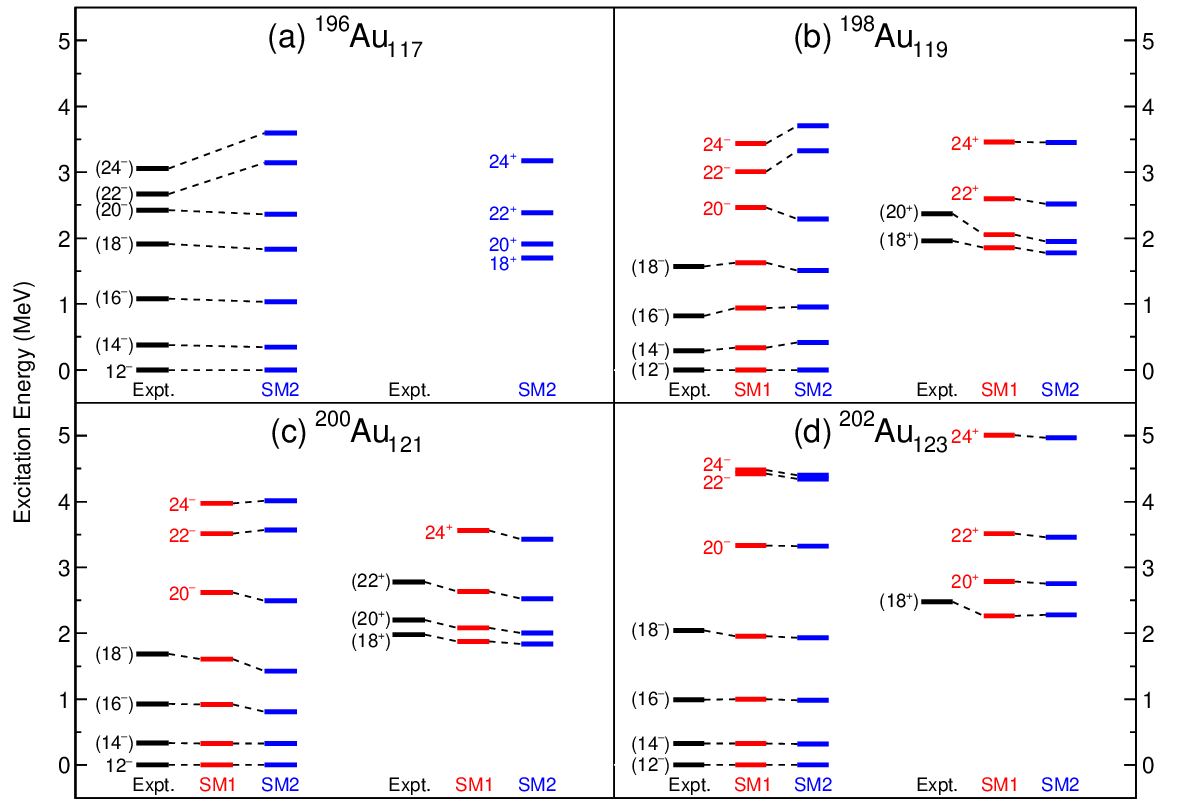}
  \caption{Same as Fig.~\ref{fig:Levels_Au_EvenN} for level schemes of even-$A$ $^{196-202}$Au isotopes. The excitation energies are plotted relative to the $12^-$ states. Experimental and shell-model states sharing the same spin–parity are connected by the dashed lines to guide the eye. 
  }
  \label{fig:Levels_Au_OddN}
\end{figure*}

\section{Shell-model calculations and Discussions}
\label{discussion}

Large-Scale Shell-Model (LSSM) calculations were used to interpret the newly observed high-spin states in neutron-rich Au isotopes. The model space included five proton orbitals ($0g_{7/2}$, $1d_{5/2}$, $1d_{3/2}$, $2s_{1/2}$, $0h_{11/2}$) between the $Z=50-82$ shells and six neutron orbitals ($0h_{9/2}$, $1f_{7/2}$, $1f_{5/2}$, $2p_{3/2}$, $2p_{1/2}$, $0i_{13/2}$) between the $N=82-126$ shells, excluding all other orbitals. Two-body matrix elements (TBMEs) were derived from the Kuo-Herling interaction \cite{herling1972}, with proton-proton, neutron-neutron, and proton-neutron interactions treated using the Kuo-Herling Hole Effective (KHHE) interaction framework \cite{warburton1991}. The interaction was subsequently refined to improve agreement with spectroscopic data near $^{208}$Pb \cite{yuan2022}. Calculations for the $^{198\text{–}205}$Au isotopes were performed using the KSHELL code \cite{shimizu2013} within the full model space (denoted as SM1 henceforth).\

\begin{figure}[t]
  \centering
  \includegraphics[width=0.5\textwidth]{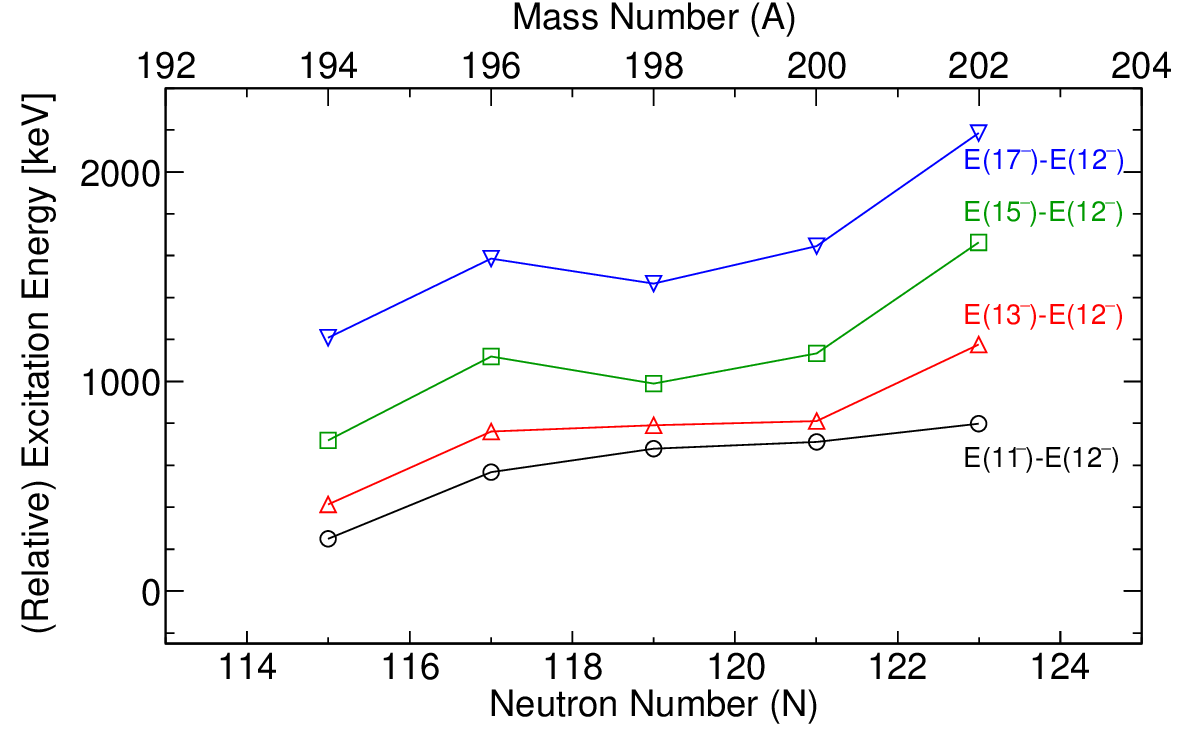}
  \caption{The shell-model calculated excitation energies (SM2) for odd-$J$ negative-parity states in even-A $^{194-202}$Au relative to the $12^-$ state.}
  \label{fig:Syst_OddJBand}
\end{figure}

\begin{figure}[t]
  \centering
  \includegraphics[width=0.5\textwidth]{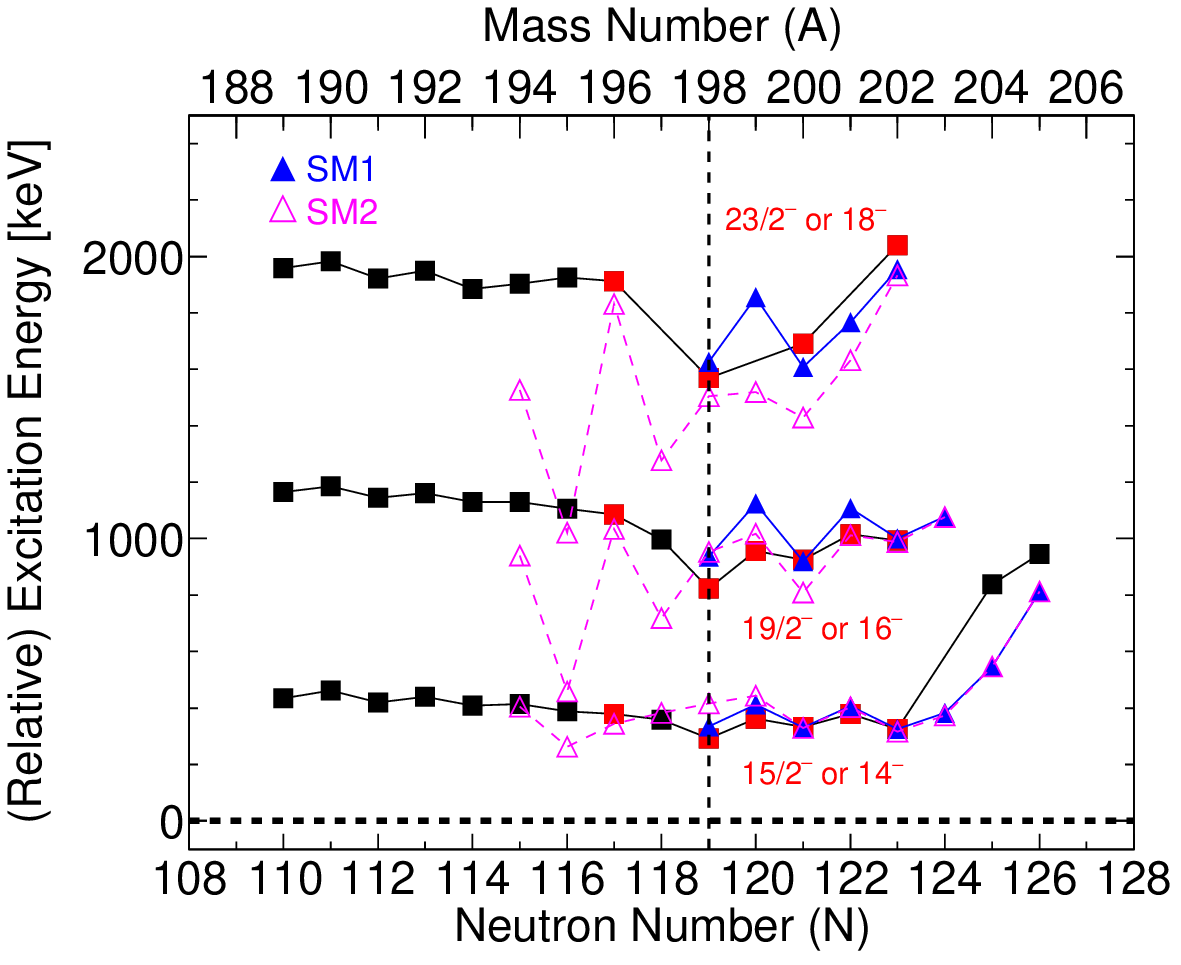}
  \caption{Comparison of LSSM calculations with the experimental level energies of Au isotopes shown in Fig.~\ref{fig:Syst_Energy_AuHg_N119}. The SM1 and SM2 calculation results are denoted by solid blue and open magenta triangles, respectively. The dashed-line indicates $N=119$, where the SM1 calculation ends.}
  \label{fig:Syst_Energy_AuSM_N119}
\end{figure}

As the neutron number decreases from the $N=126$ shell closure, the computational complexity of LSSM calculations grows rapidly due to the increasing number of active single-particle configurations. To address this, the Monopole-Based Truncation (MBT) scheme \cite{chen2023} was applied to lighter Au isotopes ($A \leq 197$), where full model space calculations become impractical. In this approach, a subset of single-particle configurations is selected based on the sum of effective single-particle energies ($E^{SC}$) \cite{chen2023}. 
For the MBT calculations in the present work, configurations with $E^{SC} < 10$ MeV are retained (denoted as SM2 henceforth). This 10 MeV truncation threshold has been adopted in earlier studies and has been shown to reasonably reproduce the $\beta$-decay properties of nuclides in the southwest region of $^{208}$Pb \cite{kumar2024} and the excited-state properties in Hg isotopes \cite{Sahoo2024}.

\subsection{Comparison with Level Schemes}
\label{sec:Discuss_LS}

Fig.~\ref{fig:Levels_Au_EvenN} presents a comparison between experimental level schemes and shell-model predictions using the SM1 and SM2 model spaces for the odd-even $^{195-201}$Au isotopes. In particular, the calculated energies of the yrast-state are compared with the corresponding experimental data. The excitation energies are plotted relative to the $11/2^-$ isomeric state. For excited states depopulated by unobserved low-energy transitions (see Fig.~\ref{fig:Au_LS_All}), the level energy is tentatively assigned to be that of the feeding state, assuming the unknown low transition energies are small. The negative-parity states in $^{199}$Au and $^{201}$Au are well reproduced by both SM1 and SM2 calculations. However, SM2 calculations for $^{195,\,197}$Au predict a more compressed level spacing compared to the experimental levels and the calculated results for the heavier isotopes. For positive-parity states of $^{195\text{–}201}$Au, the calculated excitation energies tend to be systematically lower than the experimental values.

Shell-model calculations show that the $21/2^+$ and $25/2^+$ states of $^{201}$Au are nearly degenerate, in agreement with the experimental results. On the other hand, this narrow energy gap seems to be larger in the SM2 calculations for the $^{195}$Au isotope, which is also in agreement with the absence of a corresponding isomer in the present work. 

The experimental and calculated shell-model (SM1 and SM2 model spaces) excitation energies of odd-odd $^{196-202}$Au isotopes, relative to the $12^-$ isomeric state, are shown in Fig.~\ref{fig:Levels_Au_OddN}. Both SM1 and SM2 shell-model calculations reproduce the observed energies relatively well, compared to those for the odd-even isotopes.

Fig.~\ref{fig:Syst_OddJBand} shows the calculated (SM2) excitation energies of the odd-$J$ band built on the $11^-$ state, plotted relative to the $12^-$ state, for the odd-odd isotopes $^{194-202}$Au. The excitation energies of the $11^-$ band relative to the $12^-$ state increase with the neutron number, qualitatively accounting for the lowering of the $12^-$ state below the $11^-$ state in the heavier isotopes. Consequently, inter-band transitions from even-$J$ states to $J-1$ states are suppressed, since the $J\!-\!1$ levels lie above the $J$ levels. This behavior was explained in terms of the variation of the neutron Fermi surface for the $\pi h_{11/2}\otimes \nu i_{13/2}$ configuration as a function of neutron number \cite{toki1979}. 

Fig.~\ref{fig:Syst_Energy_AuSM_N119} shows the experimental level energies for Au isotopes along with shell-model calculations. 
The SM1 results (solid triangles) reproduce the overall experimental excitation energies (solid squares), but since the results are not available below $N=119$, these calculations cannot probe the energy drop at that neutron number.
The SM2 calculations (open triangles) show fair agreement for even-$A$ Au isotopes. However, the agreement is not as good for the odd-$A$ isotopes. Therefore, the SM2 calculations that can be extended to lower-$A$ are unable to explain the observed trend.

A notable increase of roughly 500 keV in the measured two neutron separation energy~\cite{ENSDF}, $\Delta_{2n}(N)=S_{2n}(N)-S_{2n}(N+2)$, is observed at $N=119$ for ground states in the Au isotopic chain. However, neither the binding energy differences nor the pairing energy differences for the selected levels calculated by the shell-model exhibit a minimum at $N=119$. The underlying mechanism driving this energy drop, therefore, remains an open question. 

\subsection{$\pi h_{11/2}$ and $\nu i_{13/2}$ Hole Occupancies} 
\label{sec:config}

To gain a deeper understanding of the high-spin states in neutron-rich Au isotopes, their underlying shell-model configurations are systematically investigated. As the neutron number decreases below $N=126$, shell-model calculations predict that the wavefunctions in Au isotopes become more fragmented, reflecting enhanced collectivity. For example, the 11/2$^-$ level in $^{205}$Au ($N=126$) is characterized by a relatively pure single-particle configuration of $73\%$, unlike the corresponding 11/2$^-$ states in lighter isotopes ($A\leq203$), which are highly fragmented with no single configuration contributing more than $8\%$ to the wavefunction.

Despite pronounced fragmentation of the wavefunctions, systematic trends emerge in the hole occupancies of high-$j$ unique-parity orbitals. Fig.~\ref{fig:Hole_Occ_allA}(a) and (b) show the SM1 average hole occupancies of $\pi h_{11/2}$ and $\nu i_{13/2}$ as a function of the total spin ($2J$) for negative- and positive-parity states in odd-even $^{199-205}$Au isotopes, respectively. For the negative-parity states, the occupancy of $\pi h_{11/2}$ remains nearly one, independent of spin up to $2J=32\,\hbar$. However, for $^{205}$Au, an extra pair breaking occurs above $2J=23\,\hbar$ (see the jump of the green solid line in Fig.~\ref{fig:Hole_Occ_allA}(a)). 
An extra hole in the $\nu i_{13/2}$ orbital emerges at $2J=31\,\hbar$ in $^{201}$Au and $^{203}$Au, and at $2J=27\,\hbar$ in $^{199}$Au. 
For the positive-parity states, the $\pi h_{11/2}$ and $\nu i_{13/2}$ orbitals maintain roughly one hole for these spins, with the sole exception of $^{205}$Au arising from the $N=126$ shell closure.

Fig.~\ref{fig:Hole_Occ_allA}(c) illustrates the SM1 average hole occupancies in the $\pi h_{11/2}$ and $\nu i_{13/2}$ orbitals as a function of the total spin ($2J$) for negative-parity states in odd–odd $^{198–204}$Au. As in the odd–even isotopes, the $\pi h_{11/2}$ orbital carries roughly one hole for low-lying negative-parity levels. The $\nu i_{13/2}$ hole occupancy remains close to one in $^{200–204}$Au, with a slight increase in $^{198}$Au. At high-spin, pair breaking in $\pi h_{11/2}$ introduces extra holes at $2J=36\,\hbar$ in $^{204}$Au, whereas $\nu i_{13/2}$ pair breaking occurs at $2J=44\,\hbar$ in both $^{198}$Au and $^{200}$Au. Notably, $^{202}$Au exhibits additional holes in both orbitals at $2J=44\,\hbar$. 
In the positive-parity sequences shown in Fig.~\ref{fig:Hole_Occ_allA}(d), the $\pi h_{11/2}$ orbital carries two holes, while $\nu i_{13/2}$ has one hole in $^{204}$Au. The situation is reversed in $^{198}$Au and $^{200}$Au, with one hole in $\pi h_{11/2}$ and two holes in $\nu i_{13/2}$. The case of $^{202}$Au exhibits a spin-dependent evolution of both $\pi h_{11/2}$ and $\nu i_{13/2}$ occupancies. 

The low-lying negative-parity levels in both odd–even ($^{199\text{–}205}$Au) and odd–odd ($^{198\text{–}204}$Au) isotopes generally contain approximately one hole in the $\pi h_{11/2}$ orbital. The average hole occupancy of the $\nu i_{13/2}$ orbital tends to increase with decreasing neutron number, reflecting the increase in mixing of configurations with more than one broken pair. However, in the case of odd–odd isotopes, the low-lying negative-parity levels retain roughly one hole in the $\nu i_{13/2}$ orbital. These features are consistent with a particle–core–coupled structure in the neutron-rich Au isotopes involving high-$j$ unique-parity orbitals, as discussed in detail in Ref.~\cite{Cho2026_PLB}.

\subsection{Proton and Neutron Spin Contributions}
\label{sec:config}

\begin{figure*}[p]
  \centering
  \includegraphics[width=0.92\textwidth]{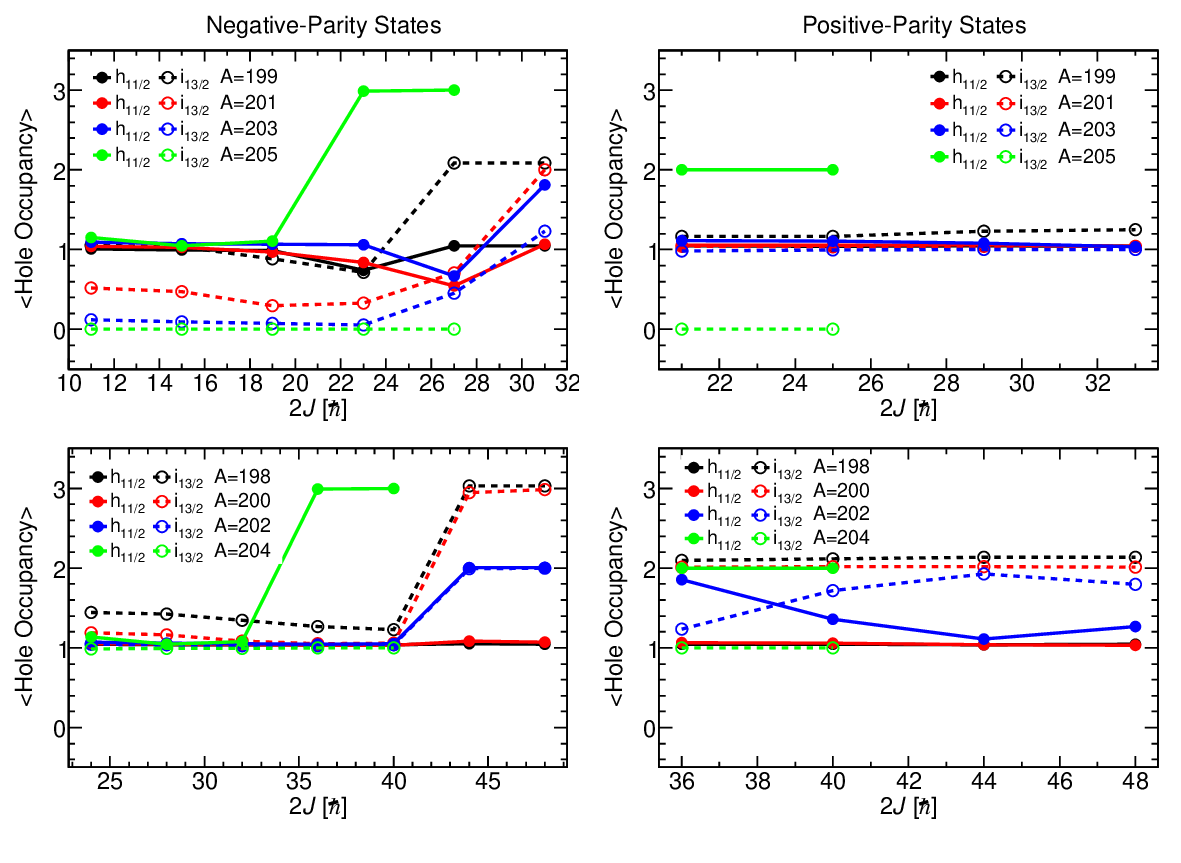}
  \vspace{-9 mm}\caption{(a)-(b) Shell-model calculated (SM1) average hole occupancies in the $\pi h_{11/2}$ (solid circles) and $\nu i_{13/2}$ (open circles) orbitals as a function of spin ($2J$) for negative-parity (a) and positive-parity (b) states in odd-even $^{199-205}$Au isotopes. (c)-(d) Same as (a) and (b) for odd-odd $^{198-204}$Au isotopes. 
  }
  \label{fig:Hole_Occ_allA}
  \centering
  \includegraphics[width=0.92\textwidth]{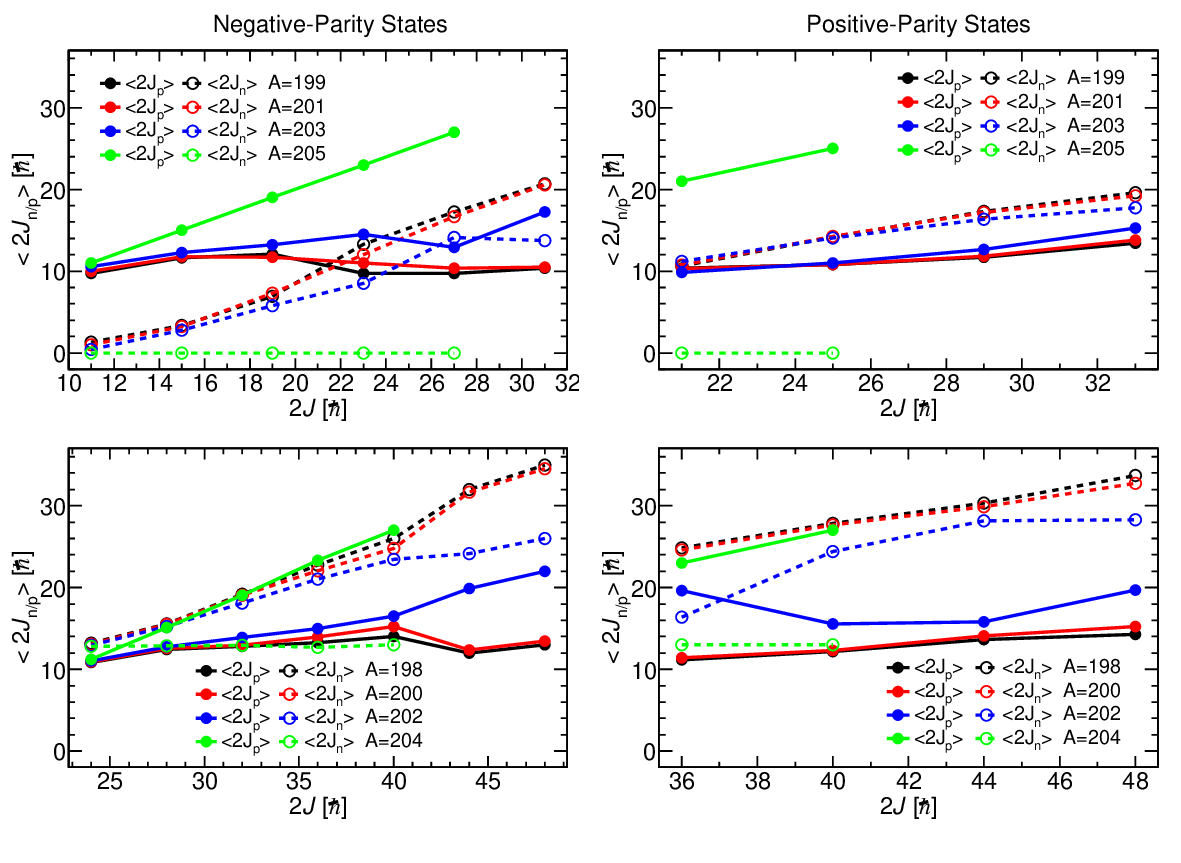}
  \vspace{-9 mm}\caption{(a)-(b) Shell-model calculated (SM1) neutron ($\langle2J_n\rangle$, open circles) and proton ($\langle2J_p\rangle$, solid circles) average spin contributions $2J_\text{n/p}$ as a function of $2J$ for negative-parity (a) and positive-parity (b) states in odd-even $^{199-205}$Au isotopes. (c)-(d) Same as (a) and (b) for odd-odd $^{198-204}$Au isotopes. 
  }
  \label{fig:JpJn_allA}
\end{figure*}
The spin contributions from protons and neutrons are investigated to see the respective role of protons and neutrons in the high-spin states of neutron-rich Au isotopes. Fig.~\ref{fig:JpJn_allA}(a) and (b) show the SM1 average spin contributions of the protons ($\langle 2J_p\rangle$) and the neutrons ($\langle 2J_n\rangle$), plotted against the total spin ($2J$) for negative- and positive-parity states in odd–even $^{199-205}$Au, respectively. In all cases except $^{205}$Au, where the contribution of the proton dominates, the value of $\langle 2J_p\rangle$ remains essentially constant at $11\,\hbar$. These results imply that the majority of the $\langle 2J_p\rangle$ is generated by a single $\pi h_{11/2}$ hole, although substantial mixing with the Hg core states is nevertheless present~\cite{Cho2026_PLB}. Meanwhile, $\langle 2J_n\rangle$ increases steadily as a function of spin. A similar behavior is observed in positive-parity states.

Fig.~\ref{fig:JpJn_allA}(c) shows the spin contributions of the negative-parity states in odd–odd $^{198–204}$Au. In $^{204}$Au, $\langle2J_n\rangle$ remains near $13\,\hbar$, corresponding to the spin carried by a single $\nu i_{13/2}$ hole. $^{198}$Au and $^{200}$Au show an increasing trend of $\langle2J_n\rangle$ with spin, whereas $\langle2J_p\rangle$ remains nearly constant, consistent with the spin contribution expected from a single $\pi h_{11/2}$ hole. In $^{202}$Au, both $\langle2J_p\rangle$ and $\langle2J_n\rangle$ gradually increase. The positive-parity states follow a similar pattern, as shown in Fig. \ref{fig:JpJn_allA}(d), except in $^{202}$Au where the contributions of proton and neutron intersect at $2J=40\,\hbar$.

In general, the average proton spin contribution exceeds that of neutrons near $N=126$, but gradually reduces with decreasing neutron number, leading to a crossover in $^{202}$Au where the neutron spin contribution becomes larger.
Across these states $\langle2J_p\rangle$ remains close to $11\,\hbar$ except for $N=126$, whereas $\langle2J_n\rangle$ increases steadily with total spin, underscoring the dominant role of neutrons in the building of high-spin excitations.

\section{Summary}
\label{summary}

Multi-nucleon transfer reactions between $^{136}$Xe and $^{198}$Pt at 7 MeV/u were used to access hard-to-reach neutron-rich Au isotopes in the vicinity of the $N=126$ shell closure. The complementary capabilities of VAMOS++, AGATA, and the CATLIFE detectors enabled prompt and delayed $\gamma$-ray spectroscopy of isotopically selected Target-Like Fragments. New and extended high-spin level schemes of $^{195-202}$Au were established above the $11/2^-$ (odd-$A$) and $12^-$ (even-$A$) isomeric states to understand the evolution of the structure of these Au isotopes as a function of neutron number.

The $\Delta J=2$ level sequences above the $11/2^-$ (odd-$A$) and $12^-$ (even-$A$) isomers are found to evolve smoothly toward the $N=126$ shell closure. 
These measured level energies exhibit dips at $N=119$ which become more pronounced in higher spin states. 
New ($25/2^+$) isomers are observed in $^{199}$Au and $^{201}$Au, while the corresponding states are non-isomeric in lighter isotopes ($N\leq118$). Additionally, odd-J levels built on the $11^-$ states seen in lighter odd-$N$ isotopes ($N\leq115$) are not observed in heavier isotopes ($N\geq117$).

Shell-model calculations are in reasonable agreement with the experimental results. 
The calculations show the near degeneracy of the $21/2^+$ and $25/2^+$ states in $^{199}$Au and $^{201}$Au, supporting the experimental observations. The calculations also predict the trend of increasing the excitation energies of the $11^-$ band relative to the $12^-$ band as a function of the neutron number, explaining the missing observation of the $11^-$ band in $N\ge117$.
The average hole occupancies show that, for negative-parity states, a single $\pi h_{11/2}$ hole persists in both even- and odd-$A$ nuclei, while a single $\nu i_{13/2}$ hole is maintained in the even-$A$ isotopes. 
For positive-parity states in odd-$A$ nuclei, both the $\pi h_{11/2}$ and $\nu i_{13/2}$ hole occupancies remain nearly constant for different spin states across isotopes.
The relative spin contributions of protons and neutrons were also analyzed. In odd-$A$ nuclei, the negative-parity states exhibit an almost constant proton spin contribution. 
In even-$A$ nuclei, a crossover occurs at $^{202}$Au. For nuclei with smaller neutron numbers, the neutron contribution becomes dominant in generating high-spin states, associated with a change in the hole occupancy of the proton $\pi h_{11/2}$ and neutron $\nu i_{13/2}$ orbitals. 

The excitation energy systematics in the Au isotopes hints towards possible structural changes at $N=119$, which is not fully reproduced by the present shell-model calculations. 
Interestingly, a similar feature was observed in the two neutron separation energy, $\Delta_{2n}(N)=S_{2n}(N)-S_{2n}(N+2)$, where a change of $\approx$500 keV appears at $N=119$ in the ground states of the Au isotopic chains. 
More experimental and theoretical effort is required to understand the evolution of the level energies at $N\approx119$. 

\section*{Acknowledgments}

The authors are grateful for the excellent support from the GANIL accelerator staff and the AGATA collaboration. 
YC, YS, SC acknowledge the support from the National Research Foundation of Korea (NRF) grants funded by the Korean government (MSIT), Grants No RS-2024-00436392, RS-2022-NR070152, 2022R1A2C2005093, No. 2019K2A9A2A10017576 and No. 2020R1A2C1005981. 
YC, YHK, KIH, YS, JP, JH and SB acknowledge the support from the Institute for Basic Science (IBS) of the Republic of Korea (Grants No.~IBS-R031-D1). 
AA acknowledges that the work was supported by the UK STFC and GANIL visiting researcher program.
CHK, MJK, GMG, and KC were supported by the National Research Foundation of Korea (RS-2024-00338255). 
JH was supported by the National Research Foundation of Korea (NRF) funded by the Ministry of Education, grant. no. 2020R1A6A3A03039081. 
CY and ML acknowledge the National Natural Science Foundation of China under Grant No. 12475129.  
RPV acknowledges work partially funded by MCIN/AEI/10.13039/501100011033, Spain, with grants PID2020-118265GB-C4, PID2023-150056NB-C4, PRTR-C17.I01, by Generalitat Valenciana, Spain, with grant CIPROM/2022/54, ASFAE/2022/031, and by the EU NextGenerationEU and FEDER funds.

\appendix*
\section{Table of transition energies and intensities of $^{195-202}$Au isotopes}
Summary of the observed $\gamma$-ray transitions in the studied nuclei. The energies ($E_\gamma$), relative intensities ($I_{\gamma+\text{ce}}$), spin and parity of the initial ($I_i$) and final ($I_f$) states along with the level energy of the initial state ($E_{level}$) are presented for prompt and delayed $\gamma$ rays. Intensities are corrected for detector efficiency and internal conversion calculated by BrIcc ~\cite{NIMA2008_KIBEDI}. Uncertainties in $\gamma$-ray energies are 1 keV unless otherwise mentioned. Some $\gamma$-ray transitions are listed under both Prompt and Delayed, as states below the isomer can be directly populated in the reaction and their de-excitation $\gamma$ rays are observed in the prompt spectrum.

\begin{table}[H]
  \centering
  \begin{minipage}{0.5\textwidth}
    \centering
    \caption{The $\gamma$-ray transitions in $^{195}$Au. The "x" denotes the unobserved low-energy transition from the ($31/2^-$) state to the ($29/2^+$) state.} 
    \scalebox{0.99}{
    \begin{tabular}{cccccc}
    \toprule
    & $E_{\text{level}}$ (keV) & $E_{\gamma}$ (keV) & $I_{\gamma+\text{ce}}$ & $I_i$ & $I_f$ \\
    \midrule
    \multirow{7}{*}{Prompt} 
    & x+2879  & 416.8  & 100(4) & $(35/2^-)$ & $(31/2^-)$ \\
    & x+3285  & 406.0  & 63(8) & $(39/2^-)$ & $(35/2^-)$ \\
    & x+3509  & 224.2  & 49(10) & $(43/2^-)$ & $(39/2^-)$ \\
    & x+4164  & 655.4  & 41(5)  & $(47/2^-)$ & $(43/2^-)$ \\
    & x+5127  & 962.6  & 26(4)  & $(51/2^-)$ & $(47/2^-)$ \\
    & x+3647  & 767.5  & 35(15)  & $\cdot$ & $(35/2^-)$ \\
    & x+4545  & 897.6  & 34(10)  & $\cdot$ & $\cdot$ \\
    \midrule
    \multirow{7}{*}{Delayed} 
    & 707   & 388.4  & 100(4) & $(15/2^-)$ & $11/2^-$ \\
    & 1425  & 718.0  & 112(6) & $(19/2^-)$ & $(15/2^-)$ \\
    & 1813  & 388.1  & 100(4) & $(21/2^+)$ & $(19/2^-)$ \\
    & 1981  & 168.1  & 53(2)  & $(25/2^+)$ & $(21/2^+)$ \\
    & 2242  & 261.2  & 43(3)  & $(27/2^+)$ & $(25/2^+)$ \\
    & 2421  & 178.9  & 18(2)  & $(29/2^+)$ & $(27/2^+)$ \\
    & 2462  & 481.2  & 11(1)  & $(29/2^+)$ & $(25/2^+)$ \\
    \bottomrule
    \end{tabular}
    }
    \label{tab:GammaSummary_195Au}
  \end{minipage}\hfill

  \begin{minipage}{0.5\textwidth}\
    \centering
    \caption{The $\gamma$-ray transitions in $^{197}$Au. The "x" denotes the unobserved low-energy transition from the ($31/2^-$) state to the ($29/2^+$) state.}
    \begin{tabular}{cccccc}
    \toprule
    & $E_{\text{level}}$ (keV) & $E_{\gamma}$ (keV) & $I_{\gamma+\text{ce}}$ & $I_i$ &  $I_f$ \\
    \midrule
    \multirow{3}{*}{Prompt} 
    & 767  & 357.9  & 100(8) & $(15/2^-)$ & $11/2^-$ \\
    & 1407  & 639.8  & 24(4) & $(19/2^-)$ & $(15/2^-)$ \\
    & 2097  & 260.9  & 58(3) & $(25/2^+)$ & $(21/2^+)$ \\
    \midrule
    \multirow{4}{*}{Prompt} 
    & x+2861  & 328.0  & 100(4) & $(35/2^-)$ & $(31/2^-)$ \\
    & x+3410  & 549.4  & 78(3)  & $(39/2^-)$ & $(35/2^-)$ \\
    & x+4179  & 769.1  & 55(18) & $(43/2^-)$ & $(39/2^-)$ \\
    & x+3876  & 465.2  & 28(2)  & $\cdot$    & $(39/2^-)$ \\
    \midrule
    \multirow{5}{*}{Delayed} 
    & 767   & 357.9  & 100(7) & $(15/2^-)$ & $11/2^-$ \\
    & 1407  & 639.7  & 104(7) & $(19/2^-)$ & $(15/2^-)$ \\
    & 1836  & 429.1  & 107(7) & $(21/2^+)$ & $(19/2^-)$ \\
    & 2097  & 260.9  & 86(5)  & $(25/2^+)$ & $(21/2^+)$ \\
    & 2533  & 435.5  & 93(6)  & $(29/2^+)$ & $(25/2^+)$ \\
    \bottomrule
    \end{tabular}
    \label{tab:GammaSummary_197Au}
  \end{minipage}\hfill
\end{table}

\begin{table}
    \centering
    \begin{minipage}{0.5\textwidth}
    \centering
    \caption{The $\gamma$-ray transitions in $^{199}$Au. The "x" denotes the unobserved low-energy transition from the ($25/2^+$) state to the ($21/2^+$) state.}
    \begin{tabular}{cccccc}
    \toprule
    & $E_{\text{level}}$ (keV) & $E_{\gamma}$ (keV) & $I_{\gamma+\text{ce}}$ & $I_i$ & $I_f$ \\
    \midrule
    \multirow{2}{*}{Prompt} 
    & 317  & 317.4  & 51(6)  & $(5/2^+)$ & $3/2^+$ \\
    & 494  & 493.8  & 100(8) & $(7/2^+)$ & $3/2^+$ \\
    \midrule
    \multirow{2}{*}{Prompt} 
    & 911  & 361.9  & 100(17) & $(15/2^-)$ & $(11/2^-)$ \\
    & 1505 & 594.4  & 79(21)  & $(19/2^-)$ & $(15/2^-)$ \\
    \midrule
    \multirow{2}{*}{Prompt} 
    & x+2439  & 358.5  & 100(19) & $(29/2^+)$ & $(25/2^+)$ \\
    & x+3037  & 597.7  & 52(18)  & $(33/2^+)$ & $(29/2^+)$ \\
    \midrule
    \multirow{1}{*}{Delayed} 
    & 494   & 493.9  & 100(5) & $(7/2^+)$ & $3/2^+$ \\
    \midrule
    \multirow{3}{*}{Delayed} 
    & 911   & 361.5  & 100(9)  & $(15/2^-)$ & $(11/2^-)$ \\
    & 1505  & 593.9  & 109(19) & $(19/2^-)$ & $(15/2^-)$ \\
    & 2080  & 574.3  & 46(11)  & $(21/2^+)$ & $(19/2^-)$ \\
    \bottomrule
    \end{tabular}
    \label{tab:GammaSummary_199Au}
  \end{minipage}\hfill

    \begin{minipage}{0.5\textwidth}
    \centering
    \caption{The $\gamma$-ray transitions in $^{201}$Au. The "x" denotes the unobserved low-energy transition from the ($25/2^+$) state to the ($21/2^+$) state.}
    \begin{tabular}{cccccc}
    \toprule
    & $E_{\text{level}}$ (keV) & $E_{\gamma}$ (keV) & $I_{\gamma+\text{ce}}$ & $I_i$ & $I_f$ \\
    \midrule
    \multirow{2}{*}{Prompt} 
    & 973   & 378.6  & 100(14) & $(15/2^-)$ & $11/2^-$ \\
    & 1612  & 638.4  & 32(7)   & $(19/2^-)$ & $(15/2^-)$ \\
    \midrule
    \multirow{2}{*}{Prompt} 
    & x+2554  & 362.7  & 100(6) & $(29/2^+)$ & $(25/2^+)$ \\
    & x+3253  & 699.2  & 57(5)  & $(33/2^+)$ & $(29/2^+)$ \\
    \midrule
    \multirow{1}{*}{Delayed} 
    & 554   & 554.1  & 100(10) & $(7/2^+)$ & $3/2^+$ \\
    \midrule
    \multirow{3}{*}{Delayed} 
    & 973   & 378.7  & 100(8)  & $(15/2^-)$ & $11/2^-$ \\
    & 1612  & 637.9  & 109(11) & $(19/2^-)$ & $(15/2^-)$ \\
    & 2192  & 579.6  & 90(8)   & $(21/2^+)$ & $(19/2^-)$ \\
    \bottomrule
    \end{tabular}
    \label{tab:GammaSummary_201Au}    
    \end{minipage}

    \begin{minipage}{0.5\textwidth}
    \centering
    \caption{The $\gamma$-ray transitions in $^{196}$Au.}
    \begin{tabular}{cccccc}
    \toprule
    & $E_{\text{level}}$ (keV) & $E_{\gamma}$ (keV) & $I_{\gamma+\text{ce}}$ & $I_i$ & $I_f$ \\
    \midrule
    \multirow{8}{*}{Prompt} 
    & 975   & 379.3  & 100(22) & $(14^-)$ & $12^-$ \\
    & 1680  & 705.5  & 66(14 ) & $(16^-)$ & $(14^-)$ \\
    & 2508  & 828    & 55(8)   & $(18^-)$ & $(16^-)$ \\
    & 3020  & 511.6  & 55(2)   & $(20^-)$ & $(18^-)$ \\
    & 3264  & 244    & 54(15)  & $(22^-)$ & $(20^-)$ \\
    & 3656  & 392.4  & 45(8)   & $(24^-)$ & $(22^-)$ \\
    & 1309  & 333.6  & 56(2)   & $\cdot$  & $(14^-)$ \\
    & 1923  & 614.6  & 27(2)   & $\cdot$  & $\cdot$ \\
    \bottomrule
    \end{tabular}
    \label{tab:GammaSummary_196Au}
    \end{minipage}

    \begin{minipage}{0.5\textwidth}
    \centering
    \caption{The $\gamma$-ray transitions in $^{198}$Au.}
      \begin{tabular}{cccccc}
      \toprule
      & $E_{\text{level}}$ (keV) & $E_{\gamma}$ (keV) & $I_{\gamma+\text{ce}}$ & $I_i$ & $I_f$ \\
      \midrule
      \multirow{3}{*}{Prompt} 
      & 492  & 179.8  & 100(15) & $(7^+)$ & $5^+$ \\
      & 696  & 203.5  & 57(3)   & $(8^+)$ & $(7^+)$ \\
      \midrule
      \multirow{7}{*}{Prompt} 
      & 1105  & 292.8  & 100(2) & $(14^-)$ & $(12^-)$ \\
      & 1636  & 530.9  & 88(1)  & $(16^-)$ & $(14^-)$ \\
      & 2383  & 746.8  & 56(2)  & $(18^-)$ & $(16^-)$ \\
      & 2410  & 774.2  & 45(5)  & $\cdot$ & $(16^-)$ \\
      & 2769  & 386.1  & 37(6)  & $(18^+)$ & $(18^-)$ \\
      & 3180  & 410.6  & 29(5)  & $(20^+)$ & $(18^+)$ \\
      & 3172  & 403.1  & 12(3)  & $\cdot$ & $(18^+)$ \\
      \midrule
      \multirow{2}{*}{Delayed} 
      & 215  & 215.0  & 100(6) & $4^-$ & $2^-$ \\
      & 312  & 97.4   & 92(5)  & $5^+$ & $4^-$ \\
      \bottomrule
      \end{tabular}
      \label{tab:GammaSummary_198Au}
    \end{minipage}
    
\end{table}

\begin{table}[H]
  \centering
  \caption{The $\gamma$-ray transitions in $^{200}$Au.}
  \begin{tabular}{cccccc}
  \toprule
  & $E_{\text{level}}$ (keV) & $E_{\gamma}$ (keV) & $I_{\gamma+\text{ce}}$ & $I_i$ & $I_f$ \\
  \midrule
  \multirow{6}{*}{Prompt} 
  & 1343  & 333.1  & 100(6) & $(14^-)$ & $12^-$ \\
  & 1936  & 592.7  & 66(38) & $(16^-)$ & $(14^-)$ \\
  & 2701  & 764.7  & 49(8)  & $(18^-)$ & $(16^-)$ \\
  & 2990  & 288.4  & 47(1)  & $(18^+)$ & $(18^-)$ \\
  & 3212  & 221.9  & 25(1)  & $(20^+)$ & $(18^+)$ \\
  & 3789  & 577.6  & 23(4)  & $(22^+)$ & $(20^+)$ \\
  \bottomrule
  \end{tabular}
  \label{tab:GammaSummary_200Au}
\end{table}

\begin{table}[H]
  \centering
  \caption{The $\gamma$-ray transitions in $^{202}$Au. The "x" denotes the level energy of the $(12^-)$ state.}
  \begin{tabular}{cccccc}
  \toprule
  & $E_{\text{level}}$ (keV) & $E_{\gamma}$ (keV) & $I_{\gamma+\text{ce}}$ & $I_i$ & $I_f$ \\
  \midrule
  \multirow{4}{*}{Prompt} 
  & x+325   & 324.8  & 100(22) & $(14^-)$ & $(12^-)$ \\
  & x+993   & 668.1  & 50(6)   & $(16^-)$ & $(14^-)$ \\
  & x+2040  & 1046.9 & 29(4)   & $(18^-)$ & $(16^-)$ \\
  & x+2475  & 435.4  & 28(5)   & $(18^+)$ & $(18^-)$ \\
  \bottomrule
  \end{tabular}
  \label{tab:GammaSummary_202Au}
\end{table}
\FloatBarrier

\input{output.bbl}
\end{document}

%% file: AuthorList.tex
\author{Y.~Cho~\orcidlink{0009-0000-1767-996X}}
\affiliation{Department of Physics and Astronomy, Seoul National University, Seoul 08826, Republic of Korea}
\affiliation{Institute for Nuclear and Particle Astrophysics, Seoul National University, Seoul 08826, Republic of Korea}
\affiliation{Center for Exotic Nuclear Studies, Institute for Basic Science, Daejeon 34126, Republic of Korea}
\author{Y.~H.~Kim~\orcidlink{0000-0002-7221-058X}}
\email{yunghee.kim@ibs.re.kr}
\affiliation{Center for Exotic Nuclear Studies, Institute for Basic Science, Daejeon 34126, Republic of Korea}
\author{A.~Navin~\orcidlink{ 0000-0003-2502-5864}}
\affiliation{Grand Acc\'el\'erateur National d'Ions Lourds, CEA/DRF-CNRS/IN2P3, 
Boulevard Henri Becquerel, F-14076 Caen, France}
\author{M.~Rejmund\orcidlink{0009-0009-8626-8756}}
\affiliation{Grand Acc\'el\'erateur National d'Ions Lourds, CEA/DRF-CNRS/IN2P3, 
Boulevard Henri Becquerel, F-14076 Caen, France}
\author{A.~Lemasson\orcidlink{0000-0002-9434-8520}}
\affiliation{Grand Acc\'el\'erateur National d'Ions Lourds, CEA/DRF-CNRS/IN2P3, 
Boulevard Henri Becquerel, F-14076 Caen, France}
\author{D.~Ramos~\orcidlink{0000-0003-1822-6537}}
\affiliation{Grand Acc\'el\'erateur National d'Ions Lourds, CEA/DRF-CNRS/IN2P3, 
Boulevard Henri Becquerel, F-14076 Caen, France}
\author{E.~Cl\'ement~\orcidlink{0000-0003-1887-717X}}
\affiliation{Grand Acc\'el\'erateur National d'Ions Lourds, CEA/DRF-CNRS/IN2P3, 
Boulevard Henri Becquerel, F-14076 Caen, France}
\author{C.~Yuan~\orcidlink{0000-0002-3495-3614}}
\affiliation{Sino-French Institute of Nuclear Engineering and Technology, Sun Yat-Sen University, Zhuhai, Guangdong 519082, China}
\author{M.~Liu~\orcidlink{0000-0001-8028-1487}}
\affiliation{Sino-French Institute of Nuclear Engineering and Technology, Sun Yat-Sen University, Zhuhai, Guangdong 519082, China}
\author{P.~van~Isacker~\orcidlink{0000-0003-3938-9220}}
\affiliation{Grand Acc\'el\'erateur National d'Ions Lourds, CEA/DRF-CNRS/IN2P3, 
Boulevard Henri Becquerel, F-14076 Caen, France}
\author{A.~N.~Andreyev~\orcidlink{0000-0003-2828-0262}}
\affiliation{School of Physics, Engineering and Technology, University of York, Heslington, York, North Yorkshire YO10 5DD, United Kingdom}
\author{J.~Dudouet~\orcidlink{0000-0002-9018-6763}}
\affiliation{Université Lyon, Université Claude Bernard Lyon 1, CNRS/IN2P3, IP2I Lyon, UMR 5822, F-69622 Villeurbanne, France} 
\author{S.~Choi\orcidlink{0000-0002-9448-969X}}
\affiliation{Department of Physics and Astronomy, Seoul National University, Seoul 08826, Republic of Korea}
\affiliation{Institute for Nuclear and Particle Astrophysics, Seoul National University, Seoul 08826, Republic of Korea}
\author{Y.~Son~\orcidlink{0009-0007-7117-4357}}
\affiliation{Department of Physics and Astronomy, Seoul National University, Seoul 08826, Republic of Korea}
\affiliation{Institute for Nuclear and Particle Astrophysics, Seoul National University, Seoul 08826, Republic of Korea}
\affiliation{Center for Exotic Nuclear Studies, Institute for Basic Science, Daejeon 34126, Republic of Korea}
\author{A.~Mukherjee~\orcidlink{0009-0009-3590-3668}}
\affiliation{Center for Exotic Nuclear Studies, Institute for Basic Science, Daejeon 34126, Republic of Korea}

\author{D.~Ackermann~\orcidlink{0000-0001-6284-1516}}
\affiliation{Grand Acc\'el\'erateur National d'Ions Lourds, CEA/DRF-CNRS/IN2P3, 
Boulevard Henri Becquerel, F-14076 Caen, France}

\author{G.~de Angelis~\orcidlink{0000-0002-0506-3815}}
\affiliation{INFN Laboratori Nazionali di Legnaro, IT-35020 Legnaro, Italy}

\author{S.~Bae~\orcidlink{0000-0003-0420-2163}}
\affiliation{Center for Exotic Nuclear Studies, Institute for Basic Science, Daejeon 34126, Republic of Korea} 

\author{R.~Banik~\orcidlink{0000-0001-6969-0458}}
\affiliation{Institute of Engineering \& Management, Kolkata 700091, India}

\author{S.~Bhattacharya}
\affiliation{Variable Energy Cyclotron Centre, 1/AF, Bidhannagar, Kolkata, West Bengal 700064, India}

\author{S.~Bhattacharyya~\orcidlink{0000-0002-1712-5825}}
\affiliation{Variable Energy Cyclotron Centre, 1/AF, Bidhannagar, Kolkata, West Bengal 700064, India}

\author{K.~Chae~\orcidlink{0000-0003-1819-079X}}
\affiliation{Department of Physics, Sungkyunkwan University, Suwon 16419, Korea}

\author{F.~Didierjean~\orcidlink{0009-0002-4950-3162}}
\affiliation{University of Strasbourg, CNRS/IPHC UMR7178, F-67000 Strasbourg, France}

\author{C.~Foug\'eres~\orcidlink{0000-0002-1236-4739}}
\affiliation{Grand Acc\'el\'erateur National d'Ions Lourds, CEA/DRF-CNRS/IN2P3, 
Boulevard Henri Becquerel, F-14076 Caen, France} 

\author{G.~de France~\orcidlink{0000-0002-7439-1759}}
\affiliation{Grand Acc\'el\'erateur National d'Ions Lourds, CEA/DRF-CNRS/IN2P3, 
Boulevard Henri Becquerel, F-14076 Caen, France}

\author{G.~Fremont}
\affiliation{Grand Acc\'el\'erateur National d'Ions Lourds, CEA/DRF-CNRS/IN2P3, 
Boulevard Henri Becquerel, F-14076 Caen, France}

\author{J.~Goupil}
\affiliation{Grand Acc\'el\'erateur National d'Ions Lourds, CEA/DRF-CNRS/IN2P3, 
Boulevard Henri Becquerel, F-14076 Caen, France}

\author{G.~M.~Gu~\orcidlink{0009-0007-5112-8505}}
\affiliation{Department of Physics, Sungkyunkwan University, Suwon 16419, Korea}

\author{J.~Ha~\orcidlink{0000-0003-4843-8573}}
\affiliation{Dipartimento di Fisica e Astronomia, Università degli Studi di Padova, Via Marzolo 8, Padova, I-35131, Italy}

\author{K.~I.~Hahn~\orcidlink{0000-0002-9250-6484}}
\affiliation{Center for Exotic Nuclear Studies, Institute for Basic Science, Daejeon 34126, Republic of Korea}

\author{Y.~Hirayama~\orcidlink{0000-0002-9485-1177}}
\affiliation{Wako Nuclear Science Center, IPNS, High Energy Accelerator Research Organization (KEK), Wako, Saitama 351-0198, Japan}

\author{S.~Jeong~\orcidlink{0000-0002-0054-2278}}
\affiliation{Wako Nuclear Science Center, IPNS, High Energy Accelerator Research Organization (KEK), Wako, Saitama 351-0198, Japan}

\author{C.~H.~Kim~\orcidlink{0000-0002-4614-4793}}
\affiliation{Department of Physics, Sungkyunkwan University, Suwon 16419, Korea}

\author{M.~J.~Kim~\orcidlink{0000-0002-4372-5592}}
\affiliation{Department of Physics, Sungkyunkwan University, Suwon 16419, Korea}

\author{S.~H.~Kim~\orcidlink{0009-0008-6013-5982}}
\affiliation{Department of Physics, Sungkyunkwan University, Suwon 16419, Korea}

\author{W.~Korten~\orcidlink{0000-0002-3940-0816}}
\affiliation{CEA/DRF/IRFU, 17 rue des Martyrs 38054 Grenoble Cedex 9 France}

\author{P.~Marini~\orcidlink{0000-0001-9537-8126}}
\affiliation{Grand Acc\'el\'erateur National d'Ions Lourds, CEA/DRF-CNRS/IN2P3, 
Boulevard Henri Becquerel, F-14076 Caen, France}
\affiliation{LP2I Bordeaux, UMR5797, Université de Bordeaux, CNRS, F-33170, Gradignan, France}

\author{H.~Miyatake~\orcidlink{0000-0002-6888-8507}}
\affiliation{Wako Nuclear Science Center, IPNS, High Energy Accelerator Research Organization (KEK), Wako, Saitama 351-0198, Japan}

\author{M.~Mukai~\orcidlink{0000-0002-7146-2961}}
\affiliation{Wako Nuclear Science Center, IPNS, High Energy Accelerator Research Organization (KEK), Wako, Saitama 351-0198, Japan}
\affiliation{RIKEN Nishina Center, 2-1 Hirosawa, Wako, Saitama 351-0198, Japan} %

\author{G.~Mukherjee~\orcidlink{0000-0003-0457-4012}}
\affiliation{Variable Energy Cyclotron Centre, 1/AF, Bidhannagar, Kolkata, West Bengal 700064, India}

\author{T.~Niwase~\orcidlink{0000-0002-0054-2278}}
\affiliation{Wako Nuclear Science Center, IPNS, High Energy Accelerator Research Organization (KEK), Wako, Saitama 351-0198, Japan}

\author{T.~Otsuka~\orcidlink{0000-0002-1593-5322}}
\affiliation{Department of Physics, University of Tokyo, Hongo, Bunkyo-ku, Tokyo 113-0033, Japan}

\author{J.~Park~\orcidlink{0000-0003-0866-7523}}
\affiliation{Center for Exotic Nuclear Studies, Institute for Basic Science, Daejeon 34126, Republic of Korea}
\affiliation{Department of Physics, Hope College, Michigan 49423, United States}

\author{R.~M.~P\'erez-Vidal~\orcidlink{0000-0002-4075-4152}}
\affiliation{Instituto de Física Corpuscular, CSIC-Universidad de Valencia, Valencia E-46980, Spain} 

\author{F.~Recchia~\orcidlink{0000-0002-8428-0112}}
\affiliation{University of Padua, Via 8 Febbraio, 1848, 2, 35122 Padua, Italy}

\author{K.~Rezynkina~\orcidlink{0000-0002-3017-6440}}
\affiliation{Istituto Nazionale di Fisica Nucleare, Sezione di Padova, I-35131 Padova, Italy}

\author{P.~Schury~\orcidlink{0000-0002-7094-6761}}
\affiliation{RIKEN Nishina Center, 2-1 Hirosawa, Wako, Saitama 351-0198, Japan} 

\author{J. C.~Thomas~\orcidlink{0000-0001-6618-9772}}
\affiliation{Grand Acc\'el\'erateur National d'Ions Lourds, CEA/DRF-CNRS/IN2P3, 
Boulevard Henri Becquerel, F-14076 Caen, France}

\author{D.~T.~Kattikat~Melcom~\orcidlink{0009-0001-0427-5584}}
\affiliation{LP2I Bordeaux, UMR5797, Université de Bordeaux, CNRS, F-33170, Gradignan, France}

\author{I.~Tsekhanovich~\orcidlink{0000-0002-8188-544X}}
\affiliation{LP2I Bordeaux, UMR5797, Université de Bordeaux, CNRS, F-33170, Gradignan, France}

\author{A.~Utepov~\orcidlink{0000-0002-1015-7285}}
\affiliation{Grand Acc\'el\'erateur National d'Ions Lourds, CEA/DRF-CNRS/IN2P3, 
Boulevard Henri Becquerel, F-14076 Caen, France} 

\author{Y.~X.~Watanabe~\orcidlink{0000-0003-2425-7188}}
\affiliation{Wako Nuclear Science Center, IPNS, High Energy Accelerator Research Organization (KEK), Wako, Saitama 351-0198, Japan}

%% file: output.bbl
%

%% file: main.bbl
\begin{thebibliography}{83}%
\makeatletter
\providecommand \@ifxundefined [1]{%
 \@ifx{#1\undefined}
}%
\providecommand \@ifnum [1]{%
 \ifnum #1\expandafter \@firstoftwo
 \else \expandafter \@secondoftwo
 \fi
}%
\providecommand \@ifx [1]{%
 \ifx #1\expandafter \@firstoftwo
 \else \expandafter \@secondoftwo
 \fi
}%
\providecommand \natexlab [1]{#1}%
\providecommand \enquote  [1]{``#1''}%
\providecommand \bibnamefont  [1]{#1}%
\providecommand \bibfnamefont [1]{#1}%
\providecommand \citenamefont [1]{#1}%
\providecommand \href@noop [0]{\@secondoftwo}%
\providecommand \href [0]{\begingroup \@sanitize@url \@href}%
\providecommand \@href[1]{\@@startlink{#1}\@@href}%
\providecommand \@@href[1]{\endgroup#1\@@endlink}%
\providecommand \@sanitize@url [0]{\catcode `\\12\catcode `\$12\catcode `\&12\catcode `\#12\catcode `\^12\catcode `\_12\catcode `\%12\relax}%
\providecommand \@@startlink[1]{}%
\providecommand \@@endlink[0]{}%
\providecommand \url  [0]{\begingroup\@sanitize@url \@url }%
\providecommand \@url [1]{\endgroup\@href {#1}{\urlprefix }}%
\providecommand \urlprefix  [0]{URL }%
\providecommand \Eprint [0]{\href }%
\providecommand \doibase [0]{https://doi.org/}%
\providecommand \selectlanguage [0]{\@gobble}%
\providecommand \bibinfo  [0]{\@secondoftwo}%
\providecommand \bibfield  [0]{\@secondoftwo}%
\providecommand \translation [1]{[#1]}%
\providecommand \BibitemOpen [0]{}%
\providecommand \bibitemStop [0]{}%
\providecommand \bibitemNoStop [0]{.\EOS\space}%
\providecommand \EOS [0]{\spacefactor3000\relax}%
\providecommand \BibitemShut  [1]{\csname bibitem#1\endcsname}%
\let\auto@bib@innerbib\@empty
\bibitem [{\citenamefont {Cho}\ \emph {et~al.}(2026)\citenamefont {Cho}, \citenamefont {Kim}, \citenamefont {Navin}, \citenamefont {Rejmund}, \citenamefont {Lemasson}, \citenamefont {Ramos}, \citenamefont {Clément}, \citenamefont {Dudouet}, \citenamefont {{Van Isacker}}, \citenamefont {Yuan}, \citenamefont {Liu}, \citenamefont {Andreyev}, \citenamefont {Son}, \citenamefont {Choi}, \citenamefont {Ackermann}, \citenamefont {{de Angelis}}, \citenamefont {Bae}, \citenamefont {Banik}, \citenamefont {Bhattacharya}, \citenamefont {Bhattacharyya}, \citenamefont {Chae}, \citenamefont {Didierjean}, \citenamefont {Fougères}, \citenamefont {{de France}}, \citenamefont {Fremont}, \citenamefont {Goupil}, \citenamefont {Gu}, \citenamefont {Ha}, \citenamefont {Hahn}, \citenamefont {Hirayama}, \citenamefont {Jeong}, \citenamefont {{Kattikat Melcom}}, \citenamefont {Kim}, \citenamefont {Kim}, \citenamefont {Kim}, \citenamefont {Korten}, \citenamefont {Marini}, \citenamefont {Miyatake}, \citenamefont {Mukai}, \citenamefont
  {Mukherjee}, \citenamefont {Mukherjee}, \citenamefont {Niwase}, \citenamefont {Otsuka}, \citenamefont {Park}, \citenamefont {Pérez-Vidal}, \citenamefont {Recchia}, \citenamefont {Rezynkina}, \citenamefont {Schury}, \citenamefont {Thomas}, \citenamefont {Tsekhanovich}, \citenamefont {Utepov},\ and\ \citenamefont {Watanabe}}]{Cho2026_PLB}%
  \BibitemOpen
  \bibfield  {author} {\bibinfo {author} {\bibfnamefont {Y.}~\bibnamefont {Cho}}, \bibinfo {author} {\bibfnamefont {Y.}~\bibnamefont {Kim}}, \bibinfo {author} {\bibfnamefont {A.}~\bibnamefont {Navin}}, \bibinfo {author} {\bibfnamefont {M.}~\bibnamefont {Rejmund}}, \bibinfo {author} {\bibfnamefont {A.}~\bibnamefont {Lemasson}}, \bibinfo {author} {\bibfnamefont {D.}~\bibnamefont {Ramos}}, \bibinfo {author} {\bibfnamefont {E.}~\bibnamefont {Clément}}, \bibinfo {author} {\bibfnamefont {J.}~\bibnamefont {Dudouet}}, \bibinfo {author} {\bibfnamefont {P.}~\bibnamefont {{Van Isacker}}}, \bibinfo {author} {\bibfnamefont {C.}~\bibnamefont {Yuan}}, \bibinfo {author} {\bibfnamefont {M.}~\bibnamefont {Liu}}, \bibinfo {author} {\bibfnamefont {A.}~\bibnamefont {Andreyev}}, \bibinfo {author} {\bibfnamefont {Y.}~\bibnamefont {Son}}, \bibinfo {author} {\bibfnamefont {S.}~\bibnamefont {Choi}}, \bibinfo {author} {\bibfnamefont {D.}~\bibnamefont {Ackermann}}, \bibinfo {author} {\bibfnamefont {G.}~\bibnamefont {{de Angelis}}},
  \bibinfo {author} {\bibfnamefont {S.}~\bibnamefont {Bae}}, \bibinfo {author} {\bibfnamefont {R.}~\bibnamefont {Banik}}, \bibinfo {author} {\bibfnamefont {S.}~\bibnamefont {Bhattacharya}}, \bibinfo {author} {\bibfnamefont {S.}~\bibnamefont {Bhattacharyya}}, \bibinfo {author} {\bibfnamefont {K.}~\bibnamefont {Chae}}, \bibinfo {author} {\bibfnamefont {F.}~\bibnamefont {Didierjean}}, \bibinfo {author} {\bibfnamefont {C.}~\bibnamefont {Fougères}}, \bibinfo {author} {\bibfnamefont {G.}~\bibnamefont {{de France}}}, \bibinfo {author} {\bibfnamefont {G.}~\bibnamefont {Fremont}}, \bibinfo {author} {\bibfnamefont {J.}~\bibnamefont {Goupil}}, \bibinfo {author} {\bibfnamefont {G.}~\bibnamefont {Gu}}, \bibinfo {author} {\bibfnamefont {J.}~\bibnamefont {Ha}}, \bibinfo {author} {\bibfnamefont {K.}~\bibnamefont {Hahn}}, \bibinfo {author} {\bibfnamefont {Y.}~\bibnamefont {Hirayama}}, \bibinfo {author} {\bibfnamefont {S.}~\bibnamefont {Jeong}}, \bibinfo {author} {\bibfnamefont {D.}~\bibnamefont {{Kattikat Melcom}}}, \bibinfo
  {author} {\bibfnamefont {C.}~\bibnamefont {Kim}}, \bibinfo {author} {\bibfnamefont {M.}~\bibnamefont {Kim}}, \bibinfo {author} {\bibfnamefont {S.}~\bibnamefont {Kim}}, \bibinfo {author} {\bibfnamefont {W.}~\bibnamefont {Korten}}, \bibinfo {author} {\bibfnamefont {P.}~\bibnamefont {Marini}}, \bibinfo {author} {\bibfnamefont {H.}~\bibnamefont {Miyatake}}, \bibinfo {author} {\bibfnamefont {M.}~\bibnamefont {Mukai}}, \bibinfo {author} {\bibfnamefont {A.}~\bibnamefont {Mukherjee}}, \bibinfo {author} {\bibfnamefont {G.}~\bibnamefont {Mukherjee}}, \bibinfo {author} {\bibfnamefont {T.}~\bibnamefont {Niwase}}, \bibinfo {author} {\bibfnamefont {T.}~\bibnamefont {Otsuka}}, \bibinfo {author} {\bibfnamefont {J.}~\bibnamefont {Park}}, \bibinfo {author} {\bibfnamefont {R.}~\bibnamefont {Pérez-Vidal}}, \bibinfo {author} {\bibfnamefont {F.}~\bibnamefont {Recchia}}, \bibinfo {author} {\bibfnamefont {K.}~\bibnamefont {Rezynkina}}, \bibinfo {author} {\bibfnamefont {P.}~\bibnamefont {Schury}}, \bibinfo {author} {\bibfnamefont
  {J.}~\bibnamefont {Thomas}}, \bibinfo {author} {\bibfnamefont {I.}~\bibnamefont {Tsekhanovich}}, \bibinfo {author} {\bibfnamefont {A.}~\bibnamefont {Utepov}},\ and\ \bibinfo {author} {\bibfnamefont {Y.}~\bibnamefont {Watanabe}},\ }\href {https://doi.org/https://doi.org/10.1016/j.physletb.2026.140558} {\bibfield  {journal} {\bibinfo  {journal} {Physics Letters B}\ }\textbf {\bibinfo {volume} {878}},\ \bibinfo {pages} {140558} (\bibinfo {year} {2026})}\BibitemShut {NoStop}%
\bibitem [{\citenamefont {Herling}\ and\ \citenamefont {Kuo}(1972)}]{herling1972}%
  \BibitemOpen
  \bibfield  {author} {\bibinfo {author} {\bibfnamefont {G.}~\bibnamefont {Herling}}\ and\ \bibinfo {author} {\bibfnamefont {T.}~\bibnamefont {Kuo}},\ }\href {https://doi.org/https://doi.org/10.1016/0375-9474(72)90905-0} {\bibfield  {journal} {\bibinfo  {journal} {Nuclear Physics A}\ }\textbf {\bibinfo {volume} {181}},\ \bibinfo {pages} {113} (\bibinfo {year} {1972})}\BibitemShut {NoStop}%
\bibitem [{\citenamefont {Warburton}\ and\ \citenamefont {Brown}(1991)}]{warburton1991}%
  \BibitemOpen
  \bibfield  {author} {\bibinfo {author} {\bibfnamefont {E.~K.}\ \bibnamefont {Warburton}}\ and\ \bibinfo {author} {\bibfnamefont {B.~A.}\ \bibnamefont {Brown}},\ }\href {https://doi.org/10.1103/PhysRevC.43.602} {\bibfield  {journal} {\bibinfo  {journal} {Phys. Rev. C}\ }\textbf {\bibinfo {volume} {43}},\ \bibinfo {pages} {602} (\bibinfo {year} {1991})}\BibitemShut {NoStop}%
\bibitem [{\citenamefont {Yuan}\ \emph {et~al.}(2022)\citenamefont {Yuan}, \citenamefont {Liu}, \citenamefont {Shimizu}, \citenamefont {Podoly\'ak}, \citenamefont {Suzuki}, \citenamefont {Otsuka},\ and\ \citenamefont {Liu}}]{yuan2022}%
  \BibitemOpen
  \bibfield  {author} {\bibinfo {author} {\bibfnamefont {C.}~\bibnamefont {Yuan}}, \bibinfo {author} {\bibfnamefont {M.}~\bibnamefont {Liu}}, \bibinfo {author} {\bibfnamefont {N.}~\bibnamefont {Shimizu}}, \bibinfo {author} {\bibfnamefont {Z.}~\bibnamefont {Podoly\'ak}}, \bibinfo {author} {\bibfnamefont {T.}~\bibnamefont {Suzuki}}, \bibinfo {author} {\bibfnamefont {T.}~\bibnamefont {Otsuka}},\ and\ \bibinfo {author} {\bibfnamefont {Z.}~\bibnamefont {Liu}},\ }\href {https://doi.org/10.1103/PhysRevC.106.044314} {\bibfield  {journal} {\bibinfo  {journal} {Phys. Rev. C}\ }\textbf {\bibinfo {volume} {106}},\ \bibinfo {pages} {044314} (\bibinfo {year} {2022})}\BibitemShut {NoStop}%
\bibitem [{\citenamefont {Steer}\ \emph {et~al.}(2011)\citenamefont {Steer}, \citenamefont {Podoly\'ak}, \citenamefont {Pietri}, \citenamefont {G\'orska}, \citenamefont {Grawe}, \citenamefont {Maier}, \citenamefont {Regan}, \citenamefont {Rudolph}, \citenamefont {Garnsworthy}, \citenamefont {Hoischen}, \citenamefont {Gerl}, \citenamefont {Wollersheim}, \citenamefont {Becker}, \citenamefont {Bednarczyk}, \citenamefont {C\'aceres}, \citenamefont {Doornenbal}, \citenamefont {Geissel}, \citenamefont {Gr\ifmmode~\mbox{\k{e}}\else \k{e}\fi{}bosz}, \citenamefont {Kelic}, \citenamefont {Kojouharov}, \citenamefont {Kurz}, \citenamefont {Montes}, \citenamefont {Prokopwicz}, \citenamefont {Saito}, \citenamefont {Schaffner}, \citenamefont {Tashenov}, \citenamefont {Heinz}, \citenamefont {Pf\"utzner}, \citenamefont {Kurtukian-Nieto}, \citenamefont {Benzoni}, \citenamefont {Jungclaus}, \citenamefont {Balabanski}, \citenamefont {Bowry}, \citenamefont {Brandau}, \citenamefont {Brown}, \citenamefont {Bruce}, \citenamefont
  {Catford}, \citenamefont {Cullen}, \citenamefont {Dombr\'adi}, \citenamefont {Estevez}, \citenamefont {Gelletly}, \citenamefont {Ilie}, \citenamefont {Jolie}, \citenamefont {Jones}, \citenamefont {Kmiecik}, \citenamefont {Kondev}, \citenamefont {Kr\"ucken}, \citenamefont {Lalkovski}, \citenamefont {Liu}, \citenamefont {Maj}, \citenamefont {Myalski}, \citenamefont {Schwertel}, \citenamefont {Shizuma}, \citenamefont {Walker}, \citenamefont {Werner-Malento},\ and\ \citenamefont {Wieland}}]{Steer2011}%
  \BibitemOpen
  \bibfield  {author} {\bibinfo {author} {\bibfnamefont {S.~J.}\ \bibnamefont {Steer}}, \bibinfo {author} {\bibfnamefont {Z.}~\bibnamefont {Podoly\'ak}}, \bibinfo {author} {\bibfnamefont {S.}~\bibnamefont {Pietri}}, \bibinfo {author} {\bibfnamefont {M.}~\bibnamefont {G\'orska}}, \bibinfo {author} {\bibfnamefont {H.}~\bibnamefont {Grawe}}, \bibinfo {author} {\bibfnamefont {K.~H.}\ \bibnamefont {Maier}}, \bibinfo {author} {\bibfnamefont {P.~H.}\ \bibnamefont {Regan}}, \bibinfo {author} {\bibfnamefont {D.}~\bibnamefont {Rudolph}}, \bibinfo {author} {\bibfnamefont {A.~B.}\ \bibnamefont {Garnsworthy}}, \bibinfo {author} {\bibfnamefont {R.}~\bibnamefont {Hoischen}}, \bibinfo {author} {\bibfnamefont {J.}~\bibnamefont {Gerl}}, \bibinfo {author} {\bibfnamefont {H.~J.}\ \bibnamefont {Wollersheim}}, \bibinfo {author} {\bibfnamefont {F.}~\bibnamefont {Becker}}, \bibinfo {author} {\bibfnamefont {P.}~\bibnamefont {Bednarczyk}}, \bibinfo {author} {\bibfnamefont {L.}~\bibnamefont {C\'aceres}}, \bibinfo {author}
  {\bibfnamefont {P.}~\bibnamefont {Doornenbal}}, \bibinfo {author} {\bibfnamefont {H.}~\bibnamefont {Geissel}}, \bibinfo {author} {\bibfnamefont {J.}~\bibnamefont {Gr\ifmmode~\mbox{\k{e}}\else \k{e}\fi{}bosz}}, \bibinfo {author} {\bibfnamefont {A.}~\bibnamefont {Kelic}}, \bibinfo {author} {\bibfnamefont {I.}~\bibnamefont {Kojouharov}}, \bibinfo {author} {\bibfnamefont {N.}~\bibnamefont {Kurz}}, \bibinfo {author} {\bibfnamefont {F.}~\bibnamefont {Montes}}, \bibinfo {author} {\bibfnamefont {W.}~\bibnamefont {Prokopwicz}}, \bibinfo {author} {\bibfnamefont {T.}~\bibnamefont {Saito}}, \bibinfo {author} {\bibfnamefont {H.}~\bibnamefont {Schaffner}}, \bibinfo {author} {\bibfnamefont {S.}~\bibnamefont {Tashenov}}, \bibinfo {author} {\bibfnamefont {A.}~\bibnamefont {Heinz}}, \bibinfo {author} {\bibfnamefont {M.}~\bibnamefont {Pf\"utzner}}, \bibinfo {author} {\bibfnamefont {T.}~\bibnamefont {Kurtukian-Nieto}}, \bibinfo {author} {\bibfnamefont {G.}~\bibnamefont {Benzoni}}, \bibinfo {author} {\bibfnamefont
  {A.}~\bibnamefont {Jungclaus}}, \bibinfo {author} {\bibfnamefont {D.~L.}\ \bibnamefont {Balabanski}}, \bibinfo {author} {\bibfnamefont {M.}~\bibnamefont {Bowry}}, \bibinfo {author} {\bibfnamefont {C.}~\bibnamefont {Brandau}}, \bibinfo {author} {\bibfnamefont {A.}~\bibnamefont {Brown}}, \bibinfo {author} {\bibfnamefont {A.~M.}\ \bibnamefont {Bruce}}, \bibinfo {author} {\bibfnamefont {W.~N.}\ \bibnamefont {Catford}}, \bibinfo {author} {\bibfnamefont {I.~J.}\ \bibnamefont {Cullen}}, \bibinfo {author} {\bibfnamefont {Z.}~\bibnamefont {Dombr\'adi}}, \bibinfo {author} {\bibfnamefont {M.~E.}\ \bibnamefont {Estevez}}, \bibinfo {author} {\bibfnamefont {W.}~\bibnamefont {Gelletly}}, \bibinfo {author} {\bibfnamefont {G.}~\bibnamefont {Ilie}}, \bibinfo {author} {\bibfnamefont {J.}~\bibnamefont {Jolie}}, \bibinfo {author} {\bibfnamefont {G.~A.}\ \bibnamefont {Jones}}, \bibinfo {author} {\bibfnamefont {M.}~\bibnamefont {Kmiecik}}, \bibinfo {author} {\bibfnamefont {F.~G.}\ \bibnamefont {Kondev}}, \bibinfo {author}
  {\bibfnamefont {R.}~\bibnamefont {Kr\"ucken}}, \bibinfo {author} {\bibfnamefont {S.}~\bibnamefont {Lalkovski}}, \bibinfo {author} {\bibfnamefont {Z.}~\bibnamefont {Liu}}, \bibinfo {author} {\bibfnamefont {A.}~\bibnamefont {Maj}}, \bibinfo {author} {\bibfnamefont {S.}~\bibnamefont {Myalski}}, \bibinfo {author} {\bibfnamefont {S.}~\bibnamefont {Schwertel}}, \bibinfo {author} {\bibfnamefont {T.}~\bibnamefont {Shizuma}}, \bibinfo {author} {\bibfnamefont {P.~M.}\ \bibnamefont {Walker}}, \bibinfo {author} {\bibfnamefont {E.}~\bibnamefont {Werner-Malento}},\ and\ \bibinfo {author} {\bibfnamefont {O.}~\bibnamefont {Wieland}},\ }\href {https://doi.org/10.1103/PhysRevC.84.044313} {\bibfield  {journal} {\bibinfo  {journal} {Phys. Rev. C}\ }\textbf {\bibinfo {volume} {84}},\ \bibinfo {pages} {044313} (\bibinfo {year} {2011})}\BibitemShut {NoStop}%
\bibitem [{\citenamefont {Podolyák}\ \emph {et~al.}(2009)\citenamefont {Podolyák}, \citenamefont {Farrelly}, \citenamefont {Regan}, \citenamefont {Garnsworthy}, \citenamefont {Steer}, \citenamefont {Górska}, \citenamefont {Benlliure}, \citenamefont {Casarejos}, \citenamefont {Pietri}, \citenamefont {Gerl}, \citenamefont {Wollersheim}, \citenamefont {Kumar}, \citenamefont {Molina}, \citenamefont {Algora}, \citenamefont {Alkhomashi}, \citenamefont {Benzoni}, \citenamefont {Blazhev}, \citenamefont {Boutachkov}, \citenamefont {Bruce}, \citenamefont {Caceres}, \citenamefont {Cullen}, \citenamefont {{Denis Bacelar}}, \citenamefont {Doornenbal}, \citenamefont {Estevez}, \citenamefont {Fujita}, \citenamefont {Gelletly}, \citenamefont {Geissel}, \citenamefont {Grawe}, \citenamefont {Grȩbosz}, \citenamefont {Hoischen}, \citenamefont {Kojouharov}, \citenamefont {Lalkovski}, \citenamefont {Liu}, \citenamefont {Maier}, \citenamefont {Mihai}, \citenamefont {Mücher}, \citenamefont {Rubio}, \citenamefont {Schaffner},
  \citenamefont {Tamii}, \citenamefont {Tashenov}, \citenamefont {Valiente-Dobón}, \citenamefont {Walker},\ and\ \citenamefont {Woods}}]{PODOLYAK2009}%
  \BibitemOpen
  \bibfield  {author} {\bibinfo {author} {\bibfnamefont {Z.}~\bibnamefont {Podolyák}}, \bibinfo {author} {\bibfnamefont {G.}~\bibnamefont {Farrelly}}, \bibinfo {author} {\bibfnamefont {P.}~\bibnamefont {Regan}}, \bibinfo {author} {\bibfnamefont {A.}~\bibnamefont {Garnsworthy}}, \bibinfo {author} {\bibfnamefont {S.}~\bibnamefont {Steer}}, \bibinfo {author} {\bibfnamefont {M.}~\bibnamefont {Górska}}, \bibinfo {author} {\bibfnamefont {J.}~\bibnamefont {Benlliure}}, \bibinfo {author} {\bibfnamefont {E.}~\bibnamefont {Casarejos}}, \bibinfo {author} {\bibfnamefont {S.}~\bibnamefont {Pietri}}, \bibinfo {author} {\bibfnamefont {J.}~\bibnamefont {Gerl}}, \bibinfo {author} {\bibfnamefont {H.}~\bibnamefont {Wollersheim}}, \bibinfo {author} {\bibfnamefont {R.}~\bibnamefont {Kumar}}, \bibinfo {author} {\bibfnamefont {F.}~\bibnamefont {Molina}}, \bibinfo {author} {\bibfnamefont {A.}~\bibnamefont {Algora}}, \bibinfo {author} {\bibfnamefont {N.}~\bibnamefont {Alkhomashi}}, \bibinfo {author} {\bibfnamefont {G.}~\bibnamefont
  {Benzoni}}, \bibinfo {author} {\bibfnamefont {A.}~\bibnamefont {Blazhev}}, \bibinfo {author} {\bibfnamefont {P.}~\bibnamefont {Boutachkov}}, \bibinfo {author} {\bibfnamefont {A.}~\bibnamefont {Bruce}}, \bibinfo {author} {\bibfnamefont {L.}~\bibnamefont {Caceres}}, \bibinfo {author} {\bibfnamefont {I.}~\bibnamefont {Cullen}}, \bibinfo {author} {\bibfnamefont {A.}~\bibnamefont {{Denis Bacelar}}}, \bibinfo {author} {\bibfnamefont {P.}~\bibnamefont {Doornenbal}}, \bibinfo {author} {\bibfnamefont {M.}~\bibnamefont {Estevez}}, \bibinfo {author} {\bibfnamefont {Y.}~\bibnamefont {Fujita}}, \bibinfo {author} {\bibfnamefont {W.}~\bibnamefont {Gelletly}}, \bibinfo {author} {\bibfnamefont {H.}~\bibnamefont {Geissel}}, \bibinfo {author} {\bibfnamefont {H.}~\bibnamefont {Grawe}}, \bibinfo {author} {\bibfnamefont {J.}~\bibnamefont {Grȩbosz}}, \bibinfo {author} {\bibfnamefont {R.}~\bibnamefont {Hoischen}}, \bibinfo {author} {\bibfnamefont {I.}~\bibnamefont {Kojouharov}}, \bibinfo {author} {\bibfnamefont {S.}~\bibnamefont
  {Lalkovski}}, \bibinfo {author} {\bibfnamefont {Z.}~\bibnamefont {Liu}}, \bibinfo {author} {\bibfnamefont {K.}~\bibnamefont {Maier}}, \bibinfo {author} {\bibfnamefont {C.}~\bibnamefont {Mihai}}, \bibinfo {author} {\bibfnamefont {D.}~\bibnamefont {Mücher}}, \bibinfo {author} {\bibfnamefont {B.}~\bibnamefont {Rubio}}, \bibinfo {author} {\bibfnamefont {H.}~\bibnamefont {Schaffner}}, \bibinfo {author} {\bibfnamefont {A.}~\bibnamefont {Tamii}}, \bibinfo {author} {\bibfnamefont {S.}~\bibnamefont {Tashenov}}, \bibinfo {author} {\bibfnamefont {J.}~\bibnamefont {Valiente-Dobón}}, \bibinfo {author} {\bibfnamefont {P.}~\bibnamefont {Walker}},\ and\ \bibinfo {author} {\bibfnamefont {P.}~\bibnamefont {Woods}},\ }\href {https://doi.org/https://doi.org/10.1016/j.physletb.2009.01.007} {\bibfield  {journal} {\bibinfo  {journal} {Physics Letters B}\ }\textbf {\bibinfo {volume} {672}},\ \bibinfo {pages} {116} (\bibinfo {year} {2009})}\BibitemShut {NoStop}%
\bibitem [{\citenamefont {Dracoulis}(2013)}]{Dracoulis2013_isomer}%
  \BibitemOpen
  \bibfield  {author} {\bibinfo {author} {\bibfnamefont {G.~D.}\ \bibnamefont {Dracoulis}},\ }\href {https://doi.org/10.1088/0031-8949/2013/T152/014015} {\bibfield  {journal} {\bibinfo  {journal} {Physica Scripta}\ }\textbf {\bibinfo {volume} {2013}},\ \bibinfo {pages} {014015} (\bibinfo {year} {2013})}\BibitemShut {NoStop}%
\bibitem [{\citenamefont {Burbidge}\ \emph {et~al.}(1957)\citenamefont {Burbidge}, \citenamefont {Burbidge}, \citenamefont {Fowler},\ and\ \citenamefont {Hoyle}}]{Burbidge1957}%
  \BibitemOpen
  \bibfield  {author} {\bibinfo {author} {\bibfnamefont {E.~M.}\ \bibnamefont {Burbidge}}, \bibinfo {author} {\bibfnamefont {G.~R.}\ \bibnamefont {Burbidge}}, \bibinfo {author} {\bibfnamefont {W.~A.}\ \bibnamefont {Fowler}},\ and\ \bibinfo {author} {\bibfnamefont {F.}~\bibnamefont {Hoyle}},\ }\href {https://doi.org/10.1103/RevModPhys.29.547} {\bibfield  {journal} {\bibinfo  {journal} {Rev. Mod. Phys.}\ }\textbf {\bibinfo {volume} {29}},\ \bibinfo {pages} {547} (\bibinfo {year} {1957})}\BibitemShut {NoStop}%
\bibitem [{\citenamefont {Mumpower}\ \emph {et~al.}(2016)\citenamefont {Mumpower}, \citenamefont {Surman}, \citenamefont {McLaughlin},\ and\ \citenamefont {Aprahamian}}]{Mumpower2016}%
  \BibitemOpen
  \bibfield  {author} {\bibinfo {author} {\bibfnamefont {M.}~\bibnamefont {Mumpower}}, \bibinfo {author} {\bibfnamefont {R.}~\bibnamefont {Surman}}, \bibinfo {author} {\bibfnamefont {G.}~\bibnamefont {McLaughlin}},\ and\ \bibinfo {author} {\bibfnamefont {A.}~\bibnamefont {Aprahamian}},\ }\href {https://doi.org/https://doi.org/10.1016/j.ppnp.2015.09.001} {\bibfield  {journal} {\bibinfo  {journal} {Progress in Particle and Nuclear Physics}\ }\textbf {\bibinfo {volume} {86}},\ \bibinfo {pages} {86} (\bibinfo {year} {2016})}\BibitemShut {NoStop}%
\bibitem [{\citenamefont {Mumpower}\ \emph {et~al.}(2024)\citenamefont {Mumpower}, \citenamefont {Sprouse}, \citenamefont {Miller}, \citenamefont {Lund}, \citenamefont {Garcia}, \citenamefont {Vassh}, \citenamefont {McLaughlin},\ and\ \citenamefont {Surman}}]{Mumpower2024}%
  \BibitemOpen
  \bibfield  {author} {\bibinfo {author} {\bibfnamefont {M.~R.}\ \bibnamefont {Mumpower}}, \bibinfo {author} {\bibfnamefont {T.~M.}\ \bibnamefont {Sprouse}}, \bibinfo {author} {\bibfnamefont {J.~M.}\ \bibnamefont {Miller}}, \bibinfo {author} {\bibfnamefont {K.~A.}\ \bibnamefont {Lund}}, \bibinfo {author} {\bibfnamefont {J.~C.}\ \bibnamefont {Garcia}}, \bibinfo {author} {\bibfnamefont {N.}~\bibnamefont {Vassh}}, \bibinfo {author} {\bibfnamefont {G.~C.}\ \bibnamefont {McLaughlin}},\ and\ \bibinfo {author} {\bibfnamefont {R.}~\bibnamefont {Surman}},\ }\href {https://doi.org/10.3847/1538-4357/ad5afc} {\bibfield  {journal} {\bibinfo  {journal} {The Astrophysical Journal}\ }\textbf {\bibinfo {volume} {970}},\ \bibinfo {pages} {173} (\bibinfo {year} {2024})}\BibitemShut {NoStop}%
\bibitem [{\citenamefont {Kumar}\ \emph {et~al.}(2024)\citenamefont {Kumar}, \citenamefont {Shimizu}, \citenamefont {Utsuno}, \citenamefont {Yuan},\ and\ \citenamefont {Srivastava}}]{kumar2024}%
  \BibitemOpen
  \bibfield  {author} {\bibinfo {author} {\bibfnamefont {A.}~\bibnamefont {Kumar}}, \bibinfo {author} {\bibfnamefont {N.}~\bibnamefont {Shimizu}}, \bibinfo {author} {\bibfnamefont {Y.}~\bibnamefont {Utsuno}}, \bibinfo {author} {\bibfnamefont {C.}~\bibnamefont {Yuan}},\ and\ \bibinfo {author} {\bibfnamefont {P.~C.}\ \bibnamefont {Srivastava}},\ }\href {https://doi.org/10.1103/PhysRevC.109.064319} {\bibfield  {journal} {\bibinfo  {journal} {Phys. Rev. C}\ }\textbf {\bibinfo {volume} {109}},\ \bibinfo {pages} {064319} (\bibinfo {year} {2024})}\BibitemShut {NoStop}%
\bibitem [{\citenamefont {Suzuki}\ \emph {et~al.}(2012)\citenamefont {Suzuki}, \citenamefont {Yoshida}, \citenamefont {Kajino},\ and\ \citenamefont {Otsuka}}]{Suzuki2012}%
  \BibitemOpen
  \bibfield  {author} {\bibinfo {author} {\bibfnamefont {T.}~\bibnamefont {Suzuki}}, \bibinfo {author} {\bibfnamefont {T.}~\bibnamefont {Yoshida}}, \bibinfo {author} {\bibfnamefont {T.}~\bibnamefont {Kajino}},\ and\ \bibinfo {author} {\bibfnamefont {T.}~\bibnamefont {Otsuka}},\ }\href {https://doi.org/10.1103/PhysRevC.85.015802} {\bibfield  {journal} {\bibinfo  {journal} {Phys. Rev. C}\ }\textbf {\bibinfo {volume} {85}},\ \bibinfo {pages} {015802} (\bibinfo {year} {2012})}\BibitemShut {NoStop}%
\bibitem [{\citenamefont {Venhart}\ \emph {et~al.}(2017)\citenamefont {Venhart}, \citenamefont {Wood}, \citenamefont {Sedlák}, \citenamefont {Balogh}, \citenamefont {Bírová}, \citenamefont {Boston}, \citenamefont {Cocolios}, \citenamefont {Harkness-Brennan}, \citenamefont {Herzberg}, \citenamefont {Holub}, \citenamefont {Joss}, \citenamefont {Judson}, \citenamefont {Kliman}, \citenamefont {Klimo}, \citenamefont {Krupa}, \citenamefont {Lušnák}, \citenamefont {Makhathini}, \citenamefont {Matoušek}, \citenamefont {Motyčák}, \citenamefont {Page}, \citenamefont {Patel}, \citenamefont {Petrík}, \citenamefont {Podshibyakin}, \citenamefont {Prajapati}, \citenamefont {Rodin}, \citenamefont {Špaček}, \citenamefont {Urban}, \citenamefont {Unsworth},\ and\ \citenamefont {Veselský}}]{Venhart2017}%
  \BibitemOpen
  \bibfield  {author} {\bibinfo {author} {\bibfnamefont {M.}~\bibnamefont {Venhart}}, \bibinfo {author} {\bibfnamefont {J.~L.}\ \bibnamefont {Wood}}, \bibinfo {author} {\bibfnamefont {M.}~\bibnamefont {Sedlák}}, \bibinfo {author} {\bibfnamefont {M.}~\bibnamefont {Balogh}}, \bibinfo {author} {\bibfnamefont {M.}~\bibnamefont {Bírová}}, \bibinfo {author} {\bibfnamefont {A.~J.}\ \bibnamefont {Boston}}, \bibinfo {author} {\bibfnamefont {T.~E.}\ \bibnamefont {Cocolios}}, \bibinfo {author} {\bibfnamefont {L.~J.}\ \bibnamefont {Harkness-Brennan}}, \bibinfo {author} {\bibfnamefont {R.-D.}\ \bibnamefont {Herzberg}}, \bibinfo {author} {\bibfnamefont {L.}~\bibnamefont {Holub}}, \bibinfo {author} {\bibfnamefont {D.~T.}\ \bibnamefont {Joss}}, \bibinfo {author} {\bibfnamefont {D.~S.}\ \bibnamefont {Judson}}, \bibinfo {author} {\bibfnamefont {J.}~\bibnamefont {Kliman}}, \bibinfo {author} {\bibfnamefont {J.}~\bibnamefont {Klimo}}, \bibinfo {author} {\bibfnamefont {L.}~\bibnamefont {Krupa}}, \bibinfo {author} {\bibfnamefont
  {J.}~\bibnamefont {LuÅ¡nák}}, \bibinfo {author} {\bibfnamefont {L.}~\bibnamefont {Makhathini}}, \bibinfo {author} {\bibfnamefont {V.}~\bibnamefont {MatouÅ¡ek}}, \bibinfo {author} {\bibfnamefont {Å.}~\bibnamefont {Motyčák}}, \bibinfo {author} {\bibfnamefont {R.~D.}\ \bibnamefont {Page}}, \bibinfo {author} {\bibfnamefont {A.}~\bibnamefont {Patel}}, \bibinfo {author} {\bibfnamefont {K.}~\bibnamefont {Petrík}}, \bibinfo {author} {\bibfnamefont {A.~V.}\ \bibnamefont {Podshibyakin}}, \bibinfo {author} {\bibfnamefont {P.~M.}\ \bibnamefont {Prajapati}}, \bibinfo {author} {\bibfnamefont {A.~M.}\ \bibnamefont {Rodin}}, \bibinfo {author} {\bibfnamefont {A.}~\bibnamefont {Å paček}}, \bibinfo {author} {\bibfnamefont {R.}~\bibnamefont {Urban}}, \bibinfo {author} {\bibfnamefont {C.}~\bibnamefont {Unsworth}},\ and\ \bibinfo {author} {\bibfnamefont {M.}~\bibnamefont {Veselský}},\ }\href {https://doi.org/10.1088/1361-6471/aa7297} {\bibfield  {journal} {\bibinfo  {journal} {Journal of Physics G: Nuclear and Particle
  Physics}\ }\textbf {\bibinfo {volume} {44}},\ \bibinfo {pages} {074003} (\bibinfo {year} {2017})}\BibitemShut {NoStop}%
\bibitem [{\citenamefont {Gono}\ \emph {et~al.}(1977)\citenamefont {Gono}, \citenamefont {Lieder}, \citenamefont {Müller-Veggian}, \citenamefont {Neskakis},\ and\ \citenamefont {Mayer-Böricke}}]{gono1977}%
  \BibitemOpen
  \bibfield  {author} {\bibinfo {author} {\bibfnamefont {Y.}~\bibnamefont {Gono}}, \bibinfo {author} {\bibfnamefont {R.}~\bibnamefont {Lieder}}, \bibinfo {author} {\bibfnamefont {M.}~\bibnamefont {Müller-Veggian}}, \bibinfo {author} {\bibfnamefont {A.}~\bibnamefont {Neskakis}},\ and\ \bibinfo {author} {\bibfnamefont {C.}~\bibnamefont {Mayer-Böricke}},\ }\href {https://doi.org/https://doi.org/10.1016/0370-2693(77)90510-X} {\bibfield  {journal} {\bibinfo  {journal} {Physics Letters B}\ }\textbf {\bibinfo {volume} {70}},\ \bibinfo {pages} {159} (\bibinfo {year} {1977})}\BibitemShut {NoStop}%
\bibitem [{\citenamefont {Gono}\ \emph {et~al.}(1979)\citenamefont {Gono}, \citenamefont {Lieder}, \citenamefont {Müller-Veggian}, \citenamefont {Neskakis},\ and\ \citenamefont {Mayer-Böricke}}]{gono1979}%
  \BibitemOpen
  \bibfield  {author} {\bibinfo {author} {\bibfnamefont {Y.}~\bibnamefont {Gono}}, \bibinfo {author} {\bibfnamefont {R.}~\bibnamefont {Lieder}}, \bibinfo {author} {\bibfnamefont {M.}~\bibnamefont {Müller-Veggian}}, \bibinfo {author} {\bibfnamefont {A.}~\bibnamefont {Neskakis}},\ and\ \bibinfo {author} {\bibfnamefont {C.}~\bibnamefont {Mayer-Böricke}},\ }\href {https://doi.org/https://doi.org/10.1016/0375-9474(79)90259-8} {\bibfield  {journal} {\bibinfo  {journal} {Nuclear Physics A}\ }\textbf {\bibinfo {volume} {327}},\ \bibinfo {pages} {269} (\bibinfo {year} {1979})}\BibitemShut {NoStop}%
\bibitem [{\citenamefont {Tjøm}\ \emph {et~al.}(1974)\citenamefont {Tjøm}, \citenamefont {Maier}, \citenamefont {Benson}, \citenamefont {Stephens},\ and\ \citenamefont {Diamond}}]{TJOM1974}%
  \BibitemOpen
  \bibfield  {author} {\bibinfo {author} {\bibfnamefont {P.}~\bibnamefont {Tjøm}}, \bibinfo {author} {\bibfnamefont {M.}~\bibnamefont {Maier}}, \bibinfo {author} {\bibfnamefont {D.}~\bibnamefont {Benson}}, \bibinfo {author} {\bibfnamefont {F.}~\bibnamefont {Stephens}},\ and\ \bibinfo {author} {\bibfnamefont {R.}~\bibnamefont {Diamond}},\ }\href {https://doi.org/https://doi.org/10.1016/0375-9474(74)90504-1} {\bibfield  {journal} {\bibinfo  {journal} {Nuclear Physics A}\ }\textbf {\bibinfo {volume} {231}},\ \bibinfo {pages} {397} (\bibinfo {year} {1974})}\BibitemShut {NoStop}%
\bibitem [{\citenamefont {Kölscbach}\ \emph {et~al.}(1985)\citenamefont {Kölscbach}, \citenamefont {Schüler}, \citenamefont {Hardt}, \citenamefont {Rosendaal}, \citenamefont {Günther}, \citenamefont {Euler}, \citenamefont {Tölle}, \citenamefont {Marten-Tölle},\ and\ \citenamefont {Zeyen}}]{kolscbach1985}%
  \BibitemOpen
  \bibfield  {author} {\bibinfo {author} {\bibfnamefont {V.}~\bibnamefont {Kölscbach}}, \bibinfo {author} {\bibfnamefont {P.}~\bibnamefont {Schüler}}, \bibinfo {author} {\bibfnamefont {K.}~\bibnamefont {Hardt}}, \bibinfo {author} {\bibfnamefont {D.}~\bibnamefont {Rosendaal}}, \bibinfo {author} {\bibfnamefont {C.}~\bibnamefont {Günther}}, \bibinfo {author} {\bibfnamefont {K.}~\bibnamefont {Euler}}, \bibinfo {author} {\bibfnamefont {R.}~\bibnamefont {Tölle}}, \bibinfo {author} {\bibfnamefont {M.}~\bibnamefont {Marten-Tölle}},\ and\ \bibinfo {author} {\bibfnamefont {P.}~\bibnamefont {Zeyen}},\ }\href {https://doi.org/https://doi.org/10.1016/0375-9474(85)90427-0} {\bibfield  {journal} {\bibinfo  {journal} {Nuclear Physics A}\ }\textbf {\bibinfo {volume} {439}},\ \bibinfo {pages} {189} (\bibinfo {year} {1985})}\BibitemShut {NoStop}%
\bibitem [{\citenamefont {Neskakis}\ \emph {et~al.}(1979)\citenamefont {Neskakis}, \citenamefont {Lieder}, \citenamefont {Beuscher}, \citenamefont {Gono}, \citenamefont {Haenni}, \citenamefont {Müller-Veggian},\ and\ \citenamefont {Mayer-Böricke}}]{neskakis1979}%
  \BibitemOpen
  \bibfield  {author} {\bibinfo {author} {\bibfnamefont {A.}~\bibnamefont {Neskakis}}, \bibinfo {author} {\bibfnamefont {R.}~\bibnamefont {Lieder}}, \bibinfo {author} {\bibfnamefont {H.}~\bibnamefont {Beuscher}}, \bibinfo {author} {\bibfnamefont {Y.}~\bibnamefont {Gono}}, \bibinfo {author} {\bibfnamefont {D.}~\bibnamefont {Haenni}}, \bibinfo {author} {\bibfnamefont {M.}~\bibnamefont {Müller-Veggian}},\ and\ \bibinfo {author} {\bibfnamefont {C.}~\bibnamefont {Mayer-Böricke}},\ }\href {https://doi.org/https://doi.org/10.1016/0370-2693(79)90196-5} {\bibfield  {journal} {\bibinfo  {journal} {Physics Letters B}\ }\textbf {\bibinfo {volume} {80}},\ \bibinfo {pages} {194} (\bibinfo {year} {1979})}\BibitemShut {NoStop}%
\bibitem [{\citenamefont {Neskakis}\ \emph {et~al.}(1982)\citenamefont {Neskakis}, \citenamefont {Lieder}, \citenamefont {Beuscher}, \citenamefont {Gono}, \citenamefont {Haenni},\ and\ \citenamefont {Müller-Veggian}}]{neskakis1982}%
  \BibitemOpen
  \bibfield  {author} {\bibinfo {author} {\bibfnamefont {A.}~\bibnamefont {Neskakis}}, \bibinfo {author} {\bibfnamefont {R.}~\bibnamefont {Lieder}}, \bibinfo {author} {\bibfnamefont {H.}~\bibnamefont {Beuscher}}, \bibinfo {author} {\bibfnamefont {Y.}~\bibnamefont {Gono}}, \bibinfo {author} {\bibfnamefont {D.}~\bibnamefont {Haenni}},\ and\ \bibinfo {author} {\bibfnamefont {M.}~\bibnamefont {Müller-Veggian}},\ }\href {https://doi.org/https://doi.org/10.1016/0375-9474(82)90606-6} {\bibfield  {journal} {\bibinfo  {journal} {Nuclear Physics A}\ }\textbf {\bibinfo {volume} {390}},\ \bibinfo {pages} {53} (\bibinfo {year} {1982})}\BibitemShut {NoStop}%
\bibitem [{\citenamefont {Bucurescu}\ and\ \citenamefont {Zamfir}(2017)}]{Bucurescu2017}%
  \BibitemOpen
  \bibfield  {author} {\bibinfo {author} {\bibfnamefont {D.}~\bibnamefont {Bucurescu}}\ and\ \bibinfo {author} {\bibfnamefont {N.~V.}\ \bibnamefont {Zamfir}},\ }\href {https://doi.org/10.1103/PhysRevC.95.014329} {\bibfield  {journal} {\bibinfo  {journal} {Phys. Rev. C}\ }\textbf {\bibinfo {volume} {95}},\ \bibinfo {pages} {014329} (\bibinfo {year} {2017})}\BibitemShut {NoStop}%
\bibitem [{\citenamefont {Guo-Jie}\ \emph {et~al.}(2006)\citenamefont {Guo-Jie}, \citenamefont {Yu-Xin}, \citenamefont {Hui},\ and\ \citenamefont {Hui-Chao}}]{Chen2006}%
  \BibitemOpen
  \bibfield  {author} {\bibinfo {author} {\bibfnamefont {C.}~\bibnamefont {Guo-Jie}}, \bibinfo {author} {\bibfnamefont {L.}~\bibnamefont {Yu-Xin}}, \bibinfo {author} {\bibfnamefont {C.}~\bibnamefont {Hui}},\ and\ \bibinfo {author} {\bibfnamefont {S.}~\bibnamefont {Hui-Chao}},\ }\href {https://doi.org/10.1088/0256-307X/23/8/022} {\bibfield  {journal} {\bibinfo  {journal} {Chinese Physics Letters}\ }\textbf {\bibinfo {volume} {23}},\ \bibinfo {pages} {2042} (\bibinfo {year} {2006})}\BibitemShut {NoStop}%
\bibitem [{\citenamefont {Gueorguieva}\ \emph {et~al.}(2003)\citenamefont {Gueorguieva}, \citenamefont {Sch\"uck}, \citenamefont {Minkova}, \citenamefont {Vieu}, \citenamefont {Hannachi}, \citenamefont {Kaci}, \citenamefont {Porquet}, \citenamefont {Wyss}, \citenamefont {Dionisi}, \citenamefont {Korichi},\ and\ \citenamefont {Lopez-Martens}}]{Gueorguieva2003}%
  \BibitemOpen
  \bibfield  {author} {\bibinfo {author} {\bibfnamefont {E.}~\bibnamefont {Gueorguieva}}, \bibinfo {author} {\bibfnamefont {C.}~\bibnamefont {Sch\"uck}}, \bibinfo {author} {\bibfnamefont {A.}~\bibnamefont {Minkova}}, \bibinfo {author} {\bibfnamefont {C.}~\bibnamefont {Vieu}}, \bibinfo {author} {\bibfnamefont {F.}~\bibnamefont {Hannachi}}, \bibinfo {author} {\bibfnamefont {M.}~\bibnamefont {Kaci}}, \bibinfo {author} {\bibfnamefont {M.-G.}\ \bibnamefont {Porquet}}, \bibinfo {author} {\bibfnamefont {R.}~\bibnamefont {Wyss}}, \bibinfo {author} {\bibfnamefont {J.~S.}\ \bibnamefont {Dionisi}}, \bibinfo {author} {\bibfnamefont {A.}~\bibnamefont {Korichi}},\ and\ \bibinfo {author} {\bibfnamefont {A.}~\bibnamefont {Lopez-Martens}},\ }\href {https://doi.org/10.1103/PhysRevC.68.054308} {\bibfield  {journal} {\bibinfo  {journal} {Phys. Rev. C}\ }\textbf {\bibinfo {volume} {68}},\ \bibinfo {pages} {054308} (\bibinfo {year} {2003})}\BibitemShut {NoStop}%
\bibitem [{\citenamefont {Oktem}\ \emph {et~al.}(2007)\citenamefont {Oktem}, \citenamefont {Balabanski}, \citenamefont {Akkus}, \citenamefont {Beausang}, \citenamefont {Bostan}, \citenamefont {Cakirli}, \citenamefont {Casten}, \citenamefont {Danchev}, \citenamefont {Djongolov}, \citenamefont {Erduran}, \citenamefont {Erturk}, \citenamefont {Gladniski}, \citenamefont {G\"urdal}, \citenamefont {Goon}, \citenamefont {Hartley}, \citenamefont {Hecht}, \citenamefont {Kr\"ucken}, \citenamefont {Nikolov}, \citenamefont {Novak}, \citenamefont {Rainovski}, \citenamefont {Riedinger}, \citenamefont {Yigitoglu}, \citenamefont {Zamfir},\ and\ \citenamefont {Zeidan}}]{Oktem2007}%
  \BibitemOpen
  \bibfield  {author} {\bibinfo {author} {\bibfnamefont {Y.}~\bibnamefont {Oktem}}, \bibinfo {author} {\bibfnamefont {D.~L.}\ \bibnamefont {Balabanski}}, \bibinfo {author} {\bibfnamefont {B.}~\bibnamefont {Akkus}}, \bibinfo {author} {\bibfnamefont {C.~W.}\ \bibnamefont {Beausang}}, \bibinfo {author} {\bibfnamefont {M.}~\bibnamefont {Bostan}}, \bibinfo {author} {\bibfnamefont {R.~B.}\ \bibnamefont {Cakirli}}, \bibinfo {author} {\bibfnamefont {R.~F.}\ \bibnamefont {Casten}}, \bibinfo {author} {\bibfnamefont {M.}~\bibnamefont {Danchev}}, \bibinfo {author} {\bibfnamefont {M.}~\bibnamefont {Djongolov}}, \bibinfo {author} {\bibfnamefont {M.~N.}\ \bibnamefont {Erduran}}, \bibinfo {author} {\bibfnamefont {S.}~\bibnamefont {Erturk}}, \bibinfo {author} {\bibfnamefont {K.~A.}\ \bibnamefont {Gladniski}}, \bibinfo {author} {\bibfnamefont {G.}~\bibnamefont {G\"urdal}}, \bibinfo {author} {\bibfnamefont {J.~T.}\ \bibnamefont {Goon}}, \bibinfo {author} {\bibfnamefont {D.~J.}\ \bibnamefont {Hartley}}, \bibinfo {author}
  {\bibfnamefont {A.~A.}\ \bibnamefont {Hecht}}, \bibinfo {author} {\bibfnamefont {R.}~\bibnamefont {Kr\"ucken}}, \bibinfo {author} {\bibfnamefont {N.}~\bibnamefont {Nikolov}}, \bibinfo {author} {\bibfnamefont {J.~R.}\ \bibnamefont {Novak}}, \bibinfo {author} {\bibfnamefont {G.}~\bibnamefont {Rainovski}}, \bibinfo {author} {\bibfnamefont {L.~L.}\ \bibnamefont {Riedinger}}, \bibinfo {author} {\bibfnamefont {I.}~\bibnamefont {Yigitoglu}}, \bibinfo {author} {\bibfnamefont {N.~V.}\ \bibnamefont {Zamfir}},\ and\ \bibinfo {author} {\bibfnamefont {O.}~\bibnamefont {Zeidan}},\ }\href {https://doi.org/10.1103/PhysRevC.76.044315} {\bibfield  {journal} {\bibinfo  {journal} {Phys. Rev. C}\ }\textbf {\bibinfo {volume} {76}},\ \bibinfo {pages} {044315} (\bibinfo {year} {2007})}\BibitemShut {NoStop}%
\bibitem [{\citenamefont {Wang}\ \emph {et~al.}(2012)\citenamefont {Wang}, \citenamefont {Zhou}, \citenamefont {Fang}, \citenamefont {Zhang}, \citenamefont {Zhang}, \citenamefont {Gao}, \citenamefont {Liu}, \citenamefont {Wang}, \citenamefont {Ma}, \citenamefont {Guo}, \citenamefont {Li}, \citenamefont {Yan}, \citenamefont {He}, \citenamefont {Wang}, \citenamefont {Fang}, \citenamefont {Wu}, \citenamefont {He}, \citenamefont {Zheng}, \citenamefont {Wang}, \citenamefont {Dong},\ and\ \citenamefont {Xu}}]{WANG2012}%
  \BibitemOpen
  \bibfield  {author} {\bibinfo {author} {\bibfnamefont {S.~C.}\ \bibnamefont {Wang}}, \bibinfo {author} {\bibfnamefont {X.~H.}\ \bibnamefont {Zhou}}, \bibinfo {author} {\bibfnamefont {Y.~D.}\ \bibnamefont {Fang}}, \bibinfo {author} {\bibfnamefont {Y.~H.}\ \bibnamefont {Zhang}}, \bibinfo {author} {\bibfnamefont {N.~T.}\ \bibnamefont {Zhang}}, \bibinfo {author} {\bibfnamefont {B.~S.}\ \bibnamefont {Gao}}, \bibinfo {author} {\bibfnamefont {M.~L.}\ \bibnamefont {Liu}}, \bibinfo {author} {\bibfnamefont {J.~G.}\ \bibnamefont {Wang}}, \bibinfo {author} {\bibfnamefont {F.}~\bibnamefont {Ma}}, \bibinfo {author} {\bibfnamefont {Y.~X.}\ \bibnamefont {Guo}}, \bibinfo {author} {\bibfnamefont {S.~C.}\ \bibnamefont {Li}}, \bibinfo {author} {\bibfnamefont {X.~L.}\ \bibnamefont {Yan}}, \bibinfo {author} {\bibfnamefont {L.}~\bibnamefont {He}}, \bibinfo {author} {\bibfnamefont {Z.~G.}\ \bibnamefont {Wang}}, \bibinfo {author} {\bibfnamefont {F.}~\bibnamefont {Fang}}, \bibinfo {author} {\bibfnamefont {X.~G.}\ \bibnamefont {Wu}},
  \bibinfo {author} {\bibfnamefont {C.~Y.}\ \bibnamefont {He}}, \bibinfo {author} {\bibfnamefont {Y.}~\bibnamefont {Zheng}}, \bibinfo {author} {\bibfnamefont {Z.~M.}\ \bibnamefont {Wang}}, \bibinfo {author} {\bibfnamefont {G.~X.}\ \bibnamefont {Dong}},\ and\ \bibinfo {author} {\bibfnamefont {F.~R.}\ \bibnamefont {Xu}},\ }\href {https://doi.org/10.1103/PhysRevC.85.027301} {\bibfield  {journal} {\bibinfo  {journal} {Phys. Rev. C}\ }\textbf {\bibinfo {volume} {85}},\ \bibinfo {pages} {027301} (\bibinfo {year} {2012})}\BibitemShut {NoStop}%
\bibitem [{\citenamefont {Gueorguieva}\ \emph {et~al.}(2004)\citenamefont {Gueorguieva}, \citenamefont {Sch\"uck}, \citenamefont {Minkova}, \citenamefont {Vieu}, \citenamefont {Kaci}, \citenamefont {Hannachi}, \citenamefont {Wyss}, \citenamefont {Dionisio}, \citenamefont {Korichi},\ and\ \citenamefont {Lopez-Martens}}]{Gueorguieva2004}%
  \BibitemOpen
  \bibfield  {author} {\bibinfo {author} {\bibfnamefont {E.}~\bibnamefont {Gueorguieva}}, \bibinfo {author} {\bibfnamefont {C.}~\bibnamefont {Sch\"uck}}, \bibinfo {author} {\bibfnamefont {A.}~\bibnamefont {Minkova}}, \bibinfo {author} {\bibfnamefont {C.}~\bibnamefont {Vieu}}, \bibinfo {author} {\bibfnamefont {M.}~\bibnamefont {Kaci}}, \bibinfo {author} {\bibfnamefont {F.}~\bibnamefont {Hannachi}}, \bibinfo {author} {\bibfnamefont {R.}~\bibnamefont {Wyss}}, \bibinfo {author} {\bibfnamefont {J.~S.}\ \bibnamefont {Dionisio}}, \bibinfo {author} {\bibfnamefont {A.}~\bibnamefont {Korichi}},\ and\ \bibinfo {author} {\bibfnamefont {A.}~\bibnamefont {Lopez-Martens}},\ }\href {https://doi.org/10.1103/PhysRevC.69.044320} {\bibfield  {journal} {\bibinfo  {journal} {Phys. Rev. C}\ }\textbf {\bibinfo {volume} {69}},\ \bibinfo {pages} {044320} (\bibinfo {year} {2004})}\BibitemShut {NoStop}%
\bibitem [{\citenamefont {Oktem}\ \emph {et~al.}(2012)\citenamefont {Oktem}, \citenamefont {Balabanski}, \citenamefont {Akkus}, \citenamefont {Susam}, \citenamefont {Atanasova}, \citenamefont {Beausang}, \citenamefont {Cakirli}, \citenamefont {Casten}, \citenamefont {Danchev}, \citenamefont {Djongolov}, \citenamefont {Ganio\ifmmode~\breve{g}\else \u{g}\fi{}lu}, \citenamefont {Gladnishki}, \citenamefont {Goon}, \citenamefont {Hartley}, \citenamefont {Hecht}, \citenamefont {Kr\"ucken}, \citenamefont {Novak}, \citenamefont {Rainovski}, \citenamefont {Riedinger}, \citenamefont {Venkova}, \citenamefont {Yigitoglu}, \citenamefont {Zamfir},\ and\ \citenamefont {Zeidan}}]{oktem2012}%
  \BibitemOpen
  \bibfield  {author} {\bibinfo {author} {\bibfnamefont {Y.}~\bibnamefont {Oktem}}, \bibinfo {author} {\bibfnamefont {D.~L.}\ \bibnamefont {Balabanski}}, \bibinfo {author} {\bibfnamefont {B.}~\bibnamefont {Akkus}}, \bibinfo {author} {\bibfnamefont {L.~A.}\ \bibnamefont {Susam}}, \bibinfo {author} {\bibfnamefont {L.}~\bibnamefont {Atanasova}}, \bibinfo {author} {\bibfnamefont {C.~W.}\ \bibnamefont {Beausang}}, \bibinfo {author} {\bibfnamefont {R.~B.}\ \bibnamefont {Cakirli}}, \bibinfo {author} {\bibfnamefont {R.~F.}\ \bibnamefont {Casten}}, \bibinfo {author} {\bibfnamefont {M.}~\bibnamefont {Danchev}}, \bibinfo {author} {\bibfnamefont {M.}~\bibnamefont {Djongolov}}, \bibinfo {author} {\bibfnamefont {E.}~\bibnamefont {Ganio\ifmmode~\breve{g}\else \u{g}\fi{}lu}}, \bibinfo {author} {\bibfnamefont {K.~A.}\ \bibnamefont {Gladnishki}}, \bibinfo {author} {\bibfnamefont {J.~T.}\ \bibnamefont {Goon}}, \bibinfo {author} {\bibfnamefont {D.~J.}\ \bibnamefont {Hartley}}, \bibinfo {author} {\bibfnamefont {A.~A.}\
  \bibnamefont {Hecht}}, \bibinfo {author} {\bibfnamefont {R.}~\bibnamefont {Kr\"ucken}}, \bibinfo {author} {\bibfnamefont {J.~R.}\ \bibnamefont {Novak}}, \bibinfo {author} {\bibfnamefont {G.}~\bibnamefont {Rainovski}}, \bibinfo {author} {\bibfnamefont {L.~L.}\ \bibnamefont {Riedinger}}, \bibinfo {author} {\bibfnamefont {T.}~\bibnamefont {Venkova}}, \bibinfo {author} {\bibfnamefont {I.}~\bibnamefont {Yigitoglu}}, \bibinfo {author} {\bibfnamefont {N.~V.}\ \bibnamefont {Zamfir}},\ and\ \bibinfo {author} {\bibfnamefont {O.}~\bibnamefont {Zeidan}},\ }\href {https://doi.org/10.1103/PhysRevC.86.054305} {\bibfield  {journal} {\bibinfo  {journal} {Phys. Rev. C}\ }\textbf {\bibinfo {volume} {86}},\ \bibinfo {pages} {054305} (\bibinfo {year} {2012})}\BibitemShut {NoStop}%
\bibitem [{\citenamefont {Gao}\ \emph {et~al.}(2012)\citenamefont {Gao}, \citenamefont {Zhou}, \citenamefont {Fang}, \citenamefont {Zhang}, \citenamefont {Wang}, \citenamefont {Zhang}, \citenamefont {Liu}, \citenamefont {Wang}, \citenamefont {Ma}, \citenamefont {Guo}, \citenamefont {Wu}, \citenamefont {He}, \citenamefont {Zheng}, \citenamefont {Wang}, \citenamefont {Li}, \citenamefont {Yan}, \citenamefont {He}, \citenamefont {Wang}, \citenamefont {Zheng}, \citenamefont {Fang},\ and\ \citenamefont {Chen}}]{gao2012}%
  \BibitemOpen
  \bibfield  {author} {\bibinfo {author} {\bibfnamefont {B.~S.}\ \bibnamefont {Gao}}, \bibinfo {author} {\bibfnamefont {X.~H.}\ \bibnamefont {Zhou}}, \bibinfo {author} {\bibfnamefont {Y.~D.}\ \bibnamefont {Fang}}, \bibinfo {author} {\bibfnamefont {Y.~H.}\ \bibnamefont {Zhang}}, \bibinfo {author} {\bibfnamefont {S.~C.}\ \bibnamefont {Wang}}, \bibinfo {author} {\bibfnamefont {N.~T.}\ \bibnamefont {Zhang}}, \bibinfo {author} {\bibfnamefont {M.~L.}\ \bibnamefont {Liu}}, \bibinfo {author} {\bibfnamefont {J.~G.}\ \bibnamefont {Wang}}, \bibinfo {author} {\bibfnamefont {F.}~\bibnamefont {Ma}}, \bibinfo {author} {\bibfnamefont {Y.~X.}\ \bibnamefont {Guo}}, \bibinfo {author} {\bibfnamefont {X.~G.}\ \bibnamefont {Wu}}, \bibinfo {author} {\bibfnamefont {C.~Y.}\ \bibnamefont {He}}, \bibinfo {author} {\bibfnamefont {Y.}~\bibnamefont {Zheng}}, \bibinfo {author} {\bibfnamefont {Z.~M.}\ \bibnamefont {Wang}}, \bibinfo {author} {\bibfnamefont {S.~C.}\ \bibnamefont {Li}}, \bibinfo {author} {\bibfnamefont {X.~L.}\ \bibnamefont
  {Yan}}, \bibinfo {author} {\bibfnamefont {L.}~\bibnamefont {He}}, \bibinfo {author} {\bibfnamefont {Z.~G.}\ \bibnamefont {Wang}}, \bibinfo {author} {\bibfnamefont {Y.}~\bibnamefont {Zheng}}, \bibinfo {author} {\bibfnamefont {F.}~\bibnamefont {Fang}},\ and\ \bibinfo {author} {\bibfnamefont {X.~M.}\ \bibnamefont {Chen}},\ }\href {https://doi.org/10.1103/PhysRevC.86.054310} {\bibfield  {journal} {\bibinfo  {journal} {Phys. Rev. C}\ }\textbf {\bibinfo {volume} {86}},\ \bibinfo {pages} {054310} (\bibinfo {year} {2012})}\BibitemShut {NoStop}%
\bibitem [{\citenamefont {Fotiades}\ \emph {et~al.}(2005)\citenamefont {Fotiades}, \citenamefont {Nelson}, \citenamefont {Devlin}, \citenamefont {Starosta}, \citenamefont {Becker}, \citenamefont {Bernstein}, \citenamefont {Garrett},\ and\ \citenamefont {Younes}}]{Fotiades2005}%
  \BibitemOpen
  \bibfield  {author} {\bibinfo {author} {\bibfnamefont {N.}~\bibnamefont {Fotiades}}, \bibinfo {author} {\bibfnamefont {R.~O.}\ \bibnamefont {Nelson}}, \bibinfo {author} {\bibfnamefont {M.}~\bibnamefont {Devlin}}, \bibinfo {author} {\bibfnamefont {K.}~\bibnamefont {Starosta}}, \bibinfo {author} {\bibfnamefont {J.~A.}\ \bibnamefont {Becker}}, \bibinfo {author} {\bibfnamefont {L.~A.}\ \bibnamefont {Bernstein}}, \bibinfo {author} {\bibfnamefont {P.~E.}\ \bibnamefont {Garrett}},\ and\ \bibinfo {author} {\bibfnamefont {W.}~\bibnamefont {Younes}},\ }\href {https://doi.org/10.1103/PhysRevC.71.064314} {\bibfield  {journal} {\bibinfo  {journal} {Phys. Rev. C}\ }\textbf {\bibinfo {volume} {71}},\ \bibinfo {pages} {064314} (\bibinfo {year} {2005})}\BibitemShut {NoStop}%
\bibitem [{\citenamefont {Miyahara}\ \emph {et~al.}(2004)\citenamefont {Miyahara}, \citenamefont {Ogata}, \citenamefont {Ishikawa}, \citenamefont {Katoh},\ and\ \citenamefont {Marnada}}]{miyahara2004}%
  \BibitemOpen
  \bibfield  {author} {\bibinfo {author} {\bibfnamefont {H.}~\bibnamefont {Miyahara}}, \bibinfo {author} {\bibfnamefont {Y.}~\bibnamefont {Ogata}}, \bibinfo {author} {\bibfnamefont {N.}~\bibnamefont {Ishikawa}}, \bibinfo {author} {\bibfnamefont {K.}~\bibnamefont {Katoh}},\ and\ \bibinfo {author} {\bibfnamefont {N.}~\bibnamefont {Marnada}},\ }\href {https://doi.org/https://doi.org/10.1016/j.apradiso.2003.11.030} {\bibfield  {journal} {\bibinfo  {journal} {Applied Radiation and Isotopes}\ }\textbf {\bibinfo {volume} {60}},\ \bibinfo {pages} {289} (\bibinfo {year} {2004})}\BibitemShut {NoStop}%
\bibitem [{\citenamefont {Munger}\ and\ \citenamefont {Peterson}(1978)}]{MUNGER1978}%
  \BibitemOpen
  \bibfield  {author} {\bibinfo {author} {\bibfnamefont {M.}~\bibnamefont {Munger}}\ and\ \bibinfo {author} {\bibfnamefont {R.}~\bibnamefont {Peterson}},\ }\href {https://doi.org/https://doi.org/10.1016/0375-9474(78)90051-9} {\bibfield  {journal} {\bibinfo  {journal} {Nuclear Physics A}\ }\textbf {\bibinfo {volume} {303}},\ \bibinfo {pages} {199} (\bibinfo {year} {1978})}\BibitemShut {NoStop}%
\bibitem [{\citenamefont {Flynn}\ \emph {et~al.}(1981)\citenamefont {Flynn}, \citenamefont {Brown}, \citenamefont {Sunier}, \citenamefont {Gursky}, \citenamefont {Cizewski},\ and\ \citenamefont {Burke}}]{flynn1981}%
  \BibitemOpen
  \bibfield  {author} {\bibinfo {author} {\bibfnamefont {E.}~\bibnamefont {Flynn}}, \bibinfo {author} {\bibfnamefont {R.~E.}\ \bibnamefont {Brown}}, \bibinfo {author} {\bibfnamefont {J.}~\bibnamefont {Sunier}}, \bibinfo {author} {\bibfnamefont {J.}~\bibnamefont {Gursky}}, \bibinfo {author} {\bibfnamefont {J.}~\bibnamefont {Cizewski}},\ and\ \bibinfo {author} {\bibfnamefont {D.}~\bibnamefont {Burke}},\ }\href {https://doi.org/https://doi.org/10.1016/0370-2693(81)91004-2} {\bibfield  {journal} {\bibinfo  {journal} {Physics Letters B}\ }\textbf {\bibinfo {volume} {105}},\ \bibinfo {pages} {125} (\bibinfo {year} {1981})}\BibitemShut {NoStop}%
\bibitem [{\citenamefont {Wapstra}\ \emph {et~al.}(1967)\citenamefont {Wapstra}, \citenamefont {Goudsmit}, \citenamefont {Jansen}, \citenamefont {Konijn}, \citenamefont {Löbner}, \citenamefont {Nijgh},\ and\ \citenamefont {{de Wit}}}]{WAPSTRA1967}%
  \BibitemOpen
  \bibfield  {author} {\bibinfo {author} {\bibfnamefont {A.}~\bibnamefont {Wapstra}}, \bibinfo {author} {\bibfnamefont {P.}~\bibnamefont {Goudsmit}}, \bibinfo {author} {\bibfnamefont {J.}~\bibnamefont {Jansen}}, \bibinfo {author} {\bibfnamefont {J.}~\bibnamefont {Konijn}}, \bibinfo {author} {\bibfnamefont {K.}~\bibnamefont {Löbner}}, \bibinfo {author} {\bibfnamefont {G.}~\bibnamefont {Nijgh}},\ and\ \bibinfo {author} {\bibfnamefont {S.}~\bibnamefont {{de Wit}}},\ }\href {https://doi.org/https://doi.org/10.1016/0375-9474(67)90305-3} {\bibfield  {journal} {\bibinfo  {journal} {Nuclear Physics A}\ }\textbf {\bibinfo {volume} {93}},\ \bibinfo {pages} {527} (\bibinfo {year} {1967})}\BibitemShut {NoStop}%
\bibitem [{\citenamefont {Cunnane}\ and\ \citenamefont {Daly}(1972)}]{Cunnane1972_198Au}%
  \BibitemOpen
  \bibfield  {author} {\bibinfo {author} {\bibfnamefont {J.~C.}\ \bibnamefont {Cunnane}}\ and\ \bibinfo {author} {\bibfnamefont {P.~J.}\ \bibnamefont {Daly}},\ }\href {https://doi.org/10.1103/PhysRevC.6.1407} {\bibfield  {journal} {\bibinfo  {journal} {Phys. Rev. C}\ }\textbf {\bibinfo {volume} {6}},\ \bibinfo {pages} {1407} (\bibinfo {year} {1972})}\BibitemShut {NoStop}%
\bibitem [{\citenamefont {Cunnane}\ \emph {et~al.}(1972)\citenamefont {Cunnane}, \citenamefont {Hochel}, \citenamefont {Yates},\ and\ \citenamefont {Daly}}]{Cunnane1972_200Au}%
  \BibitemOpen
  \bibfield  {author} {\bibinfo {author} {\bibfnamefont {J.}~\bibnamefont {Cunnane}}, \bibinfo {author} {\bibfnamefont {R.}~\bibnamefont {Hochel}}, \bibinfo {author} {\bibfnamefont {S.}~\bibnamefont {Yates}},\ and\ \bibinfo {author} {\bibfnamefont {P.}~\bibnamefont {Daly}},\ }\href {https://doi.org/https://doi.org/10.1016/0375-9474(72)90795-6} {\bibfield  {journal} {\bibinfo  {journal} {Nuclear Physics A}\ }\textbf {\bibinfo {volume} {196}},\ \bibinfo {pages} {593} (\bibinfo {year} {1972})}\BibitemShut {NoStop}%
\bibitem [{\citenamefont {Bowry}(2011)}]{Bowry2011}%
  \BibitemOpen
  \bibfield  {author} {\bibinfo {author} {\bibfnamefont {M.}~\bibnamefont {Bowry}},\ }\href@noop {} {\bibinfo {title} {{I}{S}{O}{L}{D}{E} newsletter spring 2011}} (\bibinfo {year} {2011}),\ \bibinfo {note} {[Accessed 30-07-2025]}\BibitemShut {NoStop}%
\bibitem [{\citenamefont {Barzakh}\ \emph {et~al.}(2022)\citenamefont {Barzakh}, \citenamefont {Andreyev}, \citenamefont {Atanasov}, \citenamefont {Cubiss}, \citenamefont {Harding}, \citenamefont {Monthery}, \citenamefont {Althubiti}, \citenamefont {Andel}, \citenamefont {Antalic}, \citenamefont {Ballof}, \citenamefont {Blaum}, \citenamefont {Cocolios}, \citenamefont {Duppen}, \citenamefont {Goodacre}, \citenamefont {de~Roubin}, \citenamefont {Duchemin}, \citenamefont {Farooq-Smith}, \citenamefont {Fedorov}, \citenamefont {Fedosseev}, \citenamefont {Fink}, \citenamefont {Gaffney}, \citenamefont {Ghys}, \citenamefont {Huyse}, \citenamefont {Imai}, \citenamefont {Johnson}, \citenamefont {Kreim}, \citenamefont {Lunney}, \citenamefont {Lynch}, \citenamefont {Manea}, \citenamefont {Marsh}, \citenamefont {Palenzuela}, \citenamefont {Molkanov}, \citenamefont {Neidherr}, \citenamefont {Panteleev}, \citenamefont {Rosenbusch}, \citenamefont {Rossel}, \citenamefont {Rothe}, \citenamefont {Schweikhard}, \citenamefont
  {Seliverstov}, \citenamefont {Sels}, \citenamefont {Beveren}, \citenamefont {Verstraelen}, \citenamefont {Welker}, \citenamefont {Wienholtz}, \citenamefont {Wolf},\ and\ \citenamefont {Zuber}}]{BARZAKH2022}%
  \BibitemOpen
  \bibfield  {author} {\bibinfo {author} {\bibfnamefont {A.}~\bibnamefont {Barzakh}}, \bibinfo {author} {\bibfnamefont {A.}~\bibnamefont {Andreyev}}, \bibinfo {author} {\bibfnamefont {D.}~\bibnamefont {Atanasov}}, \bibinfo {author} {\bibfnamefont {J.}~\bibnamefont {Cubiss}}, \bibinfo {author} {\bibfnamefont {R.}~\bibnamefont {Harding}}, \bibinfo {author} {\bibfnamefont {M.~A.}\ \bibnamefont {Monthery}}, \bibinfo {author} {\bibfnamefont {N.}~\bibnamefont {Althubiti}}, \bibinfo {author} {\bibfnamefont {B.}~\bibnamefont {Andel}}, \bibinfo {author} {\bibfnamefont {S.}~\bibnamefont {Antalic}}, \bibinfo {author} {\bibfnamefont {J.}~\bibnamefont {Ballof}}, \bibinfo {author} {\bibfnamefont {K.}~\bibnamefont {Blaum}}, \bibinfo {author} {\bibfnamefont {T.}~\bibnamefont {Cocolios}}, \bibinfo {author} {\bibfnamefont {P.~V.}\ \bibnamefont {Duppen}}, \bibinfo {author} {\bibfnamefont {T.~D.}\ \bibnamefont {Goodacre}}, \bibinfo {author} {\bibfnamefont {A.}~\bibnamefont {de~Roubin}}, \bibinfo {author} {\bibfnamefont
  {C.}~\bibnamefont {Duchemin}}, \bibinfo {author} {\bibfnamefont {G.}~\bibnamefont {Farooq-Smith}}, \bibinfo {author} {\bibfnamefont {D.}~\bibnamefont {Fedorov}}, \bibinfo {author} {\bibfnamefont {V.}~\bibnamefont {Fedosseev}}, \bibinfo {author} {\bibfnamefont {D.}~\bibnamefont {Fink}}, \bibinfo {author} {\bibfnamefont {L.}~\bibnamefont {Gaffney}}, \bibinfo {author} {\bibfnamefont {L.}~\bibnamefont {Ghys}}, \bibinfo {author} {\bibfnamefont {M.}~\bibnamefont {Huyse}}, \bibinfo {author} {\bibfnamefont {N.}~\bibnamefont {Imai}}, \bibinfo {author} {\bibfnamefont {J.}~\bibnamefont {Johnson}}, \bibinfo {author} {\bibfnamefont {S.}~\bibnamefont {Kreim}}, \bibinfo {author} {\bibfnamefont {D.}~\bibnamefont {Lunney}}, \bibinfo {author} {\bibfnamefont {K.}~\bibnamefont {Lynch}}, \bibinfo {author} {\bibfnamefont {V.}~\bibnamefont {Manea}}, \bibinfo {author} {\bibfnamefont {B.}~\bibnamefont {Marsh}}, \bibinfo {author} {\bibfnamefont {Y.~M.}\ \bibnamefont {Palenzuela}}, \bibinfo {author} {\bibfnamefont {P.}~\bibnamefont
  {Molkanov}}, \bibinfo {author} {\bibfnamefont {D.}~\bibnamefont {Neidherr}}, \bibinfo {author} {\bibfnamefont {V.}~\bibnamefont {Panteleev}}, \bibinfo {author} {\bibfnamefont {M.}~\bibnamefont {Rosenbusch}}, \bibinfo {author} {\bibfnamefont {R.}~\bibnamefont {Rossel}}, \bibinfo {author} {\bibfnamefont {S.}~\bibnamefont {Rothe}}, \bibinfo {author} {\bibfnamefont {L.}~\bibnamefont {Schweikhard}}, \bibinfo {author} {\bibfnamefont {M.}~\bibnamefont {Seliverstov}}, \bibinfo {author} {\bibfnamefont {S.}~\bibnamefont {Sels}}, \bibinfo {author} {\bibfnamefont {C.~V.}\ \bibnamefont {Beveren}}, \bibinfo {author} {\bibfnamefont {E.}~\bibnamefont {Verstraelen}}, \bibinfo {author} {\bibfnamefont {A.}~\bibnamefont {Welker}}, \bibinfo {author} {\bibfnamefont {F.}~\bibnamefont {Wienholtz}}, \bibinfo {author} {\bibfnamefont {R.}~\bibnamefont {Wolf}},\ and\ \bibinfo {author} {\bibfnamefont {K.}~\bibnamefont {Zuber}},\ }\href {https://doi.org/https://doi.org/10.1016/j.nimb.2021.12.011} {\bibfield  {journal} {\bibinfo
  {journal} {Nucl. Instrum. Methods Phys. Res. B}\ }\textbf {\bibinfo {volume} {513}},\ \bibinfo {pages} {26} (\bibinfo {year} {2022})}\BibitemShut {NoStop}%
\bibitem [{\citenamefont {Podolyák}(2023)}]{Podolyak2023}%
  \BibitemOpen
  \bibfield  {author} {\bibinfo {author} {\bibfnamefont {Z.}~\bibnamefont {Podolyák}},\ }\href {https://doi.org/10.1051/epjconf/202327908001} {\bibfield  {journal} {\bibinfo  {journal} {EPJ Web Conf.}\ }\textbf {\bibinfo {volume} {279}},\ \bibinfo {pages} {08001} (\bibinfo {year} {2023})}\BibitemShut {NoStop}%
\bibitem [{\citenamefont {Dasso}\ \emph {et~al.}(1994)\citenamefont {Dasso}, \citenamefont {Pollarolo},\ and\ \citenamefont {Winther}}]{Dasso1994}%
  \BibitemOpen
  \bibfield  {author} {\bibinfo {author} {\bibfnamefont {C.~H.}\ \bibnamefont {Dasso}}, \bibinfo {author} {\bibfnamefont {G.}~\bibnamefont {Pollarolo}},\ and\ \bibinfo {author} {\bibfnamefont {A.}~\bibnamefont {Winther}},\ }\href {https://doi.org/10.1103/PhysRevLett.73.1907} {\bibfield  {journal} {\bibinfo  {journal} {Phys. Rev. Lett.}\ }\textbf {\bibinfo {volume} {73}},\ \bibinfo {pages} {1907} (\bibinfo {year} {1994})}\BibitemShut {NoStop}%
\bibitem [{\citenamefont {Zagrebaev}\ and\ \citenamefont {Greiner}(2008)}]{zagrebaev2008}%
  \BibitemOpen
  \bibfield  {author} {\bibinfo {author} {\bibfnamefont {V.}~\bibnamefont {Zagrebaev}}\ and\ \bibinfo {author} {\bibfnamefont {W.}~\bibnamefont {Greiner}},\ }\href {https://doi.org/10.1103/PhysRevLett.101.122701} {\bibfield  {journal} {\bibinfo  {journal} {Phys. Rev. Lett.}\ }\textbf {\bibinfo {volume} {101}},\ \bibinfo {pages} {122701} (\bibinfo {year} {2008})}\BibitemShut {NoStop}%
\bibitem [{\citenamefont {Watanabe}\ \emph {et~al.}(2015)\citenamefont {Watanabe}, \citenamefont {Kim}, \citenamefont {Jeong}, \citenamefont {Hirayama}, \citenamefont {Imai}, \citenamefont {Ishiyama}, \citenamefont {Jung}, \citenamefont {Miyatake}, \citenamefont {Choi}, \citenamefont {Song}, \citenamefont {Clement}, \citenamefont {de~France}, \citenamefont {Navin}, \citenamefont {Rejmund}, \citenamefont {Schmitt}, \citenamefont {Pollarolo}, \citenamefont {Corradi}, \citenamefont {Fioretto}, \citenamefont {Montanari}, \citenamefont {Niikura}, \citenamefont {Suzuki}, \citenamefont {Nishibata},\ and\ \citenamefont {Takatsu}}]{watanabe2015}%
  \BibitemOpen
  \bibfield  {author} {\bibinfo {author} {\bibfnamefont {Y.~X.}\ \bibnamefont {Watanabe}}, \bibinfo {author} {\bibfnamefont {Y.~H.}\ \bibnamefont {Kim}}, \bibinfo {author} {\bibfnamefont {S.~C.}\ \bibnamefont {Jeong}}, \bibinfo {author} {\bibfnamefont {Y.}~\bibnamefont {Hirayama}}, \bibinfo {author} {\bibfnamefont {N.}~\bibnamefont {Imai}}, \bibinfo {author} {\bibfnamefont {H.}~\bibnamefont {Ishiyama}}, \bibinfo {author} {\bibfnamefont {H.~S.}\ \bibnamefont {Jung}}, \bibinfo {author} {\bibfnamefont {H.}~\bibnamefont {Miyatake}}, \bibinfo {author} {\bibfnamefont {S.}~\bibnamefont {Choi}}, \bibinfo {author} {\bibfnamefont {J.~S.}\ \bibnamefont {Song}}, \bibinfo {author} {\bibfnamefont {E.}~\bibnamefont {Clement}}, \bibinfo {author} {\bibfnamefont {G.}~\bibnamefont {de~France}}, \bibinfo {author} {\bibfnamefont {A.}~\bibnamefont {Navin}}, \bibinfo {author} {\bibfnamefont {M.}~\bibnamefont {Rejmund}}, \bibinfo {author} {\bibfnamefont {C.}~\bibnamefont {Schmitt}}, \bibinfo {author} {\bibfnamefont {G.}~\bibnamefont
  {Pollarolo}}, \bibinfo {author} {\bibfnamefont {L.}~\bibnamefont {Corradi}}, \bibinfo {author} {\bibfnamefont {E.}~\bibnamefont {Fioretto}}, \bibinfo {author} {\bibfnamefont {D.}~\bibnamefont {Montanari}}, \bibinfo {author} {\bibfnamefont {M.}~\bibnamefont {Niikura}}, \bibinfo {author} {\bibfnamefont {D.}~\bibnamefont {Suzuki}}, \bibinfo {author} {\bibfnamefont {H.}~\bibnamefont {Nishibata}},\ and\ \bibinfo {author} {\bibfnamefont {J.}~\bibnamefont {Takatsu}},\ }\href {https://doi.org/10.1103/PhysRevLett.115.172503} {\bibfield  {journal} {\bibinfo  {journal} {Phys. Rev. Lett.}\ }\textbf {\bibinfo {volume} {115}},\ \bibinfo {pages} {172503} (\bibinfo {year} {2015})}\BibitemShut {NoStop}%
\bibitem [{\citenamefont {Barrett}\ \emph {et~al.}(2015)\citenamefont {Barrett}, \citenamefont {Loveland}, \citenamefont {Yanez}, \citenamefont {Zhu}, \citenamefont {Ayangeakaa}, \citenamefont {Carpenter}, \citenamefont {Greene}, \citenamefont {Janssens}, \citenamefont {Lauritsen}, \citenamefont {McCutchan}, \citenamefont {Sonzogni}, \citenamefont {Chiara}, \citenamefont {Harker},\ and\ \citenamefont {Walters}}]{barrett2015}%
  \BibitemOpen
  \bibfield  {author} {\bibinfo {author} {\bibfnamefont {J.~S.}\ \bibnamefont {Barrett}}, \bibinfo {author} {\bibfnamefont {W.}~\bibnamefont {Loveland}}, \bibinfo {author} {\bibfnamefont {R.}~\bibnamefont {Yanez}}, \bibinfo {author} {\bibfnamefont {S.}~\bibnamefont {Zhu}}, \bibinfo {author} {\bibfnamefont {A.~D.}\ \bibnamefont {Ayangeakaa}}, \bibinfo {author} {\bibfnamefont {M.~P.}\ \bibnamefont {Carpenter}}, \bibinfo {author} {\bibfnamefont {J.~P.}\ \bibnamefont {Greene}}, \bibinfo {author} {\bibfnamefont {R.~V.~F.}\ \bibnamefont {Janssens}}, \bibinfo {author} {\bibfnamefont {T.}~\bibnamefont {Lauritsen}}, \bibinfo {author} {\bibfnamefont {E.~A.}\ \bibnamefont {McCutchan}}, \bibinfo {author} {\bibfnamefont {A.~A.}\ \bibnamefont {Sonzogni}}, \bibinfo {author} {\bibfnamefont {C.~J.}\ \bibnamefont {Chiara}}, \bibinfo {author} {\bibfnamefont {J.~L.}\ \bibnamefont {Harker}},\ and\ \bibinfo {author} {\bibfnamefont {W.~B.}\ \bibnamefont {Walters}},\ }\href {https://doi.org/10.1103/PhysRevC.91.064615} {\bibfield
  {journal} {\bibinfo  {journal} {Phys. Rev. C}\ }\textbf {\bibinfo {volume} {91}},\ \bibinfo {pages} {064615} (\bibinfo {year} {2015})}\BibitemShut {NoStop}%
\bibitem [{\citenamefont {Desai}\ \emph {et~al.}(2019)\citenamefont {Desai}, \citenamefont {Loveland}, \citenamefont {McCaleb}, \citenamefont {Yanez}, \citenamefont {Lane}, \citenamefont {Hota}, \citenamefont {Reed}, \citenamefont {Watanabe}, \citenamefont {Zhu}, \citenamefont {Auranen}, \citenamefont {Ayangeakaa}, \citenamefont {Carpenter}, \citenamefont {Greene}, \citenamefont {Kondev}, \citenamefont {Seweryniak}, \citenamefont {Janssens},\ and\ \citenamefont {Copp}}]{desai2019}%
  \BibitemOpen
  \bibfield  {author} {\bibinfo {author} {\bibfnamefont {V.~V.}\ \bibnamefont {Desai}}, \bibinfo {author} {\bibfnamefont {W.}~\bibnamefont {Loveland}}, \bibinfo {author} {\bibfnamefont {K.}~\bibnamefont {McCaleb}}, \bibinfo {author} {\bibfnamefont {R.}~\bibnamefont {Yanez}}, \bibinfo {author} {\bibfnamefont {G.}~\bibnamefont {Lane}}, \bibinfo {author} {\bibfnamefont {S.~S.}\ \bibnamefont {Hota}}, \bibinfo {author} {\bibfnamefont {M.~W.}\ \bibnamefont {Reed}}, \bibinfo {author} {\bibfnamefont {H.}~\bibnamefont {Watanabe}}, \bibinfo {author} {\bibfnamefont {S.}~\bibnamefont {Zhu}}, \bibinfo {author} {\bibfnamefont {K.}~\bibnamefont {Auranen}}, \bibinfo {author} {\bibfnamefont {A.~D.}\ \bibnamefont {Ayangeakaa}}, \bibinfo {author} {\bibfnamefont {M.~P.}\ \bibnamefont {Carpenter}}, \bibinfo {author} {\bibfnamefont {J.~P.}\ \bibnamefont {Greene}}, \bibinfo {author} {\bibfnamefont {F.~G.}\ \bibnamefont {Kondev}}, \bibinfo {author} {\bibfnamefont {D.}~\bibnamefont {Seweryniak}}, \bibinfo {author} {\bibfnamefont
  {R.~V.~F.}\ \bibnamefont {Janssens}},\ and\ \bibinfo {author} {\bibfnamefont {P.~A.}\ \bibnamefont {Copp}},\ }\href {https://doi.org/10.1103/PhysRevC.99.044604} {\bibfield  {journal} {\bibinfo  {journal} {Phys. Rev. C}\ }\textbf {\bibinfo {volume} {99}},\ \bibinfo {pages} {044604} (\bibinfo {year} {2019})}\BibitemShut {NoStop}%
\bibitem [{\citenamefont {Cocks}\ \emph {et~al.}(1997)\citenamefont {Cocks}, \citenamefont {Butler}, \citenamefont {Cann}, \citenamefont {Greenlees}, \citenamefont {Jones}, \citenamefont {Asztalos}, \citenamefont {Bhattacharyya}, \citenamefont {Broda}, \citenamefont {Clark}, \citenamefont {Deleplanque}, \citenamefont {Diamond}, \citenamefont {Fallon}, \citenamefont {Fornal}, \citenamefont {Jones}, \citenamefont {Julin}, \citenamefont {Lauritsen}, \citenamefont {Lee}, \citenamefont {Macchiavelli}, \citenamefont {MacLeod}, \citenamefont {Smith}, \citenamefont {Stephens},\ and\ \citenamefont {Zhang}}]{Cocks1997}%
  \BibitemOpen
  \bibfield  {author} {\bibinfo {author} {\bibfnamefont {J.~F.~C.}\ \bibnamefont {Cocks}}, \bibinfo {author} {\bibfnamefont {P.~A.}\ \bibnamefont {Butler}}, \bibinfo {author} {\bibfnamefont {K.~J.}\ \bibnamefont {Cann}}, \bibinfo {author} {\bibfnamefont {P.~T.}\ \bibnamefont {Greenlees}}, \bibinfo {author} {\bibfnamefont {G.~D.}\ \bibnamefont {Jones}}, \bibinfo {author} {\bibfnamefont {S.}~\bibnamefont {Asztalos}}, \bibinfo {author} {\bibfnamefont {P.}~\bibnamefont {Bhattacharyya}}, \bibinfo {author} {\bibfnamefont {R.}~\bibnamefont {Broda}}, \bibinfo {author} {\bibfnamefont {R.~M.}\ \bibnamefont {Clark}}, \bibinfo {author} {\bibfnamefont {M.~A.}\ \bibnamefont {Deleplanque}}, \bibinfo {author} {\bibfnamefont {R.~M.}\ \bibnamefont {Diamond}}, \bibinfo {author} {\bibfnamefont {P.}~\bibnamefont {Fallon}}, \bibinfo {author} {\bibfnamefont {B.}~\bibnamefont {Fornal}}, \bibinfo {author} {\bibfnamefont {P.~M.}\ \bibnamefont {Jones}}, \bibinfo {author} {\bibfnamefont {R.}~\bibnamefont {Julin}}, \bibinfo {author}
  {\bibfnamefont {T.}~\bibnamefont {Lauritsen}}, \bibinfo {author} {\bibfnamefont {I.~Y.}\ \bibnamefont {Lee}}, \bibinfo {author} {\bibfnamefont {A.~O.}\ \bibnamefont {Macchiavelli}}, \bibinfo {author} {\bibfnamefont {R.~W.}\ \bibnamefont {MacLeod}}, \bibinfo {author} {\bibfnamefont {J.~F.}\ \bibnamefont {Smith}}, \bibinfo {author} {\bibfnamefont {F.~S.}\ \bibnamefont {Stephens}},\ and\ \bibinfo {author} {\bibfnamefont {C.~T.}\ \bibnamefont {Zhang}},\ }\href {https://doi.org/10.1103/PhysRevLett.78.2920} {\bibfield  {journal} {\bibinfo  {journal} {Phys. Rev. Lett.}\ }\textbf {\bibinfo {volume} {78}},\ \bibinfo {pages} {2920} (\bibinfo {year} {1997})}\BibitemShut {NoStop}%
\bibitem [{\citenamefont {Kraus}\ \emph {et~al.}(1988)\citenamefont {Kraus}, \citenamefont {Boucenna}, \citenamefont {Linck}, \citenamefont {Lott}, \citenamefont {Rebmeister}, \citenamefont {Schulz}, \citenamefont {Sens}, \citenamefont {Mermaz}, \citenamefont {Berthier}, \citenamefont {Lucas}, \citenamefont {Gastebois}, \citenamefont {Gillibert}, \citenamefont {Miczaika}, \citenamefont {Tomasi-Gustafsson},\ and\ \citenamefont {Grunberg}}]{Kraus1988}%
  \BibitemOpen
  \bibfield  {author} {\bibinfo {author} {\bibfnamefont {L.}~\bibnamefont {Kraus}}, \bibinfo {author} {\bibfnamefont {A.}~\bibnamefont {Boucenna}}, \bibinfo {author} {\bibfnamefont {I.}~\bibnamefont {Linck}}, \bibinfo {author} {\bibfnamefont {B.}~\bibnamefont {Lott}}, \bibinfo {author} {\bibfnamefont {R.}~\bibnamefont {Rebmeister}}, \bibinfo {author} {\bibfnamefont {N.}~\bibnamefont {Schulz}}, \bibinfo {author} {\bibfnamefont {J.~C.}\ \bibnamefont {Sens}}, \bibinfo {author} {\bibfnamefont {M.~C.}\ \bibnamefont {Mermaz}}, \bibinfo {author} {\bibfnamefont {B.}~\bibnamefont {Berthier}}, \bibinfo {author} {\bibfnamefont {R.}~\bibnamefont {Lucas}}, \bibinfo {author} {\bibfnamefont {J.}~\bibnamefont {Gastebois}}, \bibinfo {author} {\bibfnamefont {A.}~\bibnamefont {Gillibert}}, \bibinfo {author} {\bibfnamefont {A.}~\bibnamefont {Miczaika}}, \bibinfo {author} {\bibfnamefont {E.}~\bibnamefont {Tomasi-Gustafsson}},\ and\ \bibinfo {author} {\bibfnamefont {C.}~\bibnamefont {Grunberg}},\ }\href
  {https://doi.org/10.1103/PhysRevC.37.2529} {\bibfield  {journal} {\bibinfo  {journal} {Phys. Rev. C}\ }\textbf {\bibinfo {volume} {37}},\ \bibinfo {pages} {2529} (\bibinfo {year} {1988})}\BibitemShut {NoStop}%
\bibitem [{\citenamefont {Guidry}\ \emph {et~al.}(1985)\citenamefont {Guidry}, \citenamefont {Juutinen}, \citenamefont {Liu}, \citenamefont {Bingham}, \citenamefont {Larabee}, \citenamefont {Riedinger}, \citenamefont {Baktash}, \citenamefont {Lee}, \citenamefont {Halbert}, \citenamefont {Cline}, \citenamefont {Kotlinski}, \citenamefont {Kernan}, \citenamefont {Semkow}, \citenamefont {Sarantites}, \citenamefont {Honkanen},\ and\ \citenamefont {Rajagopalan}}]{GUID85}%
  \BibitemOpen
  \bibfield  {author} {\bibinfo {author} {\bibfnamefont {M.}~\bibnamefont {Guidry}}, \bibinfo {author} {\bibfnamefont {S.}~\bibnamefont {Juutinen}}, \bibinfo {author} {\bibfnamefont {X.}~\bibnamefont {Liu}}, \bibinfo {author} {\bibfnamefont {C.}~\bibnamefont {Bingham}}, \bibinfo {author} {\bibfnamefont {A.}~\bibnamefont {Larabee}}, \bibinfo {author} {\bibfnamefont {L.}~\bibnamefont {Riedinger}}, \bibinfo {author} {\bibfnamefont {C.}~\bibnamefont {Baktash}}, \bibinfo {author} {\bibfnamefont {I.}~\bibnamefont {Lee}}, \bibinfo {author} {\bibfnamefont {M.}~\bibnamefont {Halbert}}, \bibinfo {author} {\bibfnamefont {D.}~\bibnamefont {Cline}}, \bibinfo {author} {\bibfnamefont {B.}~\bibnamefont {Kotlinski}}, \bibinfo {author} {\bibfnamefont {W.}~\bibnamefont {Kernan}}, \bibinfo {author} {\bibfnamefont {T.}~\bibnamefont {Semkow}}, \bibinfo {author} {\bibfnamefont {D.}~\bibnamefont {Sarantites}}, \bibinfo {author} {\bibfnamefont {K.}~\bibnamefont {Honkanen}},\ and\ \bibinfo {author} {\bibfnamefont {M.}~\bibnamefont
  {Rajagopalan}},\ }\href {https://doi.org/https://doi.org/10.1016/0370-2693(85)90196-0} {\bibfield  {journal} {\bibinfo  {journal} {Physics Letters B}\ }\textbf {\bibinfo {volume} {163}},\ \bibinfo {pages} {79} (\bibinfo {year} {1985})}\BibitemShut {NoStop}%
\bibitem [{\citenamefont {Watanabe}\ \emph {et~al.}(2021)\citenamefont {Watanabe}, \citenamefont {Watanabe}, \citenamefont {Hirayama}, \citenamefont {Andreyev}, \citenamefont {Hashimoto}, \citenamefont {Kondev}, \citenamefont {Lane}, \citenamefont {Litvinov}, \citenamefont {Liu}, \citenamefont {Miyatake}, \citenamefont {Moon}, \citenamefont {Morales}, \citenamefont {Mukai}, \citenamefont {Nishimura}, \citenamefont {Niwase}, \citenamefont {Rosenbusch}, \citenamefont {Schury}, \citenamefont {Shi}, \citenamefont {Wada},\ and\ \citenamefont {Walker}}]{WATANABE2021}%
  \BibitemOpen
  \bibfield  {author} {\bibinfo {author} {\bibfnamefont {H.}~\bibnamefont {Watanabe}}, \bibinfo {author} {\bibfnamefont {Y.}~\bibnamefont {Watanabe}}, \bibinfo {author} {\bibfnamefont {Y.}~\bibnamefont {Hirayama}}, \bibinfo {author} {\bibfnamefont {A.}~\bibnamefont {Andreyev}}, \bibinfo {author} {\bibfnamefont {T.}~\bibnamefont {Hashimoto}}, \bibinfo {author} {\bibfnamefont {F.}~\bibnamefont {Kondev}}, \bibinfo {author} {\bibfnamefont {G.}~\bibnamefont {Lane}}, \bibinfo {author} {\bibfnamefont {Y.}~\bibnamefont {Litvinov}}, \bibinfo {author} {\bibfnamefont {J.}~\bibnamefont {Liu}}, \bibinfo {author} {\bibfnamefont {H.}~\bibnamefont {Miyatake}}, \bibinfo {author} {\bibfnamefont {J.}~\bibnamefont {Moon}}, \bibinfo {author} {\bibfnamefont {A.}~\bibnamefont {Morales}}, \bibinfo {author} {\bibfnamefont {M.}~\bibnamefont {Mukai}}, \bibinfo {author} {\bibfnamefont {S.}~\bibnamefont {Nishimura}}, \bibinfo {author} {\bibfnamefont {T.}~\bibnamefont {Niwase}}, \bibinfo {author} {\bibfnamefont {M.}~\bibnamefont
  {Rosenbusch}}, \bibinfo {author} {\bibfnamefont {P.}~\bibnamefont {Schury}}, \bibinfo {author} {\bibfnamefont {Y.}~\bibnamefont {Shi}}, \bibinfo {author} {\bibfnamefont {M.}~\bibnamefont {Wada}},\ and\ \bibinfo {author} {\bibfnamefont {P.}~\bibnamefont {Walker}},\ }\href {https://doi.org/https://doi.org/10.1016/j.physletb.2021.136088} {\bibfield  {journal} {\bibinfo  {journal} {Physics Letters B}\ }\textbf {\bibinfo {volume} {814}},\ \bibinfo {pages} {136088} (\bibinfo {year} {2021})}\BibitemShut {NoStop}%
\bibitem [{\citenamefont {Hirayama}\ \emph {et~al.}(2018)\citenamefont {Hirayama}, \citenamefont {Watanabe}, \citenamefont {Mukai}, \citenamefont {Ahmed}, \citenamefont {Jeong}, \citenamefont {Kakiguchi}, \citenamefont {Kimura}, \citenamefont {Oyaizu}, \citenamefont {Park}, \citenamefont {Schury}, \citenamefont {Wada}, \citenamefont {Watanabe},\ and\ \citenamefont {Miyatake}}]{HWatanabe2018}%
  \BibitemOpen
  \bibfield  {author} {\bibinfo {author} {\bibfnamefont {Y.}~\bibnamefont {Hirayama}}, \bibinfo {author} {\bibfnamefont {Y.~X.}\ \bibnamefont {Watanabe}}, \bibinfo {author} {\bibfnamefont {M.}~\bibnamefont {Mukai}}, \bibinfo {author} {\bibfnamefont {M.}~\bibnamefont {Ahmed}}, \bibinfo {author} {\bibfnamefont {S.~C.}\ \bibnamefont {Jeong}}, \bibinfo {author} {\bibfnamefont {Y.}~\bibnamefont {Kakiguchi}}, \bibinfo {author} {\bibfnamefont {S.}~\bibnamefont {Kimura}}, \bibinfo {author} {\bibfnamefont {M.}~\bibnamefont {Oyaizu}}, \bibinfo {author} {\bibfnamefont {J.~H.}\ \bibnamefont {Park}}, \bibinfo {author} {\bibfnamefont {P.}~\bibnamefont {Schury}}, \bibinfo {author} {\bibfnamefont {M.}~\bibnamefont {Wada}}, \bibinfo {author} {\bibfnamefont {H.}~\bibnamefont {Watanabe}},\ and\ \bibinfo {author} {\bibfnamefont {H.}~\bibnamefont {Miyatake}},\ }\href {https://doi.org/10.1103/PhysRevC.98.014321} {\bibfield  {journal} {\bibinfo  {journal} {Phys. Rev. C}\ }\textbf {\bibinfo {volume} {98}},\ \bibinfo {pages} {014321}
  (\bibinfo {year} {2018})}\BibitemShut {NoStop}%
\bibitem [{\citenamefont {John}\ \emph {et~al.}(2017)\citenamefont {John}, \citenamefont {Valiente-Dob\'on}, \citenamefont {Mengoni}, \citenamefont {Modamio}, \citenamefont {Lunardi}, \citenamefont {Bazzacco}, \citenamefont {Gadea}, \citenamefont {Wheldon}, \citenamefont {Rodr\'{\i}guez}, \citenamefont {Alexander}, \citenamefont {de~Angelis}, \citenamefont {Ashwood}, \citenamefont {Barr}, \citenamefont {Benzoni}, \citenamefont {Birkenbach}, \citenamefont {Bizzeti}, \citenamefont {Bizzeti-Sona}, \citenamefont {Bottoni}, \citenamefont {Bowry}, \citenamefont {Bracco}, \citenamefont {Browne}, \citenamefont {Bunce}, \citenamefont {Camera}, \citenamefont {Corradi}, \citenamefont {Crespi}, \citenamefont {Melon}, \citenamefont {Farnea}, \citenamefont {Fioretto}, \citenamefont {Gottardo}, \citenamefont {Grente}, \citenamefont {Hess}, \citenamefont {Kokalova}, \citenamefont {Korten}, \citenamefont {Ku\ifmmode \mbox{\c{s}}\else \c{s}\fi{}o\ifmmode~\breve{g}\else \u{g}\fi{}lu}, \citenamefont {Lenzi}, \citenamefont
  {Leoni}, \citenamefont {Ljungvall}, \citenamefont {Menegazzo}, \citenamefont {Michelagnoli}, \citenamefont {Mijatovi\ifmmode~\acute{c}\else \'{c}\fi{}}, \citenamefont {Montagnoli}, \citenamefont {Montanari}, \citenamefont {Napoli}, \citenamefont {Podoly\'ak}, \citenamefont {Pollarolo}, \citenamefont {Recchia}, \citenamefont {Reiter}, \citenamefont {Roberts}, \citenamefont {\ifmmode~\mbox{\c{S}}\else \c{S}\fi{}ahin}, \citenamefont {Salsac}, \citenamefont {Scarlassara}, \citenamefont {Sferrazza}, \citenamefont {S\"oderstr\"om}, \citenamefont {Stefanini}, \citenamefont {Szilner}, \citenamefont {Ur}, \citenamefont {Vogt},\ and\ \citenamefont {Walshe}}]{John2017}%
  \BibitemOpen
  \bibfield  {author} {\bibinfo {author} {\bibfnamefont {P.~R.}\ \bibnamefont {John}}, \bibinfo {author} {\bibfnamefont {J.~J.}\ \bibnamefont {Valiente-Dob\'on}}, \bibinfo {author} {\bibfnamefont {D.}~\bibnamefont {Mengoni}}, \bibinfo {author} {\bibfnamefont {V.}~\bibnamefont {Modamio}}, \bibinfo {author} {\bibfnamefont {S.}~\bibnamefont {Lunardi}}, \bibinfo {author} {\bibfnamefont {D.}~\bibnamefont {Bazzacco}}, \bibinfo {author} {\bibfnamefont {A.}~\bibnamefont {Gadea}}, \bibinfo {author} {\bibfnamefont {C.}~\bibnamefont {Wheldon}}, \bibinfo {author} {\bibfnamefont {T.~R.}\ \bibnamefont {Rodr\'{\i}guez}}, \bibinfo {author} {\bibfnamefont {T.}~\bibnamefont {Alexander}}, \bibinfo {author} {\bibfnamefont {G.}~\bibnamefont {de~Angelis}}, \bibinfo {author} {\bibfnamefont {N.}~\bibnamefont {Ashwood}}, \bibinfo {author} {\bibfnamefont {M.}~\bibnamefont {Barr}}, \bibinfo {author} {\bibfnamefont {G.}~\bibnamefont {Benzoni}}, \bibinfo {author} {\bibfnamefont {B.}~\bibnamefont {Birkenbach}}, \bibinfo {author}
  {\bibfnamefont {P.~G.}\ \bibnamefont {Bizzeti}}, \bibinfo {author} {\bibfnamefont {A.~M.}\ \bibnamefont {Bizzeti-Sona}}, \bibinfo {author} {\bibfnamefont {S.}~\bibnamefont {Bottoni}}, \bibinfo {author} {\bibfnamefont {M.}~\bibnamefont {Bowry}}, \bibinfo {author} {\bibfnamefont {A.}~\bibnamefont {Bracco}}, \bibinfo {author} {\bibfnamefont {F.}~\bibnamefont {Browne}}, \bibinfo {author} {\bibfnamefont {M.}~\bibnamefont {Bunce}}, \bibinfo {author} {\bibfnamefont {F.}~\bibnamefont {Camera}}, \bibinfo {author} {\bibfnamefont {L.}~\bibnamefont {Corradi}}, \bibinfo {author} {\bibfnamefont {F.~C.~L.}\ \bibnamefont {Crespi}}, \bibinfo {author} {\bibfnamefont {B.}~\bibnamefont {Melon}}, \bibinfo {author} {\bibfnamefont {E.}~\bibnamefont {Farnea}}, \bibinfo {author} {\bibfnamefont {E.}~\bibnamefont {Fioretto}}, \bibinfo {author} {\bibfnamefont {A.}~\bibnamefont {Gottardo}}, \bibinfo {author} {\bibfnamefont {L.}~\bibnamefont {Grente}}, \bibinfo {author} {\bibfnamefont {H.}~\bibnamefont {Hess}}, \bibinfo {author}
  {\bibfnamefont {T.}~\bibnamefont {Kokalova}}, \bibinfo {author} {\bibfnamefont {W.}~\bibnamefont {Korten}}, \bibinfo {author} {\bibfnamefont {A.}~\bibnamefont {Ku\ifmmode \mbox{\c{s}}\else \c{s}\fi{}o\ifmmode~\breve{g}\else \u{g}\fi{}lu}}, \bibinfo {author} {\bibfnamefont {S.}~\bibnamefont {Lenzi}}, \bibinfo {author} {\bibfnamefont {S.}~\bibnamefont {Leoni}}, \bibinfo {author} {\bibfnamefont {J.}~\bibnamefont {Ljungvall}}, \bibinfo {author} {\bibfnamefont {R.}~\bibnamefont {Menegazzo}}, \bibinfo {author} {\bibfnamefont {C.}~\bibnamefont {Michelagnoli}}, \bibinfo {author} {\bibfnamefont {T.}~\bibnamefont {Mijatovi\ifmmode~\acute{c}\else \'{c}\fi{}}}, \bibinfo {author} {\bibfnamefont {G.}~\bibnamefont {Montagnoli}}, \bibinfo {author} {\bibfnamefont {D.}~\bibnamefont {Montanari}}, \bibinfo {author} {\bibfnamefont {D.~R.}\ \bibnamefont {Napoli}}, \bibinfo {author} {\bibfnamefont {Z.}~\bibnamefont {Podoly\'ak}}, \bibinfo {author} {\bibfnamefont {G.}~\bibnamefont {Pollarolo}}, \bibinfo {author} {\bibfnamefont
  {F.}~\bibnamefont {Recchia}}, \bibinfo {author} {\bibfnamefont {P.}~\bibnamefont {Reiter}}, \bibinfo {author} {\bibfnamefont {O.~J.}\ \bibnamefont {Roberts}}, \bibinfo {author} {\bibfnamefont {E.}~\bibnamefont {\ifmmode~\mbox{\c{S}}\else \c{S}\fi{}ahin}}, \bibinfo {author} {\bibfnamefont {M.-D.}\ \bibnamefont {Salsac}}, \bibinfo {author} {\bibfnamefont {F.}~\bibnamefont {Scarlassara}}, \bibinfo {author} {\bibfnamefont {M.}~\bibnamefont {Sferrazza}}, \bibinfo {author} {\bibfnamefont {P.-A.}\ \bibnamefont {S\"oderstr\"om}}, \bibinfo {author} {\bibfnamefont {A.~M.}\ \bibnamefont {Stefanini}}, \bibinfo {author} {\bibfnamefont {S.}~\bibnamefont {Szilner}}, \bibinfo {author} {\bibfnamefont {C.~A.}\ \bibnamefont {Ur}}, \bibinfo {author} {\bibfnamefont {A.}~\bibnamefont {Vogt}},\ and\ \bibinfo {author} {\bibfnamefont {J.}~\bibnamefont {Walshe}},\ }\href {https://doi.org/10.1103/PhysRevC.95.064321} {\bibfield  {journal} {\bibinfo  {journal} {Phys. Rev. C}\ }\textbf {\bibinfo {volume} {95}},\ \bibinfo {pages}
  {064321} (\bibinfo {year} {2017})}\BibitemShut {NoStop}%
\bibitem [{\citenamefont {John}\ \emph {et~al.}(2014)\citenamefont {John}, \citenamefont {Modamio}, \citenamefont {Valiente-Dob\'on}, \citenamefont {Mengoni}, \citenamefont {Lunardi}, \citenamefont {Rodr\'{\i}guez}, \citenamefont {Bazzacco}, \citenamefont {Gadea}, \citenamefont {Wheldon}, \citenamefont {Alexander}, \citenamefont {de~Angelis}, \citenamefont {Ashwood}, \citenamefont {Barr}, \citenamefont {Benzoni}, \citenamefont {Birkenbach}, \citenamefont {Bizzeti}, \citenamefont {Bizzeti-Sona}, \citenamefont {Bottoni}, \citenamefont {Bowry}, \citenamefont {Bracco}, \citenamefont {Browne}, \citenamefont {Bunce}, \citenamefont {Camera}, \citenamefont {Cederwall}, \citenamefont {Corradi}, \citenamefont {Crespi}, \citenamefont {D\'esesquelles}, \citenamefont {Eberth}, \citenamefont {Farnea}, \citenamefont {Fioretto}, \citenamefont {G\"orgen}, \citenamefont {Gottardo}, \citenamefont {Grebosz}, \citenamefont {Grente}, \citenamefont {Hess}, \citenamefont {Jungclaus}, \citenamefont {Kokalova}, \citenamefont
  {Korichi}, \citenamefont {Korten}, \citenamefont {Ku\ifmmode \mbox{\c{s}}\else \c{s}\fi{}o\ifmmode~\breve{g}\else \u{g}\fi{}lu}, \citenamefont {Lenzi}, \citenamefont {Leoni}, \citenamefont {Ljungvall}, \citenamefont {Maron}, \citenamefont {Meczynski}, \citenamefont {Melon}, \citenamefont {Menegazzo}, \citenamefont {Michelagnoli}, \citenamefont {Mijatovi\ifmmode~\acute{c}\else \'{c}\fi{}}, \citenamefont {Million}, \citenamefont {Molini}, \citenamefont {Montagnoli}, \citenamefont {Montanari}, \citenamefont {Napoli}, \citenamefont {Nolan}, \citenamefont {Oziol}, \citenamefont {Podoly\'ak}, \citenamefont {Pollarolo}, \citenamefont {Pullia}, \citenamefont {Quintana}, \citenamefont {Recchia}, \citenamefont {Reiter}, \citenamefont {Roberts}, \citenamefont {Rosso}, \citenamefont {\ifmmode~\mbox{\c{S}}\else \c{S}\fi{}ahin}, \citenamefont {Salsac}, \citenamefont {Scarlassara}, \citenamefont {Sferrazza}, \citenamefont {Simpson}, \citenamefont {S\"oderstr\"om}, \citenamefont {Stefanini}, \citenamefont {Stezowski},
  \citenamefont {Szilner}, \citenamefont {Theisen}, \citenamefont {Ur},\ and\ \citenamefont {Walshe}}]{John2014}%
  \BibitemOpen
  \bibfield  {author} {\bibinfo {author} {\bibfnamefont {P.~R.}\ \bibnamefont {John}}, \bibinfo {author} {\bibfnamefont {V.}~\bibnamefont {Modamio}}, \bibinfo {author} {\bibfnamefont {J.~J.}\ \bibnamefont {Valiente-Dob\'on}}, \bibinfo {author} {\bibfnamefont {D.}~\bibnamefont {Mengoni}}, \bibinfo {author} {\bibfnamefont {S.}~\bibnamefont {Lunardi}}, \bibinfo {author} {\bibfnamefont {T.~R.}\ \bibnamefont {Rodr\'{\i}guez}}, \bibinfo {author} {\bibfnamefont {D.}~\bibnamefont {Bazzacco}}, \bibinfo {author} {\bibfnamefont {A.}~\bibnamefont {Gadea}}, \bibinfo {author} {\bibfnamefont {C.}~\bibnamefont {Wheldon}}, \bibinfo {author} {\bibfnamefont {T.}~\bibnamefont {Alexander}}, \bibinfo {author} {\bibfnamefont {G.}~\bibnamefont {de~Angelis}}, \bibinfo {author} {\bibfnamefont {N.}~\bibnamefont {Ashwood}}, \bibinfo {author} {\bibfnamefont {M.}~\bibnamefont {Barr}}, \bibinfo {author} {\bibfnamefont {G.}~\bibnamefont {Benzoni}}, \bibinfo {author} {\bibfnamefont {B.}~\bibnamefont {Birkenbach}}, \bibinfo {author}
  {\bibfnamefont {P.~G.}\ \bibnamefont {Bizzeti}}, \bibinfo {author} {\bibfnamefont {A.~M.}\ \bibnamefont {Bizzeti-Sona}}, \bibinfo {author} {\bibfnamefont {S.}~\bibnamefont {Bottoni}}, \bibinfo {author} {\bibfnamefont {M.}~\bibnamefont {Bowry}}, \bibinfo {author} {\bibfnamefont {A.}~\bibnamefont {Bracco}}, \bibinfo {author} {\bibfnamefont {F.}~\bibnamefont {Browne}}, \bibinfo {author} {\bibfnamefont {M.}~\bibnamefont {Bunce}}, \bibinfo {author} {\bibfnamefont {F.}~\bibnamefont {Camera}}, \bibinfo {author} {\bibfnamefont {B.}~\bibnamefont {Cederwall}}, \bibinfo {author} {\bibfnamefont {L.}~\bibnamefont {Corradi}}, \bibinfo {author} {\bibfnamefont {F.~C.~L.}\ \bibnamefont {Crespi}}, \bibinfo {author} {\bibfnamefont {P.}~\bibnamefont {D\'esesquelles}}, \bibinfo {author} {\bibfnamefont {J.}~\bibnamefont {Eberth}}, \bibinfo {author} {\bibfnamefont {E.}~\bibnamefont {Farnea}}, \bibinfo {author} {\bibfnamefont {E.}~\bibnamefont {Fioretto}}, \bibinfo {author} {\bibfnamefont {A.}~\bibnamefont {G\"orgen}}, \bibinfo
  {author} {\bibfnamefont {A.}~\bibnamefont {Gottardo}}, \bibinfo {author} {\bibfnamefont {J.}~\bibnamefont {Grebosz}}, \bibinfo {author} {\bibfnamefont {L.}~\bibnamefont {Grente}}, \bibinfo {author} {\bibfnamefont {H.}~\bibnamefont {Hess}}, \bibinfo {author} {\bibfnamefont {A.}~\bibnamefont {Jungclaus}}, \bibinfo {author} {\bibfnamefont {T.}~\bibnamefont {Kokalova}}, \bibinfo {author} {\bibfnamefont {A.}~\bibnamefont {Korichi}}, \bibinfo {author} {\bibfnamefont {W.}~\bibnamefont {Korten}}, \bibinfo {author} {\bibfnamefont {A.}~\bibnamefont {Ku\ifmmode \mbox{\c{s}}\else \c{s}\fi{}o\ifmmode~\breve{g}\else \u{g}\fi{}lu}}, \bibinfo {author} {\bibfnamefont {S.}~\bibnamefont {Lenzi}}, \bibinfo {author} {\bibfnamefont {S.}~\bibnamefont {Leoni}}, \bibinfo {author} {\bibfnamefont {J.}~\bibnamefont {Ljungvall}}, \bibinfo {author} {\bibfnamefont {G.}~\bibnamefont {Maron}}, \bibinfo {author} {\bibfnamefont {W.}~\bibnamefont {Meczynski}}, \bibinfo {author} {\bibfnamefont {B.}~\bibnamefont {Melon}}, \bibinfo {author}
  {\bibfnamefont {R.}~\bibnamefont {Menegazzo}}, \bibinfo {author} {\bibfnamefont {C.}~\bibnamefont {Michelagnoli}}, \bibinfo {author} {\bibfnamefont {T.}~\bibnamefont {Mijatovi\ifmmode~\acute{c}\else \'{c}\fi{}}}, \bibinfo {author} {\bibfnamefont {B.}~\bibnamefont {Million}}, \bibinfo {author} {\bibfnamefont {P.}~\bibnamefont {Molini}}, \bibinfo {author} {\bibfnamefont {G.}~\bibnamefont {Montagnoli}}, \bibinfo {author} {\bibfnamefont {D.}~\bibnamefont {Montanari}}, \bibinfo {author} {\bibfnamefont {D.~R.}\ \bibnamefont {Napoli}}, \bibinfo {author} {\bibfnamefont {P.}~\bibnamefont {Nolan}}, \bibinfo {author} {\bibfnamefont {C.}~\bibnamefont {Oziol}}, \bibinfo {author} {\bibfnamefont {Z.}~\bibnamefont {Podoly\'ak}}, \bibinfo {author} {\bibfnamefont {G.}~\bibnamefont {Pollarolo}}, \bibinfo {author} {\bibfnamefont {A.}~\bibnamefont {Pullia}}, \bibinfo {author} {\bibfnamefont {B.}~\bibnamefont {Quintana}}, \bibinfo {author} {\bibfnamefont {F.}~\bibnamefont {Recchia}}, \bibinfo {author} {\bibfnamefont
  {P.}~\bibnamefont {Reiter}}, \bibinfo {author} {\bibfnamefont {O.~J.}\ \bibnamefont {Roberts}}, \bibinfo {author} {\bibfnamefont {D.}~\bibnamefont {Rosso}}, \bibinfo {author} {\bibfnamefont {E.}~\bibnamefont {\ifmmode~\mbox{\c{S}}\else \c{S}\fi{}ahin}}, \bibinfo {author} {\bibfnamefont {M.-D.}\ \bibnamefont {Salsac}}, \bibinfo {author} {\bibfnamefont {F.}~\bibnamefont {Scarlassara}}, \bibinfo {author} {\bibfnamefont {M.}~\bibnamefont {Sferrazza}}, \bibinfo {author} {\bibfnamefont {J.}~\bibnamefont {Simpson}}, \bibinfo {author} {\bibfnamefont {P.-A.}\ \bibnamefont {S\"oderstr\"om}}, \bibinfo {author} {\bibfnamefont {A.~M.}\ \bibnamefont {Stefanini}}, \bibinfo {author} {\bibfnamefont {O.}~\bibnamefont {Stezowski}}, \bibinfo {author} {\bibfnamefont {S.}~\bibnamefont {Szilner}}, \bibinfo {author} {\bibfnamefont {C.}~\bibnamefont {Theisen}}, \bibinfo {author} {\bibfnamefont {C.~A.}\ \bibnamefont {Ur}},\ and\ \bibinfo {author} {\bibfnamefont {J.}~\bibnamefont {Walshe}},\ }\href
  {https://doi.org/10.1103/PhysRevC.90.021301} {\bibfield  {journal} {\bibinfo  {journal} {Phys. Rev. C}\ }\textbf {\bibinfo {volume} {90}},\ \bibinfo {pages} {021301} (\bibinfo {year} {2014})}\BibitemShut {NoStop}%
\bibitem [{\citenamefont {Watanabe}\ \emph {et~al.}(2025)\citenamefont {Watanabe}, \citenamefont {Hirayama}, \citenamefont {Mukai}, \citenamefont {Schury}, \citenamefont {Niwase}, \citenamefont {Kimura}, \citenamefont {Rosenbusch}, \citenamefont {Hashimoto}, \citenamefont {Moon}, \citenamefont {Ishiyama}, \citenamefont {Iimura}, \citenamefont {Miyatake},\ and\ \citenamefont {Wada}}]{WATANABE2025}%
  \BibitemOpen
  \bibfield  {author} {\bibinfo {author} {\bibfnamefont {Y.}~\bibnamefont {Watanabe}}, \bibinfo {author} {\bibfnamefont {Y.}~\bibnamefont {Hirayama}}, \bibinfo {author} {\bibfnamefont {M.}~\bibnamefont {Mukai}}, \bibinfo {author} {\bibfnamefont {P.}~\bibnamefont {Schury}}, \bibinfo {author} {\bibfnamefont {T.}~\bibnamefont {Niwase}}, \bibinfo {author} {\bibfnamefont {S.}~\bibnamefont {Kimura}}, \bibinfo {author} {\bibfnamefont {M.}~\bibnamefont {Rosenbusch}}, \bibinfo {author} {\bibfnamefont {T.}~\bibnamefont {Hashimoto}}, \bibinfo {author} {\bibfnamefont {J.}~\bibnamefont {Moon}}, \bibinfo {author} {\bibfnamefont {H.}~\bibnamefont {Ishiyama}}, \bibinfo {author} {\bibfnamefont {S.}~\bibnamefont {Iimura}}, \bibinfo {author} {\bibfnamefont {H.}~\bibnamefont {Miyatake}},\ and\ \bibinfo {author} {\bibfnamefont {M.}~\bibnamefont {Wada}},\ }\href {https://doi.org/https://doi.org/10.1016/j.nuclphysa.2025.123140} {\bibfield  {journal} {\bibinfo  {journal} {Nuclear Physics A}\ }\textbf {\bibinfo {volume} {1061}},\
  \bibinfo {pages} {123140} (\bibinfo {year} {2025})}\BibitemShut {NoStop}%
\bibitem [{\citenamefont {Hirayama}\ \emph {et~al.}(2024)\citenamefont {Hirayama}, \citenamefont {Mukai}, \citenamefont {Watanabe}, \citenamefont {Schury}, \citenamefont {Niwase}, \citenamefont {Choi}, \citenamefont {Hashimoto}, \citenamefont {Iimura}, \citenamefont {Jeong}, \citenamefont {Miyatake}, \citenamefont {Moon}, \citenamefont {Nakada}, \citenamefont {Oyaizu}, \citenamefont {Rosenbusch}, \citenamefont {Takamine}, \citenamefont {Tajima}, \citenamefont {Taniguchi},\ and\ \citenamefont {Wada}}]{Hirayama2024}%
  \BibitemOpen
  \bibfield  {author} {\bibinfo {author} {\bibfnamefont {Y.}~\bibnamefont {Hirayama}}, \bibinfo {author} {\bibfnamefont {M.}~\bibnamefont {Mukai}}, \bibinfo {author} {\bibfnamefont {Y.}~\bibnamefont {Watanabe}}, \bibinfo {author} {\bibfnamefont {P.}~\bibnamefont {Schury}}, \bibinfo {author} {\bibfnamefont {T.}~\bibnamefont {Niwase}}, \bibinfo {author} {\bibfnamefont {H.}~\bibnamefont {Choi}}, \bibinfo {author} {\bibfnamefont {T.}~\bibnamefont {Hashimoto}}, \bibinfo {author} {\bibfnamefont {S.}~\bibnamefont {Iimura}}, \bibinfo {author} {\bibfnamefont {S.}~\bibnamefont {Jeong}}, \bibinfo {author} {\bibfnamefont {H.}~\bibnamefont {Miyatake}}, \bibinfo {author} {\bibfnamefont {J.}~\bibnamefont {Moon}}, \bibinfo {author} {\bibfnamefont {H.}~\bibnamefont {Nakada}}, \bibinfo {author} {\bibfnamefont {M.}~\bibnamefont {Oyaizu}}, \bibinfo {author} {\bibfnamefont {M.}~\bibnamefont {Rosenbusch}}, \bibinfo {author} {\bibfnamefont {A.}~\bibnamefont {Takamine}}, \bibinfo {author} {\bibfnamefont {M.}~\bibnamefont {Tajima}},
  \bibinfo {author} {\bibfnamefont {A.}~\bibnamefont {Taniguchi}},\ and\ \bibinfo {author} {\bibfnamefont {M.}~\bibnamefont {Wada}},\ }\href {https://doi.org/10.1007/s10751-024-01886-1} {\bibfield  {journal} {\bibinfo  {journal} {Interactions}\ }\textbf {\bibinfo {volume} {245}},\ \bibinfo {pages} {41} (\bibinfo {year} {2024})}\BibitemShut {NoStop}%
\bibitem [{\citenamefont {Hirayama}\ \emph {et~al.}(2022)\citenamefont {Hirayama}, \citenamefont {Mukai}, \citenamefont {Watanabe}, \citenamefont {Schury}, \citenamefont {Nakada}, \citenamefont {Moon}, \citenamefont {Hashimoto}, \citenamefont {Iimura}, \citenamefont {Jeong}, \citenamefont {Rosenbusch}, \citenamefont {Oyaizu}, \citenamefont {Niwase}, \citenamefont {Tajima}, \citenamefont {Taniguchi}, \citenamefont {Wada},\ and\ \citenamefont {Miyatake}}]{Hirayama2022}%
  \BibitemOpen
  \bibfield  {author} {\bibinfo {author} {\bibfnamefont {Y.}~\bibnamefont {Hirayama}}, \bibinfo {author} {\bibfnamefont {M.}~\bibnamefont {Mukai}}, \bibinfo {author} {\bibfnamefont {Y.~X.}\ \bibnamefont {Watanabe}}, \bibinfo {author} {\bibfnamefont {P.}~\bibnamefont {Schury}}, \bibinfo {author} {\bibfnamefont {H.}~\bibnamefont {Nakada}}, \bibinfo {author} {\bibfnamefont {J.~Y.}\ \bibnamefont {Moon}}, \bibinfo {author} {\bibfnamefont {T.}~\bibnamefont {Hashimoto}}, \bibinfo {author} {\bibfnamefont {S.}~\bibnamefont {Iimura}}, \bibinfo {author} {\bibfnamefont {S.~C.}\ \bibnamefont {Jeong}}, \bibinfo {author} {\bibfnamefont {M.}~\bibnamefont {Rosenbusch}}, \bibinfo {author} {\bibfnamefont {M.}~\bibnamefont {Oyaizu}}, \bibinfo {author} {\bibfnamefont {T.}~\bibnamefont {Niwase}}, \bibinfo {author} {\bibfnamefont {M.}~\bibnamefont {Tajima}}, \bibinfo {author} {\bibfnamefont {A.}~\bibnamefont {Taniguchi}}, \bibinfo {author} {\bibfnamefont {M.}~\bibnamefont {Wada}},\ and\ \bibinfo {author} {\bibfnamefont
  {H.}~\bibnamefont {Miyatake}},\ }\href {https://doi.org/10.1103/PhysRevC.106.034326} {\bibfield  {journal} {\bibinfo  {journal} {Phys. Rev. C}\ }\textbf {\bibinfo {volume} {106}},\ \bibinfo {pages} {034326} (\bibinfo {year} {2022})}\BibitemShut {NoStop}%
\bibitem [{\citenamefont {Choi}\ \emph {et~al.}(2020)\citenamefont {Choi}, \citenamefont {Hirayama}, \citenamefont {Choi}, \citenamefont {Hashimoto}, \citenamefont {Jeong}, \citenamefont {Miyatake}, \citenamefont {Moon}, \citenamefont {Mukai}, \citenamefont {Niwase}, \citenamefont {Oyaizu}, \citenamefont {Rosenbusch}, \citenamefont {Schury}, \citenamefont {Taniguchi}, \citenamefont {Watanabe},\ and\ \citenamefont {Wada}}]{Choi2020}%
  \BibitemOpen
  \bibfield  {author} {\bibinfo {author} {\bibfnamefont {H.}~\bibnamefont {Choi}}, \bibinfo {author} {\bibfnamefont {Y.}~\bibnamefont {Hirayama}}, \bibinfo {author} {\bibfnamefont {S.}~\bibnamefont {Choi}}, \bibinfo {author} {\bibfnamefont {T.}~\bibnamefont {Hashimoto}}, \bibinfo {author} {\bibfnamefont {S.~C.}\ \bibnamefont {Jeong}}, \bibinfo {author} {\bibfnamefont {H.}~\bibnamefont {Miyatake}}, \bibinfo {author} {\bibfnamefont {J.~Y.}\ \bibnamefont {Moon}}, \bibinfo {author} {\bibfnamefont {M.}~\bibnamefont {Mukai}}, \bibinfo {author} {\bibfnamefont {T.}~\bibnamefont {Niwase}}, \bibinfo {author} {\bibfnamefont {M.}~\bibnamefont {Oyaizu}}, \bibinfo {author} {\bibfnamefont {M.}~\bibnamefont {Rosenbusch}}, \bibinfo {author} {\bibfnamefont {P.}~\bibnamefont {Schury}}, \bibinfo {author} {\bibfnamefont {A.}~\bibnamefont {Taniguchi}}, \bibinfo {author} {\bibfnamefont {Y.~X.}\ \bibnamefont {Watanabe}},\ and\ \bibinfo {author} {\bibfnamefont {M.}~\bibnamefont {Wada}},\ }\href
  {https://doi.org/10.1103/PhysRevC.102.034309} {\bibfield  {journal} {\bibinfo  {journal} {Phys. Rev. C}\ }\textbf {\bibinfo {volume} {102}},\ \bibinfo {pages} {034309} (\bibinfo {year} {2020})}\BibitemShut {NoStop}%
\bibitem [{\citenamefont {Wheldon}\ \emph {et~al.}(2006)\citenamefont {Wheldon}, \citenamefont {Valiente-Dob\'on}, \citenamefont {Regan}, \citenamefont {Pearson}, \citenamefont {Wu}, \citenamefont {Smith}, \citenamefont {Macchiavelli}, \citenamefont {Cline}, \citenamefont {Chakrawarthy}, \citenamefont {Chapman}, \citenamefont {Cromaz}, \citenamefont {Fallon}, \citenamefont {Freeman}, \citenamefont {Gelletly}, \citenamefont {G\"orgen}, \citenamefont {Hayes}, \citenamefont {Hua}, \citenamefont {Langdown}, \citenamefont {Lee}, \citenamefont {Liang}, \citenamefont {Podoly\'ak}, \citenamefont {Sletten}, \citenamefont {Teng}, \citenamefont {Ward}, \citenamefont {Warner},\ and\ \citenamefont {Yamamoto}}]{Wheldon2006}%
  \BibitemOpen
  \bibfield  {author} {\bibinfo {author} {\bibfnamefont {C.}~\bibnamefont {Wheldon}}, \bibinfo {author} {\bibfnamefont {J.~J.}\ \bibnamefont {Valiente-Dob\'on}}, \bibinfo {author} {\bibfnamefont {P.~H.}\ \bibnamefont {Regan}}, \bibinfo {author} {\bibfnamefont {C.~J.}\ \bibnamefont {Pearson}}, \bibinfo {author} {\bibfnamefont {C.~Y.}\ \bibnamefont {Wu}}, \bibinfo {author} {\bibfnamefont {J.~F.}\ \bibnamefont {Smith}}, \bibinfo {author} {\bibfnamefont {A.~O.}\ \bibnamefont {Macchiavelli}}, \bibinfo {author} {\bibfnamefont {D.}~\bibnamefont {Cline}}, \bibinfo {author} {\bibfnamefont {R.~S.}\ \bibnamefont {Chakrawarthy}}, \bibinfo {author} {\bibfnamefont {R.}~\bibnamefont {Chapman}}, \bibinfo {author} {\bibfnamefont {M.}~\bibnamefont {Cromaz}}, \bibinfo {author} {\bibfnamefont {P.}~\bibnamefont {Fallon}}, \bibinfo {author} {\bibfnamefont {S.~J.}\ \bibnamefont {Freeman}}, \bibinfo {author} {\bibfnamefont {W.}~\bibnamefont {Gelletly}}, \bibinfo {author} {\bibfnamefont {A.}~\bibnamefont {G\"orgen}}, \bibinfo
  {author} {\bibfnamefont {A.~B.}\ \bibnamefont {Hayes}}, \bibinfo {author} {\bibfnamefont {H.}~\bibnamefont {Hua}}, \bibinfo {author} {\bibfnamefont {S.~D.}\ \bibnamefont {Langdown}}, \bibinfo {author} {\bibfnamefont {I.~Y.}\ \bibnamefont {Lee}}, \bibinfo {author} {\bibfnamefont {X.}~\bibnamefont {Liang}}, \bibinfo {author} {\bibfnamefont {Z.}~\bibnamefont {Podoly\'ak}}, \bibinfo {author} {\bibfnamefont {G.}~\bibnamefont {Sletten}}, \bibinfo {author} {\bibfnamefont {R.}~\bibnamefont {Teng}}, \bibinfo {author} {\bibfnamefont {D.}~\bibnamefont {Ward}}, \bibinfo {author} {\bibfnamefont {D.~D.}\ \bibnamefont {Warner}},\ and\ \bibinfo {author} {\bibfnamefont {A.~D.}\ \bibnamefont {Yamamoto}},\ }\href {https://doi.org/10.1103/PhysRevC.74.027303} {\bibfield  {journal} {\bibinfo  {journal} {Phys. Rev. C}\ }\textbf {\bibinfo {volume} {74}},\ \bibinfo {pages} {027303} (\bibinfo {year} {2006})}\BibitemShut {NoStop}%
\bibitem [{\citenamefont {Dracoulis}\ \emph {et~al.}(2013)\citenamefont {Dracoulis}, \citenamefont {Lane}, \citenamefont {Watanabe}, \citenamefont {Hughes}, \citenamefont {Palalani}, \citenamefont {Kondev}, \citenamefont {Carpenter}, \citenamefont {Janssens}, \citenamefont {Lauritsen}, \citenamefont {Lister}, \citenamefont {Seweryniak}, \citenamefont {Zhu}, \citenamefont {Chowdhury}, \citenamefont {Liang}, \citenamefont {Shi},\ and\ \citenamefont {Xu}}]{DRACOULIS2013}%
  \BibitemOpen
  \bibfield  {author} {\bibinfo {author} {\bibfnamefont {G.~D.}\ \bibnamefont {Dracoulis}}, \bibinfo {author} {\bibfnamefont {G.~J.}\ \bibnamefont {Lane}}, \bibinfo {author} {\bibfnamefont {H.}~\bibnamefont {Watanabe}}, \bibinfo {author} {\bibfnamefont {R.~O.}\ \bibnamefont {Hughes}}, \bibinfo {author} {\bibfnamefont {N.}~\bibnamefont {Palalani}}, \bibinfo {author} {\bibfnamefont {F.~G.}\ \bibnamefont {Kondev}}, \bibinfo {author} {\bibfnamefont {M.~P.}\ \bibnamefont {Carpenter}}, \bibinfo {author} {\bibfnamefont {R.~V.~F.}\ \bibnamefont {Janssens}}, \bibinfo {author} {\bibfnamefont {T.}~\bibnamefont {Lauritsen}}, \bibinfo {author} {\bibfnamefont {C.~J.}\ \bibnamefont {Lister}}, \bibinfo {author} {\bibfnamefont {D.}~\bibnamefont {Seweryniak}}, \bibinfo {author} {\bibfnamefont {S.}~\bibnamefont {Zhu}}, \bibinfo {author} {\bibfnamefont {P.}~\bibnamefont {Chowdhury}}, \bibinfo {author} {\bibfnamefont {W.~Y.}\ \bibnamefont {Liang}}, \bibinfo {author} {\bibfnamefont {Y.}~\bibnamefont {Shi}},\ and\ \bibinfo {author}
  {\bibfnamefont {F.~R.}\ \bibnamefont {Xu}},\ }\href {https://doi.org/10.1103/PhysRevC.87.014326} {\bibfield  {journal} {\bibinfo  {journal} {Phys. Rev. C}\ }\textbf {\bibinfo {volume} {87}},\ \bibinfo {pages} {014326} (\bibinfo {year} {2013})}\BibitemShut {NoStop}%
\bibitem [{Dat()}]{DataE806}%
  \BibitemOpen
  \href {https://doi.org/10.26143/GANIL-2021-E806_20} {}\bibinfo {note} {E806-20 Collaboration, E806-20 GANIL dataset (2021)}\BibitemShut {NoStop}%
\bibitem [{\citenamefont {Rejmund}\ \emph {et~al.}(2011)\citenamefont {Rejmund}, \citenamefont {Lecornu}, \citenamefont {Navin}, \citenamefont {Schmitt}, \citenamefont {Damoy}, \citenamefont {Delaune}, \citenamefont {Enguerrand}, \citenamefont {Fremont}, \citenamefont {Gangnant}, \citenamefont {Gaudefroy}, \citenamefont {Jacquot}, \citenamefont {Pancin}, \citenamefont {Pullanhiotan},\ and\ \citenamefont {Spitaels}}]{REJMUND_NIMA2011}%
  \BibitemOpen
  \bibfield  {author} {\bibinfo {author} {\bibfnamefont {M.}~\bibnamefont {Rejmund}}, \bibinfo {author} {\bibfnamefont {B.}~\bibnamefont {Lecornu}}, \bibinfo {author} {\bibfnamefont {A.}~\bibnamefont {Navin}}, \bibinfo {author} {\bibfnamefont {C.}~\bibnamefont {Schmitt}}, \bibinfo {author} {\bibfnamefont {S.}~\bibnamefont {Damoy}}, \bibinfo {author} {\bibfnamefont {O.}~\bibnamefont {Delaune}}, \bibinfo {author} {\bibfnamefont {J.}~\bibnamefont {Enguerrand}}, \bibinfo {author} {\bibfnamefont {G.}~\bibnamefont {Fremont}}, \bibinfo {author} {\bibfnamefont {P.}~\bibnamefont {Gangnant}}, \bibinfo {author} {\bibfnamefont {L.}~\bibnamefont {Gaudefroy}}, \bibinfo {author} {\bibfnamefont {B.}~\bibnamefont {Jacquot}}, \bibinfo {author} {\bibfnamefont {J.}~\bibnamefont {Pancin}}, \bibinfo {author} {\bibfnamefont {S.}~\bibnamefont {Pullanhiotan}},\ and\ \bibinfo {author} {\bibfnamefont {C.}~\bibnamefont {Spitaels}},\ }\href {https://doi.org/https://doi.org/10.1016/j.nima.2011.05.007} {\bibfield  {journal} {\bibinfo
  {journal} {Nucl. Instrum. Methods Phys. Res. A}\ }\textbf {\bibinfo {volume} {646}},\ \bibinfo {pages} {184} (\bibinfo {year} {2011})}\BibitemShut {NoStop}%
\bibitem [{\citenamefont {Vandebrouck}\ \emph {et~al.}(2016)\citenamefont {Vandebrouck}, \citenamefont {Lemasson}, \citenamefont {Rejmund}, \citenamefont {Fremont}, \citenamefont {Pancin}, \citenamefont {Navin}, \citenamefont {Michelagnoli}, \citenamefont {Goupil}, \citenamefont {Spitaels},\ and\ \citenamefont {Jacquot}}]{VANDEBROUCK_NIMA2016}%
  \BibitemOpen
  \bibfield  {author} {\bibinfo {author} {\bibfnamefont {M.}~\bibnamefont {Vandebrouck}}, \bibinfo {author} {\bibfnamefont {A.}~\bibnamefont {Lemasson}}, \bibinfo {author} {\bibfnamefont {M.}~\bibnamefont {Rejmund}}, \bibinfo {author} {\bibfnamefont {G.}~\bibnamefont {Fremont}}, \bibinfo {author} {\bibfnamefont {J.}~\bibnamefont {Pancin}}, \bibinfo {author} {\bibfnamefont {A.}~\bibnamefont {Navin}}, \bibinfo {author} {\bibfnamefont {C.}~\bibnamefont {Michelagnoli}}, \bibinfo {author} {\bibfnamefont {J.}~\bibnamefont {Goupil}}, \bibinfo {author} {\bibfnamefont {C.}~\bibnamefont {Spitaels}},\ and\ \bibinfo {author} {\bibfnamefont {B.}~\bibnamefont {Jacquot}},\ }\href {https://doi.org/https://doi.org/10.1016/j.nima.2015.12.040} {\bibfield  {journal} {\bibinfo  {journal} {Nucl. Instrum. Methods Phys. Res. A}\ }\textbf {\bibinfo {volume} {812}},\ \bibinfo {pages} {112} (\bibinfo {year} {2016})}\BibitemShut {NoStop}%
\bibitem [{\citenamefont {Lemasson}\ and\ \citenamefont {Rejmund}(2023)}]{LEMASSON_NIMA2023}%
  \BibitemOpen
  \bibfield  {author} {\bibinfo {author} {\bibfnamefont {A.}~\bibnamefont {Lemasson}}\ and\ \bibinfo {author} {\bibfnamefont {M.}~\bibnamefont {Rejmund}},\ }\href {https://doi.org/https://doi.org/10.1016/j.nima.2023.168407} {\bibfield  {journal} {\bibinfo  {journal} {Nucl. Instrum. Methods Phys. Res. A}\ }\textbf {\bibinfo {volume} {1054}},\ \bibinfo {pages} {168407} (\bibinfo {year} {2023})}\BibitemShut {NoStop}%
\bibitem [{\citenamefont {Cho}\ \emph {et~al.}(2023)\citenamefont {Cho}, \citenamefont {Kim}, \citenamefont {Choi}, \citenamefont {Park}, \citenamefont {Bae}, \citenamefont {Hahn}, \citenamefont {Son}, \citenamefont {Navin}, \citenamefont {Lemasson}, \citenamefont {Rejmund}, \citenamefont {Ramos}, \citenamefont {Ackermann}, \citenamefont {Utepov}, \citenamefont {Fourgeres}, \citenamefont {Thomas}, \citenamefont {Goupil}, \citenamefont {Fremont}, \citenamefont {{de France}}, \citenamefont {Watanabe}, \citenamefont {Hirayama}, \citenamefont {Jeong}, \citenamefont {Niwase}, \citenamefont {Miyatake}, \citenamefont {Schury}, \citenamefont {Rosenbusch}, \citenamefont {Chae}, \citenamefont {Kim}, \citenamefont {Kim}, \citenamefont {Gu}, \citenamefont {Kim}, \citenamefont {John}, \citenamefont {Andreyev}, \citenamefont {Korten}, \citenamefont {Recchia}, \citenamefont {de~Angelis}, \citenamefont {Vidal}, \citenamefont {Rezynkina}, \citenamefont {Ha}, \citenamefont {Didierjean}, \citenamefont {Marini}, \citenamefont
  {Treasa}, \citenamefont {Tsekhanovich}, \citenamefont {Dudouet}, \citenamefont {Bhattacharyya}, \citenamefont {Mukherjee}, \citenamefont {Banik}, \citenamefont {Bhattacharya},\ and\ \citenamefont {Mukai}}]{CHO2023}%
  \BibitemOpen
  \bibfield  {author} {\bibinfo {author} {\bibfnamefont {Y.}~\bibnamefont {Cho}}, \bibinfo {author} {\bibfnamefont {Y.}~\bibnamefont {Kim}}, \bibinfo {author} {\bibfnamefont {S.}~\bibnamefont {Choi}}, \bibinfo {author} {\bibfnamefont {J.}~\bibnamefont {Park}}, \bibinfo {author} {\bibfnamefont {S.}~\bibnamefont {Bae}}, \bibinfo {author} {\bibfnamefont {K.}~\bibnamefont {Hahn}}, \bibinfo {author} {\bibfnamefont {Y.}~\bibnamefont {Son}}, \bibinfo {author} {\bibfnamefont {A.}~\bibnamefont {Navin}}, \bibinfo {author} {\bibfnamefont {A.}~\bibnamefont {Lemasson}}, \bibinfo {author} {\bibfnamefont {M.}~\bibnamefont {Rejmund}}, \bibinfo {author} {\bibfnamefont {D.}~\bibnamefont {Ramos}}, \bibinfo {author} {\bibfnamefont {D.}~\bibnamefont {Ackermann}}, \bibinfo {author} {\bibfnamefont {A.}~\bibnamefont {Utepov}}, \bibinfo {author} {\bibfnamefont {C.}~\bibnamefont {Fourgeres}}, \bibinfo {author} {\bibfnamefont {J.}~\bibnamefont {Thomas}}, \bibinfo {author} {\bibfnamefont {J.}~\bibnamefont {Goupil}}, \bibinfo {author}
  {\bibfnamefont {G.}~\bibnamefont {Fremont}}, \bibinfo {author} {\bibfnamefont {G.}~\bibnamefont {{de France}}}, \bibinfo {author} {\bibfnamefont {Y.}~\bibnamefont {Watanabe}}, \bibinfo {author} {\bibfnamefont {Y.}~\bibnamefont {Hirayama}}, \bibinfo {author} {\bibfnamefont {S.}~\bibnamefont {Jeong}}, \bibinfo {author} {\bibfnamefont {T.}~\bibnamefont {Niwase}}, \bibinfo {author} {\bibfnamefont {H.}~\bibnamefont {Miyatake}}, \bibinfo {author} {\bibfnamefont {P.}~\bibnamefont {Schury}}, \bibinfo {author} {\bibfnamefont {M.}~\bibnamefont {Rosenbusch}}, \bibinfo {author} {\bibfnamefont {K.}~\bibnamefont {Chae}}, \bibinfo {author} {\bibfnamefont {C.}~\bibnamefont {Kim}}, \bibinfo {author} {\bibfnamefont {S.}~\bibnamefont {Kim}}, \bibinfo {author} {\bibfnamefont {G.}~\bibnamefont {Gu}}, \bibinfo {author} {\bibfnamefont {M.}~\bibnamefont {Kim}}, \bibinfo {author} {\bibfnamefont {P.}~\bibnamefont {John}}, \bibinfo {author} {\bibfnamefont {A.}~\bibnamefont {Andreyev}}, \bibinfo {author} {\bibfnamefont
  {W.}~\bibnamefont {Korten}}, \bibinfo {author} {\bibfnamefont {F.}~\bibnamefont {Recchia}}, \bibinfo {author} {\bibfnamefont {G.}~\bibnamefont {de~Angelis}}, \bibinfo {author} {\bibfnamefont {R.~P.}\ \bibnamefont {Vidal}}, \bibinfo {author} {\bibfnamefont {K.}~\bibnamefont {Rezynkina}}, \bibinfo {author} {\bibfnamefont {J.}~\bibnamefont {Ha}}, \bibinfo {author} {\bibfnamefont {F.}~\bibnamefont {Didierjean}}, \bibinfo {author} {\bibfnamefont {P.}~\bibnamefont {Marini}}, \bibinfo {author} {\bibfnamefont {D.}~\bibnamefont {Treasa}}, \bibinfo {author} {\bibfnamefont {I.}~\bibnamefont {Tsekhanovich}}, \bibinfo {author} {\bibfnamefont {J.}~\bibnamefont {Dudouet}}, \bibinfo {author} {\bibfnamefont {S.}~\bibnamefont {Bhattacharyya}}, \bibinfo {author} {\bibfnamefont {G.}~\bibnamefont {Mukherjee}}, \bibinfo {author} {\bibfnamefont {R.}~\bibnamefont {Banik}}, \bibinfo {author} {\bibfnamefont {S.}~\bibnamefont {Bhattacharya}},\ and\ \bibinfo {author} {\bibfnamefont {M.}~\bibnamefont {Mukai}},\ }\href
  {https://doi.org/https://doi.org/10.1016/j.nimb.2023.05.053} {\bibfield  {journal} {\bibinfo  {journal} {Nucl. Instrum. Methods Phys. Res. B}\ }\textbf {\bibinfo {volume} {541}},\ \bibinfo {pages} {240} (\bibinfo {year} {2023})}\BibitemShut {NoStop}%
\bibitem [{\citenamefont {Son}\ \emph {et~al.}(2023)\citenamefont {Son}, \citenamefont {Kim}, \citenamefont {Cho}, \citenamefont {Choi}, \citenamefont {Park}, \citenamefont {Bae}, \citenamefont {Hahn}, \citenamefont {Navin}, \citenamefont {Lemasson}, \citenamefont {Rejmund}, \citenamefont {Ramos}, \citenamefont {Clément}, \citenamefont {Ackermann}, \citenamefont {Utepov}, \citenamefont {Fougeres}, \citenamefont {Thomas}, \citenamefont {Goupil}, \citenamefont {Fremont},\ and\ \citenamefont {{de France}}}]{NIMB2023_SON}%
  \BibitemOpen
  \bibfield  {author} {\bibinfo {author} {\bibfnamefont {Y.}~\bibnamefont {Son}}, \bibinfo {author} {\bibfnamefont {Y.}~\bibnamefont {Kim}}, \bibinfo {author} {\bibfnamefont {Y.}~\bibnamefont {Cho}}, \bibinfo {author} {\bibfnamefont {S.}~\bibnamefont {Choi}}, \bibinfo {author} {\bibfnamefont {J.}~\bibnamefont {Park}}, \bibinfo {author} {\bibfnamefont {S.}~\bibnamefont {Bae}}, \bibinfo {author} {\bibfnamefont {K.}~\bibnamefont {Hahn}}, \bibinfo {author} {\bibfnamefont {A.}~\bibnamefont {Navin}}, \bibinfo {author} {\bibfnamefont {A.}~\bibnamefont {Lemasson}}, \bibinfo {author} {\bibfnamefont {M.}~\bibnamefont {Rejmund}}, \bibinfo {author} {\bibfnamefont {D.}~\bibnamefont {Ramos}}, \bibinfo {author} {\bibfnamefont {E.}~\bibnamefont {Clément}}, \bibinfo {author} {\bibfnamefont {D.}~\bibnamefont {Ackermann}}, \bibinfo {author} {\bibfnamefont {A.}~\bibnamefont {Utepov}}, \bibinfo {author} {\bibfnamefont {C.}~\bibnamefont {Fougeres}}, \bibinfo {author} {\bibfnamefont {J.}~\bibnamefont {Thomas}}, \bibinfo {author}
  {\bibfnamefont {J.}~\bibnamefont {Goupil}}, \bibinfo {author} {\bibfnamefont {G.}~\bibnamefont {Fremont}},\ and\ \bibinfo {author} {\bibfnamefont {G.}~\bibnamefont {{de France}}},\ }\href {https://doi.org/https://doi.org/10.1016/j.nimb.2023.04.024} {\bibfield  {journal} {\bibinfo  {journal} {Nucl. Instrum. Methods Phys. Res. B}\ }\textbf {\bibinfo {volume} {540}},\ \bibinfo {pages} {234} (\bibinfo {year} {2023})}\BibitemShut {NoStop}%
\bibitem [{\citenamefont {Simpson}\ \emph {et~al.}(2000)\citenamefont {Simpson}, \citenamefont {Azaiez}, \citenamefont {de~France}, \citenamefont {Fouan}, \citenamefont {Gerl}, \citenamefont {Julin}, \citenamefont {Korten}, \citenamefont {Nolan}, \citenamefont {Nyako}, \citenamefont {Sletten},\ and\ \citenamefont {Walker}}]{SIMPSON_Acta2000}%
  \BibitemOpen
  \bibfield  {author} {\bibinfo {author} {\bibfnamefont {J.}~\bibnamefont {Simpson}}, \bibinfo {author} {\bibfnamefont {F.}~\bibnamefont {Azaiez}}, \bibinfo {author} {\bibfnamefont {G.}~\bibnamefont {de~France}}, \bibinfo {author} {\bibfnamefont {G.}~\bibnamefont {Fouan}}, \bibinfo {author} {\bibfnamefont {J.}~\bibnamefont {Gerl}}, \bibinfo {author} {\bibfnamefont {R.}~\bibnamefont {Julin}}, \bibinfo {author} {\bibfnamefont {W.}~\bibnamefont {Korten}}, \bibinfo {author} {\bibfnamefont {P.}~\bibnamefont {Nolan}}, \bibinfo {author} {\bibfnamefont {B.}~\bibnamefont {Nyako}}, \bibinfo {author} {\bibfnamefont {G.}~\bibnamefont {Sletten}},\ and\ \bibinfo {author} {\bibfnamefont {P.}~\bibnamefont {Walker}},\ }\href {https://hal.in2p3.fr/in2p3-00438782} {\bibfield  {journal} {\bibinfo  {journal} {Acta Physica Hungarica New Series-Heavy ion}\ }\textbf {\bibinfo {volume} {11}},\ \bibinfo {pages} {159} (\bibinfo {year} {2000})}\BibitemShut {NoStop}%
\bibitem [{\citenamefont {Brown}\ \emph {et~al.}(1951)\citenamefont {Brown}, \citenamefont {Snyder}, \citenamefont {Fowler},\ and\ \citenamefont {Lauritsen}}]{Brown1951}%
  \BibitemOpen
  \bibfield  {author} {\bibinfo {author} {\bibfnamefont {A.~B.}\ \bibnamefont {Brown}}, \bibinfo {author} {\bibfnamefont {C.~W.}\ \bibnamefont {Snyder}}, \bibinfo {author} {\bibfnamefont {W.~A.}\ \bibnamefont {Fowler}},\ and\ \bibinfo {author} {\bibfnamefont {C.~C.}\ \bibnamefont {Lauritsen}},\ }\href {https://doi.org/10.1103/PhysRev.82.159} {\bibfield  {journal} {\bibinfo  {journal} {Phys. Rev.}\ }\textbf {\bibinfo {volume} {82}},\ \bibinfo {pages} {159} (\bibinfo {year} {1951})}\BibitemShut {NoStop}%
\bibitem [{\citenamefont {Szilner}\ \emph {et~al.}(2007)\citenamefont {Szilner}, \citenamefont {Ur}, \citenamefont {Corradi}, \citenamefont {M\ifmmode~\u{a}\else \u{a}\fi{}rginean}, \citenamefont {Pollarolo}, \citenamefont {Stefanini}, \citenamefont {Beghini}, \citenamefont {Behera}, \citenamefont {Fioretto}, \citenamefont {Gadea}, \citenamefont {Guiot}, \citenamefont {Latina}, \citenamefont {Mason}, \citenamefont {Montagnoli}, \citenamefont {Scarlassara}, \citenamefont {Trotta}, \citenamefont {Angelis}, \citenamefont {Vedova}, \citenamefont {Farnea}, \citenamefont {Haas}, \citenamefont {Lenzi}, \citenamefont {Lunardi}, \citenamefont {M\ifmmode~\u{a}\else \u{a}\fi{}rginean}, \citenamefont {Menegazzo}, \citenamefont {Napoli}, \citenamefont {Nespolo}, \citenamefont {Pokrovsky}, \citenamefont {Recchia}, \citenamefont {Romoli}, \citenamefont {Salsac}, \citenamefont {Soi\ifmmode~\acute{c}\else \'{c}\fi{}},\ and\ \citenamefont {Valiente-Dob\'on}}]{Szilner2007}%
  \BibitemOpen
  \bibfield  {author} {\bibinfo {author} {\bibfnamefont {S.}~\bibnamefont {Szilner}}, \bibinfo {author} {\bibfnamefont {C.~A.}\ \bibnamefont {Ur}}, \bibinfo {author} {\bibfnamefont {L.}~\bibnamefont {Corradi}}, \bibinfo {author} {\bibfnamefont {N.}~\bibnamefont {M\ifmmode~\u{a}\else \u{a}\fi{}rginean}}, \bibinfo {author} {\bibfnamefont {G.}~\bibnamefont {Pollarolo}}, \bibinfo {author} {\bibfnamefont {A.~M.}\ \bibnamefont {Stefanini}}, \bibinfo {author} {\bibfnamefont {S.}~\bibnamefont {Beghini}}, \bibinfo {author} {\bibfnamefont {B.~R.}\ \bibnamefont {Behera}}, \bibinfo {author} {\bibfnamefont {E.}~\bibnamefont {Fioretto}}, \bibinfo {author} {\bibfnamefont {A.}~\bibnamefont {Gadea}}, \bibinfo {author} {\bibfnamefont {B.}~\bibnamefont {Guiot}}, \bibinfo {author} {\bibfnamefont {A.}~\bibnamefont {Latina}}, \bibinfo {author} {\bibfnamefont {P.}~\bibnamefont {Mason}}, \bibinfo {author} {\bibfnamefont {G.}~\bibnamefont {Montagnoli}}, \bibinfo {author} {\bibfnamefont {F.}~\bibnamefont {Scarlassara}}, \bibinfo
  {author} {\bibfnamefont {M.}~\bibnamefont {Trotta}}, \bibinfo {author} {\bibfnamefont {G.~d.}\ \bibnamefont {Angelis}}, \bibinfo {author} {\bibfnamefont {F.~D.}\ \bibnamefont {Vedova}}, \bibinfo {author} {\bibfnamefont {E.}~\bibnamefont {Farnea}}, \bibinfo {author} {\bibfnamefont {F.}~\bibnamefont {Haas}}, \bibinfo {author} {\bibfnamefont {S.}~\bibnamefont {Lenzi}}, \bibinfo {author} {\bibfnamefont {S.}~\bibnamefont {Lunardi}}, \bibinfo {author} {\bibfnamefont {R.}~\bibnamefont {M\ifmmode~\u{a}\else \u{a}\fi{}rginean}}, \bibinfo {author} {\bibfnamefont {R.}~\bibnamefont {Menegazzo}}, \bibinfo {author} {\bibfnamefont {D.~R.}\ \bibnamefont {Napoli}}, \bibinfo {author} {\bibfnamefont {M.}~\bibnamefont {Nespolo}}, \bibinfo {author} {\bibfnamefont {I.~V.}\ \bibnamefont {Pokrovsky}}, \bibinfo {author} {\bibfnamefont {F.}~\bibnamefont {Recchia}}, \bibinfo {author} {\bibfnamefont {M.}~\bibnamefont {Romoli}}, \bibinfo {author} {\bibfnamefont {M.-D.}\ \bibnamefont {Salsac}}, \bibinfo {author} {\bibfnamefont
  {N.}~\bibnamefont {Soi\ifmmode~\acute{c}\else \'{c}\fi{}}},\ and\ \bibinfo {author} {\bibfnamefont {J.~J.}\ \bibnamefont {Valiente-Dob\'on}},\ }\href {https://doi.org/10.1103/PhysRevC.76.024604} {\bibfield  {journal} {\bibinfo  {journal} {Phys. Rev. C}\ }\textbf {\bibinfo {volume} {76}},\ \bibinfo {pages} {024604} (\bibinfo {year} {2007})}\BibitemShut {NoStop}%
\bibitem [{\citenamefont {Akkoyun}\ \emph {et~al.}(2012)\citenamefont {Akkoyun}, \citenamefont {Algora}, \citenamefont {Alikhani}, \citenamefont {Ameil}, \citenamefont {{de Angelis}}, \citenamefont {Arnold}, \citenamefont {Astier}, \citenamefont {Ataç}, \citenamefont {Aubert}, \citenamefont {Aufranc}, \citenamefont {Austin}, \citenamefont {Aydin}, \citenamefont {Azaiez}, \citenamefont {Badoer}, \citenamefont {Balabanski}, \citenamefont {Barrientos}, \citenamefont {Baulieu}, \citenamefont {Baumann}, \citenamefont {Bazzacco}, \citenamefont {Beck}, \citenamefont {Beck}, \citenamefont {Bednarczyk}, \citenamefont {Bellato}, \citenamefont {Bentley}, \citenamefont {Benzoni}, \citenamefont {Berthier}, \citenamefont {Berti}, \citenamefont {Beunard}, \citenamefont {{Lo Bianco}}, \citenamefont {Birkenbach}, \citenamefont {Bizzeti}, \citenamefont {Bizzeti-Sona}, \citenamefont {{Le Blanc}}, \citenamefont {Blasco}, \citenamefont {Blasi}, \citenamefont {Bloor}, \citenamefont {Boiano}, \citenamefont {Borsato}, \citenamefont
  {Bortolato}, \citenamefont {Boston}, \citenamefont {Boston}, \citenamefont {Bourgault}, \citenamefont {Boutachkov}, \citenamefont {Bouty}, \citenamefont {Bracco}, \citenamefont {Brambilla}, \citenamefont {Brawn}, \citenamefont {Brondi}, \citenamefont {Broussard}, \citenamefont {Bruyneel}, \citenamefont {Bucurescu}, \citenamefont {Burrows}, \citenamefont {Bürger}, \citenamefont {Cabaret}, \citenamefont {Cahan}, \citenamefont {Calore}, \citenamefont {Camera}, \citenamefont {Capsoni}, \citenamefont {Carrió}, \citenamefont {Casati}, \citenamefont {Castoldi}, \citenamefont {Cederwall}, \citenamefont {Cercus}, \citenamefont {Chambert}, \citenamefont {{El Chambit}}, \citenamefont {Chapman}, \citenamefont {Charles}, \citenamefont {Chavas}, \citenamefont {Clément}, \citenamefont {Cocconi}, \citenamefont {Coelli}, \citenamefont {Coleman-Smith}, \citenamefont {Colombo}, \citenamefont {Colosimo}, \citenamefont {Commeaux}, \citenamefont {Conventi}, \citenamefont {Cooper}, \citenamefont {Corsi}, \citenamefont
  {Cortesi}, \citenamefont {Costa}, \citenamefont {Crespi}, \citenamefont {Cresswell}, \citenamefont {Cullen}, \citenamefont {Curien}, \citenamefont {Czermak}, \citenamefont {Delbourg}, \citenamefont {Depalo}, \citenamefont {Descombes}, \citenamefont {Désesquelles}, \citenamefont {Detistov}, \citenamefont {Diarra}, \citenamefont {Didierjean}, \citenamefont {Dimmock}, \citenamefont {Doan}, \citenamefont {Domingo-Pardo}, \citenamefont {Doncel}, \citenamefont {Dorangeville}, \citenamefont {Dosme}, \citenamefont {Drouen}, \citenamefont {Duchêne}, \citenamefont {Dulny}, \citenamefont {Eberth}, \citenamefont {Edelbruck}, \citenamefont {Egea}, \citenamefont {Engert}, \citenamefont {Erduran}, \citenamefont {Ertürk}, \citenamefont {Fanin}, \citenamefont {Fantinel}, \citenamefont {Farnea}, \citenamefont {Faul}, \citenamefont {Filliger}, \citenamefont {Filmer}, \citenamefont {Finck}, \citenamefont {{de France}}, \citenamefont {Gadea}, \citenamefont {Gast}, \citenamefont {Geraci}, \citenamefont {Gerl}, \citenamefont
  {Gernhäuser}, \citenamefont {Giannatiempo}, \citenamefont {Giaz}, \citenamefont {Gibelin}, \citenamefont {Givechev}, \citenamefont {Goel}, \citenamefont {González}, \citenamefont {Gottardo}, \citenamefont {Grave}, \citenamefont {Gre¸bosz}, \citenamefont {Griffiths}, \citenamefont {Grint}, \citenamefont {Gros}, \citenamefont {Guevara}, \citenamefont {Gulmini}, \citenamefont {Görgen}, \citenamefont {Ha}, \citenamefont {Habermann}, \citenamefont {Harkness}, \citenamefont {Harroch}, \citenamefont {Hauschild}, \citenamefont {He}, \citenamefont {Hernández-Prieto}, \citenamefont {Hervieu}, \citenamefont {Hess}, \citenamefont {Hüyük}, \citenamefont {Ince}, \citenamefont {Isocrate}, \citenamefont {Jaworski}, \citenamefont {Johnson}, \citenamefont {Jolie}, \citenamefont {Jones}, \citenamefont {Jonson}, \citenamefont {Joshi}, \citenamefont {Judson}, \citenamefont {Jungclaus}, \citenamefont {Kaci}, \citenamefont {Karkour}, \citenamefont {Karolak}, \citenamefont {Kaşkaş}, \citenamefont {Kebbiri}, \citenamefont
  {Kempley}, \citenamefont {Khaplanov}, \citenamefont {Klupp}, \citenamefont {Kogimtzis}, \citenamefont {Kojouharov}, \citenamefont {Korichi}, \citenamefont {Korten}, \citenamefont {Kröll}, \citenamefont {Krücken}, \citenamefont {Kurz}, \citenamefont {Ky}, \citenamefont {Labiche}, \citenamefont {Lafay}, \citenamefont {Lavergne}, \citenamefont {Lazarus}, \citenamefont {Leboutelier}, \citenamefont {Lefebvre}, \citenamefont {Legay}, \citenamefont {Legeard}, \citenamefont {Lelli}, \citenamefont {Lenzi}, \citenamefont {Leoni}, \citenamefont {Lermitage}, \citenamefont {Lersch}, \citenamefont {Leske}, \citenamefont {Letts}, \citenamefont {Lhenoret}, \citenamefont {Lieder}, \citenamefont {Linget}, \citenamefont {Ljungvall}, \citenamefont {Lopez-Martens}, \citenamefont {Lotodé}, \citenamefont {Lunardi}, \citenamefont {Maj}, \citenamefont {{van der Marel}}, \citenamefont {Mariette}, \citenamefont {Marginean}, \citenamefont {Marginean}, \citenamefont {Maron}, \citenamefont {Mather}, \citenamefont {Me¸czyński},
  \citenamefont {Mendéz}, \citenamefont {Medina}, \citenamefont {Melon}, \citenamefont {Menegazzo}, \citenamefont {Mengoni}, \citenamefont {Merchan}, \citenamefont {Mihailescu}, \citenamefont {Michelagnoli}, \citenamefont {Mierzejewski}, \citenamefont {Milechina}, \citenamefont {Million}, \citenamefont {Mitev}, \citenamefont {Molini}, \citenamefont {Montanari}, \citenamefont {Moon}, \citenamefont {Morbiducci}, \citenamefont {Moro}, \citenamefont {Morrall}, \citenamefont {Möller}, \citenamefont {Nannini}, \citenamefont {Napoli}, \citenamefont {Nelson}, \citenamefont {Nespolo}, \citenamefont {Ngo}, \citenamefont {Nicoletto}, \citenamefont {Nicolini}, \citenamefont {{Le Noa}}, \citenamefont {Nolan}, \citenamefont {Norman}, \citenamefont {Nyberg}, \citenamefont {Obertelli}, \citenamefont {Olariu}, \citenamefont {Orlandi}, \citenamefont {Oxley}, \citenamefont {Özben}, \citenamefont {Ozille}, \citenamefont {Oziol}, \citenamefont {Pachoud}, \citenamefont {Palacz}, \citenamefont {Palin}, \citenamefont {Pancin},
  \citenamefont {Parisel}, \citenamefont {Pariset}, \citenamefont {Pascovici}, \citenamefont {Peghin}, \citenamefont {Pellegri}, \citenamefont {Perego}, \citenamefont {Perrier}, \citenamefont {Petcu}, \citenamefont {Petkov}, \citenamefont {Petrache}, \citenamefont {Pierre}, \citenamefont {Pietralla}, \citenamefont {Pietri}, \citenamefont {Pignanelli}, \citenamefont {Piqueras}, \citenamefont {Podolyak}, \citenamefont {{Le Pouhalec}}, \citenamefont {Pouthas}, \citenamefont {Pugnére}, \citenamefont {Pucknell}, \citenamefont {Pullia}, \citenamefont {Quintana}, \citenamefont {Raine}, \citenamefont {Rainovski}, \citenamefont {Ramina}, \citenamefont {Rampazzo}, \citenamefont {{La Rana}}, \citenamefont {Rebeschini}, \citenamefont {Recchia}, \citenamefont {Redon}, \citenamefont {Reese}, \citenamefont {Reiter}, \citenamefont {Regan}, \citenamefont {Riboldi}, \citenamefont {Richer}, \citenamefont {Rigato}, \citenamefont {Rigby}, \citenamefont {Ripamonti}, \citenamefont {Robinson}, \citenamefont {Robin}, \citenamefont
  {Roccaz}, \citenamefont {Ropert}, \citenamefont {Rossé}, \citenamefont {{Rossi Alvarez}}, \citenamefont {Rosso}, \citenamefont {Rubio}, \citenamefont {Rudolph}, \citenamefont {Saillant}, \citenamefont {Şahin}, \citenamefont {Salomon}, \citenamefont {Salsac}, \citenamefont {Salt}, \citenamefont {Salvato}, \citenamefont {Sampson}, \citenamefont {Sanchis}, \citenamefont {Santos}, \citenamefont {Schaffner}, \citenamefont {Schlarb}, \citenamefont {Scraggs}, \citenamefont {Seddon}, \citenamefont {Şenyiğit}, \citenamefont {Sigward}, \citenamefont {Simpson}, \citenamefont {Simpson}, \citenamefont {Slee}, \citenamefont {Smith}, \citenamefont {Sona}, \citenamefont {Sowicki}, \citenamefont {Spolaore}, \citenamefont {Stahl}, \citenamefont {Stanios}, \citenamefont {Stefanova}, \citenamefont {Stézowski}, \citenamefont {Strachan}, \citenamefont {Suliman}, \citenamefont {Söderström}, \citenamefont {Tain}, \citenamefont {Tanguy}, \citenamefont {Tashenov}, \citenamefont {Theisen}, \citenamefont {Thornhill},
  \citenamefont {Tomasi}, \citenamefont {Toniolo}, \citenamefont {Touzery}, \citenamefont {Travers}, \citenamefont {Triossi}, \citenamefont {Tripon}, \citenamefont {Tun-Lanoë}, \citenamefont {Turcato}, \citenamefont {Unsworth}, \citenamefont {Ur}, \citenamefont {Valiente-Dobon}, \citenamefont {Vandone}, \citenamefont {Vardaci}, \citenamefont {Venturelli}, \citenamefont {Veronese}, \citenamefont {Veyssiere}, \citenamefont {Viscione}, \citenamefont {Wadsworth}, \citenamefont {Walker}, \citenamefont {Warr}, \citenamefont {Weber}, \citenamefont {Weisshaar}, \citenamefont {Wells}, \citenamefont {Wieland}, \citenamefont {Wiens}, \citenamefont {Wittwer}, \citenamefont {Wollersheim}, \citenamefont {Zocca}, \citenamefont {Zamfir}, \citenamefont {Zie¸bliński},\ and\ \citenamefont {Zucchiatti}}]{AKKOYUN_NIMA2012}%
  \BibitemOpen
  \bibfield  {author} {\bibinfo {author} {\bibfnamefont {S.}~\bibnamefont {Akkoyun}}, \bibinfo {author} {\bibfnamefont {A.}~\bibnamefont {Algora}}, \bibinfo {author} {\bibfnamefont {B.}~\bibnamefont {Alikhani}}, \bibinfo {author} {\bibfnamefont {F.}~\bibnamefont {Ameil}}, \bibinfo {author} {\bibfnamefont {G.}~\bibnamefont {{de Angelis}}}, \bibinfo {author} {\bibfnamefont {L.}~\bibnamefont {Arnold}}, \bibinfo {author} {\bibfnamefont {A.}~\bibnamefont {Astier}}, \bibinfo {author} {\bibfnamefont {A.}~\bibnamefont {Ataç}}, \bibinfo {author} {\bibfnamefont {Y.}~\bibnamefont {Aubert}}, \bibinfo {author} {\bibfnamefont {C.}~\bibnamefont {Aufranc}}, \bibinfo {author} {\bibfnamefont {A.}~\bibnamefont {Austin}}, \bibinfo {author} {\bibfnamefont {S.}~\bibnamefont {Aydin}}, \bibinfo {author} {\bibfnamefont {F.}~\bibnamefont {Azaiez}}, \bibinfo {author} {\bibfnamefont {S.}~\bibnamefont {Badoer}}, \bibinfo {author} {\bibfnamefont {D.}~\bibnamefont {Balabanski}}, \bibinfo {author} {\bibfnamefont {D.}~\bibnamefont
  {Barrientos}}, \bibinfo {author} {\bibfnamefont {G.}~\bibnamefont {Baulieu}}, \bibinfo {author} {\bibfnamefont {R.}~\bibnamefont {Baumann}}, \bibinfo {author} {\bibfnamefont {D.}~\bibnamefont {Bazzacco}}, \bibinfo {author} {\bibfnamefont {F.}~\bibnamefont {Beck}}, \bibinfo {author} {\bibfnamefont {T.}~\bibnamefont {Beck}}, \bibinfo {author} {\bibfnamefont {P.}~\bibnamefont {Bednarczyk}}, \bibinfo {author} {\bibfnamefont {M.}~\bibnamefont {Bellato}}, \bibinfo {author} {\bibfnamefont {M.}~\bibnamefont {Bentley}}, \bibinfo {author} {\bibfnamefont {G.}~\bibnamefont {Benzoni}}, \bibinfo {author} {\bibfnamefont {R.}~\bibnamefont {Berthier}}, \bibinfo {author} {\bibfnamefont {L.}~\bibnamefont {Berti}}, \bibinfo {author} {\bibfnamefont {R.}~\bibnamefont {Beunard}}, \bibinfo {author} {\bibfnamefont {G.}~\bibnamefont {{Lo Bianco}}}, \bibinfo {author} {\bibfnamefont {B.}~\bibnamefont {Birkenbach}}, \bibinfo {author} {\bibfnamefont {P.}~\bibnamefont {Bizzeti}}, \bibinfo {author} {\bibfnamefont {A.}~\bibnamefont
  {Bizzeti-Sona}}, \bibinfo {author} {\bibfnamefont {F.}~\bibnamefont {{Le Blanc}}}, \bibinfo {author} {\bibfnamefont {J.}~\bibnamefont {Blasco}}, \bibinfo {author} {\bibfnamefont {N.}~\bibnamefont {Blasi}}, \bibinfo {author} {\bibfnamefont {D.}~\bibnamefont {Bloor}}, \bibinfo {author} {\bibfnamefont {C.}~\bibnamefont {Boiano}}, \bibinfo {author} {\bibfnamefont {M.}~\bibnamefont {Borsato}}, \bibinfo {author} {\bibfnamefont {D.}~\bibnamefont {Bortolato}}, \bibinfo {author} {\bibfnamefont {A.}~\bibnamefont {Boston}}, \bibinfo {author} {\bibfnamefont {H.}~\bibnamefont {Boston}}, \bibinfo {author} {\bibfnamefont {P.}~\bibnamefont {Bourgault}}, \bibinfo {author} {\bibfnamefont {P.}~\bibnamefont {Boutachkov}}, \bibinfo {author} {\bibfnamefont {A.}~\bibnamefont {Bouty}}, \bibinfo {author} {\bibfnamefont {A.}~\bibnamefont {Bracco}}, \bibinfo {author} {\bibfnamefont {S.}~\bibnamefont {Brambilla}}, \bibinfo {author} {\bibfnamefont {I.}~\bibnamefont {Brawn}}, \bibinfo {author} {\bibfnamefont {A.}~\bibnamefont {Brondi}},
  \bibinfo {author} {\bibfnamefont {S.}~\bibnamefont {Broussard}}, \bibinfo {author} {\bibfnamefont {B.}~\bibnamefont {Bruyneel}}, \bibinfo {author} {\bibfnamefont {D.}~\bibnamefont {Bucurescu}}, \bibinfo {author} {\bibfnamefont {I.}~\bibnamefont {Burrows}}, \bibinfo {author} {\bibfnamefont {A.}~\bibnamefont {Bürger}}, \bibinfo {author} {\bibfnamefont {S.}~\bibnamefont {Cabaret}}, \bibinfo {author} {\bibfnamefont {B.}~\bibnamefont {Cahan}}, \bibinfo {author} {\bibfnamefont {E.}~\bibnamefont {Calore}}, \bibinfo {author} {\bibfnamefont {F.}~\bibnamefont {Camera}}, \bibinfo {author} {\bibfnamefont {A.}~\bibnamefont {Capsoni}}, \bibinfo {author} {\bibfnamefont {F.}~\bibnamefont {Carrió}}, \bibinfo {author} {\bibfnamefont {G.}~\bibnamefont {Casati}}, \bibinfo {author} {\bibfnamefont {M.}~\bibnamefont {Castoldi}}, \bibinfo {author} {\bibfnamefont {B.}~\bibnamefont {Cederwall}}, \bibinfo {author} {\bibfnamefont {J.-L.}\ \bibnamefont {Cercus}}, \bibinfo {author} {\bibfnamefont {V.}~\bibnamefont {Chambert}},
  \bibinfo {author} {\bibfnamefont {M.}~\bibnamefont {{El Chambit}}}, \bibinfo {author} {\bibfnamefont {R.}~\bibnamefont {Chapman}}, \bibinfo {author} {\bibfnamefont {L.}~\bibnamefont {Charles}}, \bibinfo {author} {\bibfnamefont {J.}~\bibnamefont {Chavas}}, \bibinfo {author} {\bibfnamefont {E.}~\bibnamefont {Clément}}, \bibinfo {author} {\bibfnamefont {P.}~\bibnamefont {Cocconi}}, \bibinfo {author} {\bibfnamefont {S.}~\bibnamefont {Coelli}}, \bibinfo {author} {\bibfnamefont {P.}~\bibnamefont {Coleman-Smith}}, \bibinfo {author} {\bibfnamefont {A.}~\bibnamefont {Colombo}}, \bibinfo {author} {\bibfnamefont {S.}~\bibnamefont {Colosimo}}, \bibinfo {author} {\bibfnamefont {C.}~\bibnamefont {Commeaux}}, \bibinfo {author} {\bibfnamefont {D.}~\bibnamefont {Conventi}}, \bibinfo {author} {\bibfnamefont {R.}~\bibnamefont {Cooper}}, \bibinfo {author} {\bibfnamefont {A.}~\bibnamefont {Corsi}}, \bibinfo {author} {\bibfnamefont {A.}~\bibnamefont {Cortesi}}, \bibinfo {author} {\bibfnamefont {L.}~\bibnamefont {Costa}},
  \bibinfo {author} {\bibfnamefont {F.}~\bibnamefont {Crespi}}, \bibinfo {author} {\bibfnamefont {J.}~\bibnamefont {Cresswell}}, \bibinfo {author} {\bibfnamefont {D.}~\bibnamefont {Cullen}}, \bibinfo {author} {\bibfnamefont {D.}~\bibnamefont {Curien}}, \bibinfo {author} {\bibfnamefont {A.}~\bibnamefont {Czermak}}, \bibinfo {author} {\bibfnamefont {D.}~\bibnamefont {Delbourg}}, \bibinfo {author} {\bibfnamefont {R.}~\bibnamefont {Depalo}}, \bibinfo {author} {\bibfnamefont {T.}~\bibnamefont {Descombes}}, \bibinfo {author} {\bibfnamefont {P.}~\bibnamefont {Désesquelles}}, \bibinfo {author} {\bibfnamefont {P.}~\bibnamefont {Detistov}}, \bibinfo {author} {\bibfnamefont {C.}~\bibnamefont {Diarra}}, \bibinfo {author} {\bibfnamefont {F.}~\bibnamefont {Didierjean}}, \bibinfo {author} {\bibfnamefont {M.}~\bibnamefont {Dimmock}}, \bibinfo {author} {\bibfnamefont {Q.}~\bibnamefont {Doan}}, \bibinfo {author} {\bibfnamefont {C.}~\bibnamefont {Domingo-Pardo}}, \bibinfo {author} {\bibfnamefont {M.}~\bibnamefont {Doncel}},
  \bibinfo {author} {\bibfnamefont {F.}~\bibnamefont {Dorangeville}}, \bibinfo {author} {\bibfnamefont {N.}~\bibnamefont {Dosme}}, \bibinfo {author} {\bibfnamefont {Y.}~\bibnamefont {Drouen}}, \bibinfo {author} {\bibfnamefont {G.}~\bibnamefont {Duchêne}}, \bibinfo {author} {\bibfnamefont {B.}~\bibnamefont {Dulny}}, \bibinfo {author} {\bibfnamefont {J.}~\bibnamefont {Eberth}}, \bibinfo {author} {\bibfnamefont {P.}~\bibnamefont {Edelbruck}}, \bibinfo {author} {\bibfnamefont {J.}~\bibnamefont {Egea}}, \bibinfo {author} {\bibfnamefont {T.}~\bibnamefont {Engert}}, \bibinfo {author} {\bibfnamefont {M.}~\bibnamefont {Erduran}}, \bibinfo {author} {\bibfnamefont {S.}~\bibnamefont {Ertürk}}, \bibinfo {author} {\bibfnamefont {C.}~\bibnamefont {Fanin}}, \bibinfo {author} {\bibfnamefont {S.}~\bibnamefont {Fantinel}}, \bibinfo {author} {\bibfnamefont {E.}~\bibnamefont {Farnea}}, \bibinfo {author} {\bibfnamefont {T.}~\bibnamefont {Faul}}, \bibinfo {author} {\bibfnamefont {M.}~\bibnamefont {Filliger}}, \bibinfo {author}
  {\bibfnamefont {F.}~\bibnamefont {Filmer}}, \bibinfo {author} {\bibfnamefont {C.}~\bibnamefont {Finck}}, \bibinfo {author} {\bibfnamefont {G.}~\bibnamefont {{de France}}}, \bibinfo {author} {\bibfnamefont {A.}~\bibnamefont {Gadea}}, \bibinfo {author} {\bibfnamefont {W.}~\bibnamefont {Gast}}, \bibinfo {author} {\bibfnamefont {A.}~\bibnamefont {Geraci}}, \bibinfo {author} {\bibfnamefont {J.}~\bibnamefont {Gerl}}, \bibinfo {author} {\bibfnamefont {R.}~\bibnamefont {Gernhäuser}}, \bibinfo {author} {\bibfnamefont {A.}~\bibnamefont {Giannatiempo}}, \bibinfo {author} {\bibfnamefont {A.}~\bibnamefont {Giaz}}, \bibinfo {author} {\bibfnamefont {L.}~\bibnamefont {Gibelin}}, \bibinfo {author} {\bibfnamefont {A.}~\bibnamefont {Givechev}}, \bibinfo {author} {\bibfnamefont {N.}~\bibnamefont {Goel}}, \bibinfo {author} {\bibfnamefont {V.}~\bibnamefont {González}}, \bibinfo {author} {\bibfnamefont {A.}~\bibnamefont {Gottardo}}, \bibinfo {author} {\bibfnamefont {X.}~\bibnamefont {Grave}}, \bibinfo {author} {\bibfnamefont
  {J.}~\bibnamefont {Gre¸bosz}}, \bibinfo {author} {\bibfnamefont {R.}~\bibnamefont {Griffiths}}, \bibinfo {author} {\bibfnamefont {A.}~\bibnamefont {Grint}}, \bibinfo {author} {\bibfnamefont {P.}~\bibnamefont {Gros}}, \bibinfo {author} {\bibfnamefont {L.}~\bibnamefont {Guevara}}, \bibinfo {author} {\bibfnamefont {M.}~\bibnamefont {Gulmini}}, \bibinfo {author} {\bibfnamefont {A.}~\bibnamefont {Görgen}}, \bibinfo {author} {\bibfnamefont {H.}~\bibnamefont {Ha}}, \bibinfo {author} {\bibfnamefont {T.}~\bibnamefont {Habermann}}, \bibinfo {author} {\bibfnamefont {L.}~\bibnamefont {Harkness}}, \bibinfo {author} {\bibfnamefont {H.}~\bibnamefont {Harroch}}, \bibinfo {author} {\bibfnamefont {K.}~\bibnamefont {Hauschild}}, \bibinfo {author} {\bibfnamefont {C.}~\bibnamefont {He}}, \bibinfo {author} {\bibfnamefont {A.}~\bibnamefont {Hernández-Prieto}}, \bibinfo {author} {\bibfnamefont {B.}~\bibnamefont {Hervieu}}, \bibinfo {author} {\bibfnamefont {H.}~\bibnamefont {Hess}}, \bibinfo {author} {\bibfnamefont
  {T.}~\bibnamefont {Hüyük}}, \bibinfo {author} {\bibfnamefont {E.}~\bibnamefont {Ince}}, \bibinfo {author} {\bibfnamefont {R.}~\bibnamefont {Isocrate}}, \bibinfo {author} {\bibfnamefont {G.}~\bibnamefont {Jaworski}}, \bibinfo {author} {\bibfnamefont {A.}~\bibnamefont {Johnson}}, \bibinfo {author} {\bibfnamefont {J.}~\bibnamefont {Jolie}}, \bibinfo {author} {\bibfnamefont {P.}~\bibnamefont {Jones}}, \bibinfo {author} {\bibfnamefont {B.}~\bibnamefont {Jonson}}, \bibinfo {author} {\bibfnamefont {P.}~\bibnamefont {Joshi}}, \bibinfo {author} {\bibfnamefont {D.}~\bibnamefont {Judson}}, \bibinfo {author} {\bibfnamefont {A.}~\bibnamefont {Jungclaus}}, \bibinfo {author} {\bibfnamefont {M.}~\bibnamefont {Kaci}}, \bibinfo {author} {\bibfnamefont {N.}~\bibnamefont {Karkour}}, \bibinfo {author} {\bibfnamefont {M.}~\bibnamefont {Karolak}}, \bibinfo {author} {\bibfnamefont {A.}~\bibnamefont {Kaşkaş}}, \bibinfo {author} {\bibfnamefont {M.}~\bibnamefont {Kebbiri}}, \bibinfo {author} {\bibfnamefont {R.}~\bibnamefont
  {Kempley}}, \bibinfo {author} {\bibfnamefont {A.}~\bibnamefont {Khaplanov}}, \bibinfo {author} {\bibfnamefont {S.}~\bibnamefont {Klupp}}, \bibinfo {author} {\bibfnamefont {M.}~\bibnamefont {Kogimtzis}}, \bibinfo {author} {\bibfnamefont {I.}~\bibnamefont {Kojouharov}}, \bibinfo {author} {\bibfnamefont {A.}~\bibnamefont {Korichi}}, \bibinfo {author} {\bibfnamefont {W.}~\bibnamefont {Korten}}, \bibinfo {author} {\bibfnamefont {T.}~\bibnamefont {Kröll}}, \bibinfo {author} {\bibfnamefont {R.}~\bibnamefont {Krücken}}, \bibinfo {author} {\bibfnamefont {N.}~\bibnamefont {Kurz}}, \bibinfo {author} {\bibfnamefont {B.}~\bibnamefont {Ky}}, \bibinfo {author} {\bibfnamefont {M.}~\bibnamefont {Labiche}}, \bibinfo {author} {\bibfnamefont {X.}~\bibnamefont {Lafay}}, \bibinfo {author} {\bibfnamefont {L.}~\bibnamefont {Lavergne}}, \bibinfo {author} {\bibfnamefont {I.}~\bibnamefont {Lazarus}}, \bibinfo {author} {\bibfnamefont {S.}~\bibnamefont {Leboutelier}}, \bibinfo {author} {\bibfnamefont {F.}~\bibnamefont {Lefebvre}},
  \bibinfo {author} {\bibfnamefont {E.}~\bibnamefont {Legay}}, \bibinfo {author} {\bibfnamefont {L.}~\bibnamefont {Legeard}}, \bibinfo {author} {\bibfnamefont {F.}~\bibnamefont {Lelli}}, \bibinfo {author} {\bibfnamefont {S.}~\bibnamefont {Lenzi}}, \bibinfo {author} {\bibfnamefont {S.}~\bibnamefont {Leoni}}, \bibinfo {author} {\bibfnamefont {A.}~\bibnamefont {Lermitage}}, \bibinfo {author} {\bibfnamefont {D.}~\bibnamefont {Lersch}}, \bibinfo {author} {\bibfnamefont {J.}~\bibnamefont {Leske}}, \bibinfo {author} {\bibfnamefont {S.}~\bibnamefont {Letts}}, \bibinfo {author} {\bibfnamefont {S.}~\bibnamefont {Lhenoret}}, \bibinfo {author} {\bibfnamefont {R.}~\bibnamefont {Lieder}}, \bibinfo {author} {\bibfnamefont {D.}~\bibnamefont {Linget}}, \bibinfo {author} {\bibfnamefont {J.}~\bibnamefont {Ljungvall}}, \bibinfo {author} {\bibfnamefont {A.}~\bibnamefont {Lopez-Martens}}, \bibinfo {author} {\bibfnamefont {A.}~\bibnamefont {Lotodé}}, \bibinfo {author} {\bibfnamefont {S.}~\bibnamefont {Lunardi}}, \bibinfo {author}
  {\bibfnamefont {A.}~\bibnamefont {Maj}}, \bibinfo {author} {\bibfnamefont {J.}~\bibnamefont {{van der Marel}}}, \bibinfo {author} {\bibfnamefont {Y.}~\bibnamefont {Mariette}}, \bibinfo {author} {\bibfnamefont {N.}~\bibnamefont {Marginean}}, \bibinfo {author} {\bibfnamefont {R.}~\bibnamefont {Marginean}}, \bibinfo {author} {\bibfnamefont {G.}~\bibnamefont {Maron}}, \bibinfo {author} {\bibfnamefont {A.}~\bibnamefont {Mather}}, \bibinfo {author} {\bibfnamefont {W.}~\bibnamefont {Me¸czyński}}, \bibinfo {author} {\bibfnamefont {V.}~\bibnamefont {Mendéz}}, \bibinfo {author} {\bibfnamefont {P.}~\bibnamefont {Medina}}, \bibinfo {author} {\bibfnamefont {B.}~\bibnamefont {Melon}}, \bibinfo {author} {\bibfnamefont {R.}~\bibnamefont {Menegazzo}}, \bibinfo {author} {\bibfnamefont {D.}~\bibnamefont {Mengoni}}, \bibinfo {author} {\bibfnamefont {E.}~\bibnamefont {Merchan}}, \bibinfo {author} {\bibfnamefont {L.}~\bibnamefont {Mihailescu}}, \bibinfo {author} {\bibfnamefont {C.}~\bibnamefont {Michelagnoli}}, \bibinfo
  {author} {\bibfnamefont {J.}~\bibnamefont {Mierzejewski}}, \bibinfo {author} {\bibfnamefont {L.}~\bibnamefont {Milechina}}, \bibinfo {author} {\bibfnamefont {B.}~\bibnamefont {Million}}, \bibinfo {author} {\bibfnamefont {K.}~\bibnamefont {Mitev}}, \bibinfo {author} {\bibfnamefont {P.}~\bibnamefont {Molini}}, \bibinfo {author} {\bibfnamefont {D.}~\bibnamefont {Montanari}}, \bibinfo {author} {\bibfnamefont {S.}~\bibnamefont {Moon}}, \bibinfo {author} {\bibfnamefont {F.}~\bibnamefont {Morbiducci}}, \bibinfo {author} {\bibfnamefont {R.}~\bibnamefont {Moro}}, \bibinfo {author} {\bibfnamefont {P.}~\bibnamefont {Morrall}}, \bibinfo {author} {\bibfnamefont {O.}~\bibnamefont {Möller}}, \bibinfo {author} {\bibfnamefont {A.}~\bibnamefont {Nannini}}, \bibinfo {author} {\bibfnamefont {D.}~\bibnamefont {Napoli}}, \bibinfo {author} {\bibfnamefont {L.}~\bibnamefont {Nelson}}, \bibinfo {author} {\bibfnamefont {M.}~\bibnamefont {Nespolo}}, \bibinfo {author} {\bibfnamefont {V.}~\bibnamefont {Ngo}}, \bibinfo {author}
  {\bibfnamefont {M.}~\bibnamefont {Nicoletto}}, \bibinfo {author} {\bibfnamefont {R.}~\bibnamefont {Nicolini}}, \bibinfo {author} {\bibfnamefont {Y.}~\bibnamefont {{Le Noa}}}, \bibinfo {author} {\bibfnamefont {P.}~\bibnamefont {Nolan}}, \bibinfo {author} {\bibfnamefont {M.}~\bibnamefont {Norman}}, \bibinfo {author} {\bibfnamefont {J.}~\bibnamefont {Nyberg}}, \bibinfo {author} {\bibfnamefont {A.}~\bibnamefont {Obertelli}}, \bibinfo {author} {\bibfnamefont {A.}~\bibnamefont {Olariu}}, \bibinfo {author} {\bibfnamefont {R.}~\bibnamefont {Orlandi}}, \bibinfo {author} {\bibfnamefont {D.}~\bibnamefont {Oxley}}, \bibinfo {author} {\bibfnamefont {C.}~\bibnamefont {Özben}}, \bibinfo {author} {\bibfnamefont {M.}~\bibnamefont {Ozille}}, \bibinfo {author} {\bibfnamefont {C.}~\bibnamefont {Oziol}}, \bibinfo {author} {\bibfnamefont {E.}~\bibnamefont {Pachoud}}, \bibinfo {author} {\bibfnamefont {M.}~\bibnamefont {Palacz}}, \bibinfo {author} {\bibfnamefont {J.}~\bibnamefont {Palin}}, \bibinfo {author} {\bibfnamefont
  {J.}~\bibnamefont {Pancin}}, \bibinfo {author} {\bibfnamefont {C.}~\bibnamefont {Parisel}}, \bibinfo {author} {\bibfnamefont {P.}~\bibnamefont {Pariset}}, \bibinfo {author} {\bibfnamefont {G.}~\bibnamefont {Pascovici}}, \bibinfo {author} {\bibfnamefont {R.}~\bibnamefont {Peghin}}, \bibinfo {author} {\bibfnamefont {L.}~\bibnamefont {Pellegri}}, \bibinfo {author} {\bibfnamefont {A.}~\bibnamefont {Perego}}, \bibinfo {author} {\bibfnamefont {S.}~\bibnamefont {Perrier}}, \bibinfo {author} {\bibfnamefont {M.}~\bibnamefont {Petcu}}, \bibinfo {author} {\bibfnamefont {P.}~\bibnamefont {Petkov}}, \bibinfo {author} {\bibfnamefont {C.}~\bibnamefont {Petrache}}, \bibinfo {author} {\bibfnamefont {E.}~\bibnamefont {Pierre}}, \bibinfo {author} {\bibfnamefont {N.}~\bibnamefont {Pietralla}}, \bibinfo {author} {\bibfnamefont {S.}~\bibnamefont {Pietri}}, \bibinfo {author} {\bibfnamefont {M.}~\bibnamefont {Pignanelli}}, \bibinfo {author} {\bibfnamefont {I.}~\bibnamefont {Piqueras}}, \bibinfo {author} {\bibfnamefont
  {Z.}~\bibnamefont {Podolyak}}, \bibinfo {author} {\bibfnamefont {P.}~\bibnamefont {{Le Pouhalec}}}, \bibinfo {author} {\bibfnamefont {J.}~\bibnamefont {Pouthas}}, \bibinfo {author} {\bibfnamefont {D.}~\bibnamefont {Pugnére}}, \bibinfo {author} {\bibfnamefont {V.}~\bibnamefont {Pucknell}}, \bibinfo {author} {\bibfnamefont {A.}~\bibnamefont {Pullia}}, \bibinfo {author} {\bibfnamefont {B.}~\bibnamefont {Quintana}}, \bibinfo {author} {\bibfnamefont {R.}~\bibnamefont {Raine}}, \bibinfo {author} {\bibfnamefont {G.}~\bibnamefont {Rainovski}}, \bibinfo {author} {\bibfnamefont {L.}~\bibnamefont {Ramina}}, \bibinfo {author} {\bibfnamefont {G.}~\bibnamefont {Rampazzo}}, \bibinfo {author} {\bibfnamefont {G.}~\bibnamefont {{La Rana}}}, \bibinfo {author} {\bibfnamefont {M.}~\bibnamefont {Rebeschini}}, \bibinfo {author} {\bibfnamefont {F.}~\bibnamefont {Recchia}}, \bibinfo {author} {\bibfnamefont {N.}~\bibnamefont {Redon}}, \bibinfo {author} {\bibfnamefont {M.}~\bibnamefont {Reese}}, \bibinfo {author} {\bibfnamefont
  {P.}~\bibnamefont {Reiter}}, \bibinfo {author} {\bibfnamefont {P.}~\bibnamefont {Regan}}, \bibinfo {author} {\bibfnamefont {S.}~\bibnamefont {Riboldi}}, \bibinfo {author} {\bibfnamefont {M.}~\bibnamefont {Richer}}, \bibinfo {author} {\bibfnamefont {M.}~\bibnamefont {Rigato}}, \bibinfo {author} {\bibfnamefont {S.}~\bibnamefont {Rigby}}, \bibinfo {author} {\bibfnamefont {G.}~\bibnamefont {Ripamonti}}, \bibinfo {author} {\bibfnamefont {A.}~\bibnamefont {Robinson}}, \bibinfo {author} {\bibfnamefont {J.}~\bibnamefont {Robin}}, \bibinfo {author} {\bibfnamefont {J.}~\bibnamefont {Roccaz}}, \bibinfo {author} {\bibfnamefont {J.-A.}\ \bibnamefont {Ropert}}, \bibinfo {author} {\bibfnamefont {B.}~\bibnamefont {Rossé}}, \bibinfo {author} {\bibfnamefont {C.}~\bibnamefont {{Rossi Alvarez}}}, \bibinfo {author} {\bibfnamefont {D.}~\bibnamefont {Rosso}}, \bibinfo {author} {\bibfnamefont {B.}~\bibnamefont {Rubio}}, \bibinfo {author} {\bibfnamefont {D.}~\bibnamefont {Rudolph}}, \bibinfo {author} {\bibfnamefont
  {F.}~\bibnamefont {Saillant}}, \bibinfo {author} {\bibfnamefont {E.}~\bibnamefont {Şahin}}, \bibinfo {author} {\bibfnamefont {F.}~\bibnamefont {Salomon}}, \bibinfo {author} {\bibfnamefont {M.-D.}\ \bibnamefont {Salsac}}, \bibinfo {author} {\bibfnamefont {J.}~\bibnamefont {Salt}}, \bibinfo {author} {\bibfnamefont {G.}~\bibnamefont {Salvato}}, \bibinfo {author} {\bibfnamefont {J.}~\bibnamefont {Sampson}}, \bibinfo {author} {\bibfnamefont {E.}~\bibnamefont {Sanchis}}, \bibinfo {author} {\bibfnamefont {C.}~\bibnamefont {Santos}}, \bibinfo {author} {\bibfnamefont {H.}~\bibnamefont {Schaffner}}, \bibinfo {author} {\bibfnamefont {M.}~\bibnamefont {Schlarb}}, \bibinfo {author} {\bibfnamefont {D.}~\bibnamefont {Scraggs}}, \bibinfo {author} {\bibfnamefont {D.}~\bibnamefont {Seddon}}, \bibinfo {author} {\bibfnamefont {M.}~\bibnamefont {Şenyiğit}}, \bibinfo {author} {\bibfnamefont {M.-H.}\ \bibnamefont {Sigward}}, \bibinfo {author} {\bibfnamefont {G.}~\bibnamefont {Simpson}}, \bibinfo {author} {\bibfnamefont
  {J.}~\bibnamefont {Simpson}}, \bibinfo {author} {\bibfnamefont {M.}~\bibnamefont {Slee}}, \bibinfo {author} {\bibfnamefont {J.}~\bibnamefont {Smith}}, \bibinfo {author} {\bibfnamefont {P.}~\bibnamefont {Sona}}, \bibinfo {author} {\bibfnamefont {B.}~\bibnamefont {Sowicki}}, \bibinfo {author} {\bibfnamefont {P.}~\bibnamefont {Spolaore}}, \bibinfo {author} {\bibfnamefont {C.}~\bibnamefont {Stahl}}, \bibinfo {author} {\bibfnamefont {T.}~\bibnamefont {Stanios}}, \bibinfo {author} {\bibfnamefont {E.}~\bibnamefont {Stefanova}}, \bibinfo {author} {\bibfnamefont {O.}~\bibnamefont {Stézowski}}, \bibinfo {author} {\bibfnamefont {J.}~\bibnamefont {Strachan}}, \bibinfo {author} {\bibfnamefont {G.}~\bibnamefont {Suliman}}, \bibinfo {author} {\bibfnamefont {P.-A.}\ \bibnamefont {Söderström}}, \bibinfo {author} {\bibfnamefont {J.}~\bibnamefont {Tain}}, \bibinfo {author} {\bibfnamefont {S.}~\bibnamefont {Tanguy}}, \bibinfo {author} {\bibfnamefont {S.}~\bibnamefont {Tashenov}}, \bibinfo {author} {\bibfnamefont
  {C.}~\bibnamefont {Theisen}}, \bibinfo {author} {\bibfnamefont {J.}~\bibnamefont {Thornhill}}, \bibinfo {author} {\bibfnamefont {F.}~\bibnamefont {Tomasi}}, \bibinfo {author} {\bibfnamefont {N.}~\bibnamefont {Toniolo}}, \bibinfo {author} {\bibfnamefont {R.}~\bibnamefont {Touzery}}, \bibinfo {author} {\bibfnamefont {B.}~\bibnamefont {Travers}}, \bibinfo {author} {\bibfnamefont {A.}~\bibnamefont {Triossi}}, \bibinfo {author} {\bibfnamefont {M.}~\bibnamefont {Tripon}}, \bibinfo {author} {\bibfnamefont {K.}~\bibnamefont {Tun-Lanoë}}, \bibinfo {author} {\bibfnamefont {M.}~\bibnamefont {Turcato}}, \bibinfo {author} {\bibfnamefont {C.}~\bibnamefont {Unsworth}}, \bibinfo {author} {\bibfnamefont {C.}~\bibnamefont {Ur}}, \bibinfo {author} {\bibfnamefont {J.}~\bibnamefont {Valiente-Dobon}}, \bibinfo {author} {\bibfnamefont {V.}~\bibnamefont {Vandone}}, \bibinfo {author} {\bibfnamefont {E.}~\bibnamefont {Vardaci}}, \bibinfo {author} {\bibfnamefont {R.}~\bibnamefont {Venturelli}}, \bibinfo {author} {\bibfnamefont
  {F.}~\bibnamefont {Veronese}}, \bibinfo {author} {\bibfnamefont {C.}~\bibnamefont {Veyssiere}}, \bibinfo {author} {\bibfnamefont {E.}~\bibnamefont {Viscione}}, \bibinfo {author} {\bibfnamefont {R.}~\bibnamefont {Wadsworth}}, \bibinfo {author} {\bibfnamefont {P.}~\bibnamefont {Walker}}, \bibinfo {author} {\bibfnamefont {N.}~\bibnamefont {Warr}}, \bibinfo {author} {\bibfnamefont {C.}~\bibnamefont {Weber}}, \bibinfo {author} {\bibfnamefont {D.}~\bibnamefont {Weisshaar}}, \bibinfo {author} {\bibfnamefont {D.}~\bibnamefont {Wells}}, \bibinfo {author} {\bibfnamefont {O.}~\bibnamefont {Wieland}}, \bibinfo {author} {\bibfnamefont {A.}~\bibnamefont {Wiens}}, \bibinfo {author} {\bibfnamefont {G.}~\bibnamefont {Wittwer}}, \bibinfo {author} {\bibfnamefont {H.}~\bibnamefont {Wollersheim}}, \bibinfo {author} {\bibfnamefont {F.}~\bibnamefont {Zocca}}, \bibinfo {author} {\bibfnamefont {N.}~\bibnamefont {Zamfir}}, \bibinfo {author} {\bibfnamefont {M.}~\bibnamefont {Zie¸bliński}},\ and\ \bibinfo {author} {\bibfnamefont
  {A.}~\bibnamefont {Zucchiatti}},\ }\href {https://doi.org/https://doi.org/10.1016/j.nima.2011.11.081} {\bibfield  {journal} {\bibinfo  {journal} {Nucl. Instrum. Methods Phys. Res. A}\ }\textbf {\bibinfo {volume} {668}},\ \bibinfo {pages} {26} (\bibinfo {year} {2012})}\BibitemShut {NoStop}%
\bibitem [{\citenamefont {Clément}\ \emph {et~al.}(2017)\citenamefont {Clément}, \citenamefont {Michelagnoli}, \citenamefont {{de France}}, \citenamefont {Li}, \citenamefont {Lemasson}, \citenamefont {{Barthe Dejean}}, \citenamefont {Beuzard}, \citenamefont {Bougault}, \citenamefont {Cacitti}, \citenamefont {Foucher}, \citenamefont {Fremont}, \citenamefont {Gangnant}, \citenamefont {Goupil}, \citenamefont {Houarner}, \citenamefont {Jean}, \citenamefont {Lefevre}, \citenamefont {Legeard}, \citenamefont {Legruel}, \citenamefont {Maugeais}, \citenamefont {Ménager}, \citenamefont {Ménard}, \citenamefont {Munoz}, \citenamefont {Ozille}, \citenamefont {Raine}, \citenamefont {Ropert}, \citenamefont {Saillant}, \citenamefont {Spitaels}, \citenamefont {Tripon}, \citenamefont {Vallerand}, \citenamefont {Voltolini}, \citenamefont {Korten}, \citenamefont {Salsac}, \citenamefont {Theisen}, \citenamefont {Zielińska}, \citenamefont {Joannem}, \citenamefont {Karolak}, \citenamefont {Kebbiri}, \citenamefont {Lotode},
  \citenamefont {Touzery}, \citenamefont {Walter}, \citenamefont {Korichi}, \citenamefont {Ljungvall}, \citenamefont {Lopez-Martens}, \citenamefont {Ralet}, \citenamefont {Dosme}, \citenamefont {Grave}, \citenamefont {Karkour}, \citenamefont {Lafay}, \citenamefont {Legay}, \citenamefont {Kojouharov}, \citenamefont {Domingo-Pardo}, \citenamefont {Gadea}, \citenamefont {Pérez-Vidal}, \citenamefont {Civera}, \citenamefont {Birkenbach}, \citenamefont {Eberth}, \citenamefont {Hess}, \citenamefont {Lewandowski}, \citenamefont {Reiter}, \citenamefont {Nannini}, \citenamefont {{De Angelis}}, \citenamefont {Jaworski}, \citenamefont {John}, \citenamefont {Napoli}, \citenamefont {Valiente-Dobón}, \citenamefont {Barrientos}, \citenamefont {Bortolato}, \citenamefont {Benzoni}, \citenamefont {Bracco}, \citenamefont {Brambilla}, \citenamefont {Camera}, \citenamefont {Crespi}, \citenamefont {Leoni}, \citenamefont {Million}, \citenamefont {Pullia}, \citenamefont {Wieland}, \citenamefont {Bazzacco}, \citenamefont {Lenzi},
  \citenamefont {Lunardi}, \citenamefont {Menegazzo}, \citenamefont {Mengoni}, \citenamefont {Recchia}, \citenamefont {Bellato}, \citenamefont {Isocrate}, \citenamefont {{Egea Canet}}, \citenamefont {Didierjean}, \citenamefont {Duchêne}, \citenamefont {Baumann}, \citenamefont {Brucker}, \citenamefont {Dangelser}, \citenamefont {Filliger}, \citenamefont {Friedmann}, \citenamefont {Gaudiot}, \citenamefont {Grapton}, \citenamefont {Kocher}, \citenamefont {Mathieu}, \citenamefont {Sigward}, \citenamefont {Thomas}, \citenamefont {Veeramootoo}, \citenamefont {Dudouet}, \citenamefont {Stézowski}, \citenamefont {Aufranc}, \citenamefont {Aubert}, \citenamefont {Labiche}, \citenamefont {Simpson}, \citenamefont {Burrows}, \citenamefont {Coleman-Smith}, \citenamefont {Grant}, \citenamefont {Lazarus}, \citenamefont {Morrall}, \citenamefont {Pucknell}, \citenamefont {Boston}, \citenamefont {Judson}, \citenamefont {Lalović}, \citenamefont {Nyberg}, \citenamefont {Collado}, \citenamefont {González}, \citenamefont {Kuti},
  \citenamefont {Nyakó}, \citenamefont {Maj},\ and\ \citenamefont {Rudigier}}]{NIMA2017_CLEMENT}%
  \BibitemOpen
  \bibfield  {author} {\bibinfo {author} {\bibfnamefont {E.}~\bibnamefont {Clément}}, \bibinfo {author} {\bibfnamefont {C.}~\bibnamefont {Michelagnoli}}, \bibinfo {author} {\bibfnamefont {G.}~\bibnamefont {{de France}}}, \bibinfo {author} {\bibfnamefont {H.}~\bibnamefont {Li}}, \bibinfo {author} {\bibfnamefont {A.}~\bibnamefont {Lemasson}}, \bibinfo {author} {\bibfnamefont {C.}~\bibnamefont {{Barthe Dejean}}}, \bibinfo {author} {\bibfnamefont {M.}~\bibnamefont {Beuzard}}, \bibinfo {author} {\bibfnamefont {P.}~\bibnamefont {Bougault}}, \bibinfo {author} {\bibfnamefont {J.}~\bibnamefont {Cacitti}}, \bibinfo {author} {\bibfnamefont {J.-L.}\ \bibnamefont {Foucher}}, \bibinfo {author} {\bibfnamefont {G.}~\bibnamefont {Fremont}}, \bibinfo {author} {\bibfnamefont {P.}~\bibnamefont {Gangnant}}, \bibinfo {author} {\bibfnamefont {J.}~\bibnamefont {Goupil}}, \bibinfo {author} {\bibfnamefont {C.}~\bibnamefont {Houarner}}, \bibinfo {author} {\bibfnamefont {M.}~\bibnamefont {Jean}}, \bibinfo {author} {\bibfnamefont
  {A.}~\bibnamefont {Lefevre}}, \bibinfo {author} {\bibfnamefont {L.}~\bibnamefont {Legeard}}, \bibinfo {author} {\bibfnamefont {F.}~\bibnamefont {Legruel}}, \bibinfo {author} {\bibfnamefont {C.}~\bibnamefont {Maugeais}}, \bibinfo {author} {\bibfnamefont {L.}~\bibnamefont {Ménager}}, \bibinfo {author} {\bibfnamefont {N.}~\bibnamefont {Ménard}}, \bibinfo {author} {\bibfnamefont {H.}~\bibnamefont {Munoz}}, \bibinfo {author} {\bibfnamefont {M.}~\bibnamefont {Ozille}}, \bibinfo {author} {\bibfnamefont {B.}~\bibnamefont {Raine}}, \bibinfo {author} {\bibfnamefont {J.}~\bibnamefont {Ropert}}, \bibinfo {author} {\bibfnamefont {F.}~\bibnamefont {Saillant}}, \bibinfo {author} {\bibfnamefont {C.}~\bibnamefont {Spitaels}}, \bibinfo {author} {\bibfnamefont {M.}~\bibnamefont {Tripon}}, \bibinfo {author} {\bibfnamefont {P.}~\bibnamefont {Vallerand}}, \bibinfo {author} {\bibfnamefont {G.}~\bibnamefont {Voltolini}}, \bibinfo {author} {\bibfnamefont {W.}~\bibnamefont {Korten}}, \bibinfo {author} {\bibfnamefont {M.-D.}\
  \bibnamefont {Salsac}}, \bibinfo {author} {\bibfnamefont {C.}~\bibnamefont {Theisen}}, \bibinfo {author} {\bibfnamefont {M.}~\bibnamefont {Zielińska}}, \bibinfo {author} {\bibfnamefont {T.}~\bibnamefont {Joannem}}, \bibinfo {author} {\bibfnamefont {M.}~\bibnamefont {Karolak}}, \bibinfo {author} {\bibfnamefont {M.}~\bibnamefont {Kebbiri}}, \bibinfo {author} {\bibfnamefont {A.}~\bibnamefont {Lotode}}, \bibinfo {author} {\bibfnamefont {R.}~\bibnamefont {Touzery}}, \bibinfo {author} {\bibfnamefont {C.}~\bibnamefont {Walter}}, \bibinfo {author} {\bibfnamefont {A.}~\bibnamefont {Korichi}}, \bibinfo {author} {\bibfnamefont {J.}~\bibnamefont {Ljungvall}}, \bibinfo {author} {\bibfnamefont {A.}~\bibnamefont {Lopez-Martens}}, \bibinfo {author} {\bibfnamefont {D.}~\bibnamefont {Ralet}}, \bibinfo {author} {\bibfnamefont {N.}~\bibnamefont {Dosme}}, \bibinfo {author} {\bibfnamefont {X.}~\bibnamefont {Grave}}, \bibinfo {author} {\bibfnamefont {N.}~\bibnamefont {Karkour}}, \bibinfo {author} {\bibfnamefont {X.}~\bibnamefont
  {Lafay}}, \bibinfo {author} {\bibfnamefont {E.}~\bibnamefont {Legay}}, \bibinfo {author} {\bibfnamefont {I.}~\bibnamefont {Kojouharov}}, \bibinfo {author} {\bibfnamefont {C.}~\bibnamefont {Domingo-Pardo}}, \bibinfo {author} {\bibfnamefont {A.}~\bibnamefont {Gadea}}, \bibinfo {author} {\bibfnamefont {R.}~\bibnamefont {Pérez-Vidal}}, \bibinfo {author} {\bibfnamefont {J.}~\bibnamefont {Civera}}, \bibinfo {author} {\bibfnamefont {B.}~\bibnamefont {Birkenbach}}, \bibinfo {author} {\bibfnamefont {J.}~\bibnamefont {Eberth}}, \bibinfo {author} {\bibfnamefont {H.}~\bibnamefont {Hess}}, \bibinfo {author} {\bibfnamefont {L.}~\bibnamefont {Lewandowski}}, \bibinfo {author} {\bibfnamefont {P.}~\bibnamefont {Reiter}}, \bibinfo {author} {\bibfnamefont {A.}~\bibnamefont {Nannini}}, \bibinfo {author} {\bibfnamefont {G.}~\bibnamefont {{De Angelis}}}, \bibinfo {author} {\bibfnamefont {G.}~\bibnamefont {Jaworski}}, \bibinfo {author} {\bibfnamefont {P.}~\bibnamefont {John}}, \bibinfo {author} {\bibfnamefont {D.}~\bibnamefont
  {Napoli}}, \bibinfo {author} {\bibfnamefont {J.}~\bibnamefont {Valiente-Dobón}}, \bibinfo {author} {\bibfnamefont {D.}~\bibnamefont {Barrientos}}, \bibinfo {author} {\bibfnamefont {D.}~\bibnamefont {Bortolato}}, \bibinfo {author} {\bibfnamefont {G.}~\bibnamefont {Benzoni}}, \bibinfo {author} {\bibfnamefont {A.}~\bibnamefont {Bracco}}, \bibinfo {author} {\bibfnamefont {S.}~\bibnamefont {Brambilla}}, \bibinfo {author} {\bibfnamefont {F.}~\bibnamefont {Camera}}, \bibinfo {author} {\bibfnamefont {F.}~\bibnamefont {Crespi}}, \bibinfo {author} {\bibfnamefont {S.}~\bibnamefont {Leoni}}, \bibinfo {author} {\bibfnamefont {B.}~\bibnamefont {Million}}, \bibinfo {author} {\bibfnamefont {A.}~\bibnamefont {Pullia}}, \bibinfo {author} {\bibfnamefont {O.}~\bibnamefont {Wieland}}, \bibinfo {author} {\bibfnamefont {D.}~\bibnamefont {Bazzacco}}, \bibinfo {author} {\bibfnamefont {S.}~\bibnamefont {Lenzi}}, \bibinfo {author} {\bibfnamefont {S.}~\bibnamefont {Lunardi}}, \bibinfo {author} {\bibfnamefont {R.}~\bibnamefont
  {Menegazzo}}, \bibinfo {author} {\bibfnamefont {D.}~\bibnamefont {Mengoni}}, \bibinfo {author} {\bibfnamefont {F.}~\bibnamefont {Recchia}}, \bibinfo {author} {\bibfnamefont {M.}~\bibnamefont {Bellato}}, \bibinfo {author} {\bibfnamefont {R.}~\bibnamefont {Isocrate}}, \bibinfo {author} {\bibfnamefont {F.}~\bibnamefont {{Egea Canet}}}, \bibinfo {author} {\bibfnamefont {F.}~\bibnamefont {Didierjean}}, \bibinfo {author} {\bibfnamefont {G.}~\bibnamefont {Duchêne}}, \bibinfo {author} {\bibfnamefont {R.}~\bibnamefont {Baumann}}, \bibinfo {author} {\bibfnamefont {M.}~\bibnamefont {Brucker}}, \bibinfo {author} {\bibfnamefont {E.}~\bibnamefont {Dangelser}}, \bibinfo {author} {\bibfnamefont {M.}~\bibnamefont {Filliger}}, \bibinfo {author} {\bibfnamefont {H.}~\bibnamefont {Friedmann}}, \bibinfo {author} {\bibfnamefont {G.}~\bibnamefont {Gaudiot}}, \bibinfo {author} {\bibfnamefont {J.-N.}\ \bibnamefont {Grapton}}, \bibinfo {author} {\bibfnamefont {H.}~\bibnamefont {Kocher}}, \bibinfo {author} {\bibfnamefont
  {C.}~\bibnamefont {Mathieu}}, \bibinfo {author} {\bibfnamefont {M.-H.}\ \bibnamefont {Sigward}}, \bibinfo {author} {\bibfnamefont {D.}~\bibnamefont {Thomas}}, \bibinfo {author} {\bibfnamefont {S.}~\bibnamefont {Veeramootoo}}, \bibinfo {author} {\bibfnamefont {J.}~\bibnamefont {Dudouet}}, \bibinfo {author} {\bibfnamefont {O.}~\bibnamefont {Stézowski}}, \bibinfo {author} {\bibfnamefont {C.}~\bibnamefont {Aufranc}}, \bibinfo {author} {\bibfnamefont {Y.}~\bibnamefont {Aubert}}, \bibinfo {author} {\bibfnamefont {M.}~\bibnamefont {Labiche}}, \bibinfo {author} {\bibfnamefont {J.}~\bibnamefont {Simpson}}, \bibinfo {author} {\bibfnamefont {I.}~\bibnamefont {Burrows}}, \bibinfo {author} {\bibfnamefont {P.}~\bibnamefont {Coleman-Smith}}, \bibinfo {author} {\bibfnamefont {A.}~\bibnamefont {Grant}}, \bibinfo {author} {\bibfnamefont {I.}~\bibnamefont {Lazarus}}, \bibinfo {author} {\bibfnamefont {P.}~\bibnamefont {Morrall}}, \bibinfo {author} {\bibfnamefont {V.}~\bibnamefont {Pucknell}}, \bibinfo {author} {\bibfnamefont
  {A.}~\bibnamefont {Boston}}, \bibinfo {author} {\bibfnamefont {D.}~\bibnamefont {Judson}}, \bibinfo {author} {\bibfnamefont {N.}~\bibnamefont {Lalović}}, \bibinfo {author} {\bibfnamefont {J.}~\bibnamefont {Nyberg}}, \bibinfo {author} {\bibfnamefont {J.}~\bibnamefont {Collado}}, \bibinfo {author} {\bibfnamefont {V.}~\bibnamefont {González}}, \bibinfo {author} {\bibfnamefont {I.}~\bibnamefont {Kuti}}, \bibinfo {author} {\bibfnamefont {B.}~\bibnamefont {Nyakó}}, \bibinfo {author} {\bibfnamefont {A.}~\bibnamefont {Maj}},\ and\ \bibinfo {author} {\bibfnamefont {M.}~\bibnamefont {Rudigier}},\ }\href {https://doi.org/https://doi.org/10.1016/j.nima.2017.02.063} {\bibfield  {journal} {\bibinfo  {journal} {Nucl. Instrum. Methods Phys. Res. A}\ }\textbf {\bibinfo {volume} {855}},\ \bibinfo {pages} {1} (\bibinfo {year} {2017})}\BibitemShut {NoStop}%
\bibitem [{\citenamefont {Dudouet}(2024)}]{Cubix}%
  \BibitemOpen
  \bibfield  {author} {\bibinfo {author} {\bibfnamefont {J.}~\bibnamefont {Dudouet}},\ }\href {https://doi.org/10.5281/zenodo.15100409} {\bibinfo {title} {Cubix (1.3)}} (\bibinfo {year} {2024})\BibitemShut {NoStop}%
\bibitem [{ROO(2024)}]{ROOT}%
  \BibitemOpen
  \href@noop {} {\bibinfo {title} {{R}{O}{O}{T}: analyzing petabytes of data, scientifically. --- root.cern}},\ \bibinfo {howpublished} {\url{https://root.cern}} (\bibinfo {year} {2024})\BibitemShut {NoStop}%
\bibitem [{\citenamefont {Kibédi}\ \emph {et~al.}(2008)\citenamefont {Kibédi}, \citenamefont {Burrows}, \citenamefont {Trzhaskovskaya}, \citenamefont {Davidson},\ and\ \citenamefont {Nestor}}]{NIMA2008_KIBEDI}%
  \BibitemOpen
  \bibfield  {author} {\bibinfo {author} {\bibfnamefont {T.}~\bibnamefont {Kibédi}}, \bibinfo {author} {\bibfnamefont {T.}~\bibnamefont {Burrows}}, \bibinfo {author} {\bibfnamefont {M.}~\bibnamefont {Trzhaskovskaya}}, \bibinfo {author} {\bibfnamefont {P.}~\bibnamefont {Davidson}},\ and\ \bibinfo {author} {\bibfnamefont {C.}~\bibnamefont {Nestor}},\ }\href {https://doi.org/https://doi.org/10.1016/j.nima.2008.02.051} {\bibfield  {journal} {\bibinfo  {journal} {Nucl. Instrum. Methods Phys. Res. A}\ }\textbf {\bibinfo {volume} {589}},\ \bibinfo {pages} {202} (\bibinfo {year} {2008})}\BibitemShut {NoStop}%
\bibitem [{\citenamefont {{ENSDF database at National Nuclear Data Center}}()}]{ENSDF}%
  \BibitemOpen
  \bibfield  {author} {\bibinfo {author} {\bibnamefont {{ENSDF database at National Nuclear Data Center}}},\ }\href {https://www.nndc.bnl.gov/ensdf/} {}\bibinfo {note} {Available at \url{https://www.nndc.bnl.gov/ensdf/}}\BibitemShut {NoStop}%
\bibitem [{\citenamefont {Borggreen}\ and\ \citenamefont {Naumann}(1968)}]{BORGGREEN1968}%
  \BibitemOpen
  \bibfield  {author} {\bibinfo {author} {\bibfnamefont {J.}~\bibnamefont {Borggreen}}\ and\ \bibinfo {author} {\bibfnamefont {R.}~\bibnamefont {Naumann}},\ }\href {https://doi.org/https://doi.org/10.1016/0375-9474(68)90638-6} {\bibfield  {journal} {\bibinfo  {journal} {Nuclear Physics A}\ }\textbf {\bibinfo {volume} {115}},\ \bibinfo {pages} {14} (\bibinfo {year} {1968})}\BibitemShut {NoStop}%
\bibitem [{\citenamefont {Okano}\ \emph {et~al.}(1969)\citenamefont {Okano}, \citenamefont {Kawase}, \citenamefont {Uehara},\ and\ \citenamefont {Hayashi}}]{OKANO1969}%
  \BibitemOpen
  \bibfield  {author} {\bibinfo {author} {\bibfnamefont {K.}~\bibnamefont {Okano}}, \bibinfo {author} {\bibfnamefont {Y.}~\bibnamefont {Kawase}}, \bibinfo {author} {\bibfnamefont {S.}~\bibnamefont {Uehara}},\ and\ \bibinfo {author} {\bibfnamefont {T.}~\bibnamefont {Hayashi}},\ }\href {https://doi.org/https://doi.org/10.1016/0375-9474(69)90057-8} {\bibfield  {journal} {\bibinfo  {journal} {Nuclear Physics A}\ }\textbf {\bibinfo {volume} {136}},\ \bibinfo {pages} {321} (\bibinfo {year} {1969})}\BibitemShut {NoStop}%
\bibitem [{\citenamefont {Moody}\ \emph {et~al.}(2020)\citenamefont {Moody}, \citenamefont {Gharibyan}, \citenamefont {Shaughnessy}, \citenamefont {Grant}, \citenamefont {Gostic}, \citenamefont {Cerjan}, \citenamefont {Yeamans}, \citenamefont {Despotopulos},\ and\ \citenamefont {Faye}}]{Moody_2020}%
  \BibitemOpen
  \bibfield  {author} {\bibinfo {author} {\bibfnamefont {K.~J.}\ \bibnamefont {Moody}}, \bibinfo {author} {\bibfnamefont {N.}~\bibnamefont {Gharibyan}}, \bibinfo {author} {\bibfnamefont {D.~A.}\ \bibnamefont {Shaughnessy}}, \bibinfo {author} {\bibfnamefont {P.~M.}\ \bibnamefont {Grant}}, \bibinfo {author} {\bibfnamefont {J.~M.}\ \bibnamefont {Gostic}}, \bibinfo {author} {\bibfnamefont {C.~J.}\ \bibnamefont {Cerjan}}, \bibinfo {author} {\bibfnamefont {C.~B.}\ \bibnamefont {Yeamans}}, \bibinfo {author} {\bibfnamefont {J.~D.}\ \bibnamefont {Despotopulos}},\ and\ \bibinfo {author} {\bibfnamefont {S.~A.}\ \bibnamefont {Faye}},\ }\href {https://doi.org/10.1088/1361-6471/ab67e9} {\bibfield  {journal} {\bibinfo  {journal} {Journal of Physics G: Nuclear and Particle Physics}\ }\textbf {\bibinfo {volume} {47}},\ \bibinfo {pages} {045116} (\bibinfo {year} {2020})}\BibitemShut {NoStop}%
\bibitem [{\citenamefont {Metz}\ \emph {et~al.}(1999)\citenamefont {Metz}, \citenamefont {Jolie}, \citenamefont {Graw}, \citenamefont {Hertenberger}, \citenamefont {Gr\"oger}, \citenamefont {G\"unther}, \citenamefont {Warr},\ and\ \citenamefont {Eisermann}}]{Metz1999}%
  \BibitemOpen
  \bibfield  {author} {\bibinfo {author} {\bibfnamefont {A.}~\bibnamefont {Metz}}, \bibinfo {author} {\bibfnamefont {J.}~\bibnamefont {Jolie}}, \bibinfo {author} {\bibfnamefont {G.}~\bibnamefont {Graw}}, \bibinfo {author} {\bibfnamefont {R.}~\bibnamefont {Hertenberger}}, \bibinfo {author} {\bibfnamefont {J.}~\bibnamefont {Gr\"oger}}, \bibinfo {author} {\bibfnamefont {C.}~\bibnamefont {G\"unther}}, \bibinfo {author} {\bibfnamefont {N.}~\bibnamefont {Warr}},\ and\ \bibinfo {author} {\bibfnamefont {Y.}~\bibnamefont {Eisermann}},\ }\href {https://doi.org/10.1103/PhysRevLett.83.1542} {\bibfield  {journal} {\bibinfo  {journal} {Phys. Rev. Lett.}\ }\textbf {\bibinfo {volume} {83}},\ \bibinfo {pages} {1542} (\bibinfo {year} {1999})}\BibitemShut {NoStop}%
\bibitem [{\citenamefont {Mirza}\ \emph {et~al.}(1975)\citenamefont {Mirza}, \citenamefont {L{\textsurd}∂bner}, \citenamefont {Breitig}, \citenamefont {Baader}, \citenamefont {Koch},\ and\ \citenamefont {Schult}}]{Mirza1975}%
  \BibitemOpen
  \bibfield  {author} {\bibinfo {author} {\bibfnamefont {J.~A.}\ \bibnamefont {Mirza}}, \bibinfo {author} {\bibfnamefont {K.~E.~G.}\ \bibnamefont {L{\textsurd}∂bner}}, \bibinfo {author} {\bibfnamefont {D.}~\bibnamefont {Breitig}}, \bibinfo {author} {\bibfnamefont {H.~A.}\ \bibnamefont {Baader}}, \bibinfo {author} {\bibfnamefont {H.~R.}\ \bibnamefont {Koch}},\ and\ \bibinfo {author} {\bibfnamefont {O.~W.~B.}\ \bibnamefont {Schult}},\ }\href {https://doi.org/10.1007/BF01408145} {\bibfield  {journal} {\bibinfo  {journal} {Zeitschrift f{\"u}r Physik A Atoms and Nuclei}\ }\textbf {\bibinfo {volume} {272}},\ \bibinfo {pages} {175} (\bibinfo {year} {1975})}\BibitemShut {NoStop}%
\bibitem [{\citenamefont {Negi}\ \emph {et~al.}(2019)\citenamefont {Negi}, \citenamefont {Tandel}, \citenamefont {Chauhan}, \citenamefont {Chowdhury}, \citenamefont {Janssens}, \citenamefont {Carpenter}, \citenamefont {Khoo}, \citenamefont {Kondev}, \citenamefont {Lauritsen}, \citenamefont {Lister}, \citenamefont {Seweryniak},\ and\ \citenamefont {Zhu}}]{negi2019}%
  \BibitemOpen
  \bibfield  {author} {\bibinfo {author} {\bibfnamefont {D.}~\bibnamefont {Negi}}, \bibinfo {author} {\bibfnamefont {S.~K.}\ \bibnamefont {Tandel}}, \bibinfo {author} {\bibfnamefont {P.}~\bibnamefont {Chauhan}}, \bibinfo {author} {\bibfnamefont {P.}~\bibnamefont {Chowdhury}}, \bibinfo {author} {\bibfnamefont {R.~V.~F.}\ \bibnamefont {Janssens}}, \bibinfo {author} {\bibfnamefont {M.~P.}\ \bibnamefont {Carpenter}}, \bibinfo {author} {\bibfnamefont {T.~L.}\ \bibnamefont {Khoo}}, \bibinfo {author} {\bibfnamefont {F.~G.}\ \bibnamefont {Kondev}}, \bibinfo {author} {\bibfnamefont {T.}~\bibnamefont {Lauritsen}}, \bibinfo {author} {\bibfnamefont {C.~J.}\ \bibnamefont {Lister}}, \bibinfo {author} {\bibfnamefont {D.}~\bibnamefont {Seweryniak}},\ and\ \bibinfo {author} {\bibfnamefont {S.}~\bibnamefont {Zhu}},\ }\href {https://doi.org/10.1103/PhysRevC.100.014329} {\bibfield  {journal} {\bibinfo  {journal} {Phys. Rev. C}\ }\textbf {\bibinfo {volume} {100}},\ \bibinfo {pages} {014329} (\bibinfo {year} {2019})}\BibitemShut
  {NoStop}%
\bibitem [{\citenamefont {Sakai}\ and\ \citenamefont {Daly}(1968)}]{sakai_1968}%
  \BibitemOpen
  \bibfield  {author} {\bibinfo {author} {\bibfnamefont {K.}~\bibnamefont {Sakai}}\ and\ \bibinfo {author} {\bibfnamefont {P.}~\bibnamefont {Daly}},\ }\href {https://doi.org/https://doi.org/10.1016/0375-9474(68)90342-4} {\bibfield  {journal} {\bibinfo  {journal} {Nuclear Physics A}\ }\textbf {\bibinfo {volume} {118}},\ \bibinfo {pages} {361} (\bibinfo {year} {1968})}\BibitemShut {NoStop}%
\bibitem [{\citenamefont {Hill}\ \emph {et~al.}(1976)\citenamefont {Hill}, \citenamefont {Shirk}, \citenamefont {Petry},\ and\ \citenamefont {Wang}}]{JCHill1976}%
  \BibitemOpen
  \bibfield  {author} {\bibinfo {author} {\bibfnamefont {J.~C.}\ \bibnamefont {Hill}}, \bibinfo {author} {\bibfnamefont {D.~G.}\ \bibnamefont {Shirk}}, \bibinfo {author} {\bibfnamefont {R.~F.}\ \bibnamefont {Petry}},\ and\ \bibinfo {author} {\bibfnamefont {K.~H.}\ \bibnamefont {Wang}},\ }\href {https://doi.org/10.1103/PhysRevC.13.2484} {\bibfield  {journal} {\bibinfo  {journal} {Phys. Rev. C}\ }\textbf {\bibinfo {volume} {13}},\ \bibinfo {pages} {2484} (\bibinfo {year} {1976})}\BibitemShut {NoStop}%
\bibitem [{\citenamefont {Suman}\ \emph {et~al.}(2021)\citenamefont {Suman}, \citenamefont {Tandel}, \citenamefont {Kumawat}, \citenamefont {Wahid}, \citenamefont {Hemalatha}, \citenamefont {Chowdhury}, \citenamefont {Janssens}, \citenamefont {Carpenter}, \citenamefont {Khoo}, \citenamefont {Kondev}, \citenamefont {Lauritsen}, \citenamefont {Lister}, \citenamefont {Seweryniak},\ and\ \citenamefont {Zhu}}]{suman2021}%
  \BibitemOpen
  \bibfield  {author} {\bibinfo {author} {\bibfnamefont {S.}~\bibnamefont {Suman}}, \bibinfo {author} {\bibfnamefont {S.~K.}\ \bibnamefont {Tandel}}, \bibinfo {author} {\bibfnamefont {A.}~\bibnamefont {Kumawat}}, \bibinfo {author} {\bibfnamefont {S.~G.}\ \bibnamefont {Wahid}}, \bibinfo {author} {\bibfnamefont {M.}~\bibnamefont {Hemalatha}}, \bibinfo {author} {\bibfnamefont {P.}~\bibnamefont {Chowdhury}}, \bibinfo {author} {\bibfnamefont {R.~V.~F.}\ \bibnamefont {Janssens}}, \bibinfo {author} {\bibfnamefont {M.~P.}\ \bibnamefont {Carpenter}}, \bibinfo {author} {\bibfnamefont {T.~L.}\ \bibnamefont {Khoo}}, \bibinfo {author} {\bibfnamefont {F.~G.}\ \bibnamefont {Kondev}}, \bibinfo {author} {\bibfnamefont {T.}~\bibnamefont {Lauritsen}}, \bibinfo {author} {\bibfnamefont {C.~J.}\ \bibnamefont {Lister}}, \bibinfo {author} {\bibfnamefont {D.}~\bibnamefont {Seweryniak}},\ and\ \bibinfo {author} {\bibfnamefont {S.}~\bibnamefont {Zhu}},\ }\href {https://doi.org/10.1103/PhysRevC.103.014319} {\bibfield  {journal}
  {\bibinfo  {journal} {Phys. Rev. C}\ }\textbf {\bibinfo {volume} {103}},\ \bibinfo {pages} {014319} (\bibinfo {year} {2021})}\BibitemShut {NoStop}%
\bibitem [{\citenamefont {Shimizu}(2013)}]{shimizu2013}%
  \BibitemOpen
  \bibfield  {author} {\bibinfo {author} {\bibfnamefont {N.}~\bibnamefont {Shimizu}},\ }\href {https://arxiv.org/abs/1310.5431} {} (\bibinfo {year} {2013}),\ \Eprint {https://arxiv.org/abs/1310.5431} {arXiv:1310.5431 [nucl-th]} \BibitemShut {NoStop}%
\bibitem [{\citenamefont {Chen}\ \emph {et~al.}(2023)\citenamefont {Chen}, \citenamefont {Liu}, \citenamefont {Yuan}, \citenamefont {Chen}, \citenamefont {Shimizu}, \citenamefont {Sun}, \citenamefont {Xu},\ and\ \citenamefont {Tian}}]{chen2023}%
  \BibitemOpen
  \bibfield  {author} {\bibinfo {author} {\bibfnamefont {J.}~\bibnamefont {Chen}}, \bibinfo {author} {\bibfnamefont {M.}~\bibnamefont {Liu}}, \bibinfo {author} {\bibfnamefont {C.}~\bibnamefont {Yuan}}, \bibinfo {author} {\bibfnamefont {S.}~\bibnamefont {Chen}}, \bibinfo {author} {\bibfnamefont {N.}~\bibnamefont {Shimizu}}, \bibinfo {author} {\bibfnamefont {X.}~\bibnamefont {Sun}}, \bibinfo {author} {\bibfnamefont {R.}~\bibnamefont {Xu}},\ and\ \bibinfo {author} {\bibfnamefont {Y.}~\bibnamefont {Tian}},\ }\href {https://doi.org/10.1103/PhysRevC.107.054306} {\bibfield  {journal} {\bibinfo  {journal} {Phys. Rev. C}\ }\textbf {\bibinfo {volume} {107}},\ \bibinfo {pages} {054306} (\bibinfo {year} {2023})}\BibitemShut {NoStop}%
\bibitem [{\citenamefont {Sahoo}\ \emph {et~al.}(2024)\citenamefont {Sahoo}, \citenamefont {Srivastava}, \citenamefont {Shimizu},\ and\ \citenamefont {Utsuno}}]{Sahoo2024}%
  \BibitemOpen
  \bibfield  {author} {\bibinfo {author} {\bibfnamefont {S.}~\bibnamefont {Sahoo}}, \bibinfo {author} {\bibfnamefont {P.~C.}\ \bibnamefont {Srivastava}}, \bibinfo {author} {\bibfnamefont {N.}~\bibnamefont {Shimizu}},\ and\ \bibinfo {author} {\bibfnamefont {Y.}~\bibnamefont {Utsuno}},\ }\href {https://doi.org/10.1103/PhysRevC.110.024306} {\bibfield  {journal} {\bibinfo  {journal} {Phys. Rev. C}\ }\textbf {\bibinfo {volume} {110}},\ \bibinfo {pages} {024306} (\bibinfo {year} {2024})}\BibitemShut {NoStop}%
\bibitem [{\citenamefont {Toki}\ \emph {et~al.}(1979)\citenamefont {Toki}, \citenamefont {Yadav},\ and\ \citenamefont {Faessler}}]{toki1979}%
  \BibitemOpen
  \bibfield  {author} {\bibinfo {author} {\bibfnamefont {H.}~\bibnamefont {Toki}}, \bibinfo {author} {\bibfnamefont {H.~L.}\ \bibnamefont {Yadav}},\ and\ \bibinfo {author} {\bibfnamefont {A.}~\bibnamefont {Faessler}},\ }\href {https://doi.org/10.1007/BF01434634} {\bibfield  {journal} {\bibinfo  {journal} {Zeitschrift f{\"u}r Physik A Atoms and Nuclei}\ }\textbf {\bibinfo {volume} {292}},\ \bibinfo {pages} {79} (\bibinfo {year} {1979})}\BibitemShut {NoStop}%
\end{thebibliography}
